\documentclass[
    reprint,
    amsmath,
    amssymb,
    prstab,
    showkeys,
    nofootinbib,
    longbibliography]{revtex4-2}

\usepackage{hyperref}
\usepackage[english]{babel}
\usepackage[separate-uncertainty=true]{siunitx}
\usepackage{float}
\usepackage{graphicx}
\usepackage[justification=raggedright,singlelinecheck=false]{caption}
\usepackage{mathtools}

\usepackage{tikz}
\usetikzlibrary{patterns}
\usetikzlibrary{arrows}
\usetikzlibrary{plotmarks}
\usetikzlibrary{spy}
\usepackage{pgfplots}
\pgfplotsset{compat=newest}
\usepackage{pgfplotstable}
\usepackage{tuda-pgfplots}
\usepackage{multirow}
\usepackage{placeins}
\usepackage{orcidlink}

\begin{document}

\title{Cs-O$_2$-Li as enhanced NEA surface layer with increased lifetime for GaAs photocathodes}

\author{Maximilian Herbert\orcidlink{https://orcid.org/0000-0003-4103-6547}}
\email{mherbert@ikp.tu-darmstadt.de}

\author{Tobias Eggert}
\author{Joachim Enders\orcidlink{https://orcid.org/0000-0002-8441-378X}}
\author{Markus Engart\orcidlink{https://orcid.org/0000-0004-6081-576X}}
\author{Yuliya Fritzsche}
\author{Maximilian Meier\orcidlink{https://orcid.org/0009-0005-4021-1720}}
\author{Julian Schulze\orcidlink{https://orcid.org/0009-0005-5978-4043}}
\author{Vincent Wende\orcidlink{https://orcid.org/0009-0003-6082-3949}}

\affiliation{Institut f\"ur Kernphysik, Fachbereich Physik, Technische
Universit\"at Darmstadt, Schlossgartenstrasse 9, 64289 Darmstadt, Germany}

\begin{abstract}
GaAs-based photocathodes are the only viable source capable of providing spin-polarized electrons for accelerator applications. This type of photocathode requires a thin surface layer, in order to achieve negative electron affinity (NEA) for efficient photo-emission. However, this layer is vulnerable to environmental and operational effects, leading to a decay of the quantum efficiency $\eta$ characterized by a decay constant or lifetime $\tau$. In order to increase $\tau$, additional agents can be introduced during the activation procedure to improve the chemical robustness of the surface layer. This paper presents the results of recent research on Li as enhancement agent for photocathode activation using Cs and $\text{O}_2$, forming Cs-O$_2$-Li as enhanced NEA layer. Measurements yielded an increase in lifetime by a factor of up to \num{19(2)} and an increase in extracted charge by a factor of up to \num{16.5(2.4)}, without significant reduction of $\eta$. This performance is equal to or better than that reported for other enhanced NEA layers so far.
\end{abstract}

\maketitle

\section{Introduction} \label{sec:introduction}
Spin-polarized electron sources are in high demand for a multitude of advanced physics applications such as energy recovery linacs (ERL, \cite{sinclair2006, rao2006, heine2021}), colliders \cite{brachmann2007, moortgat-pick2008, skaritka2018}, positron production \cite{abbott2016, cardman2018}, electron cooling at high-energy storage rings \cite{orlov2009}, and electron microscopy \cite{vollmer2003, suzuki2010, kuwahara2012}. Such applications require high-intensity electron beams. \\
The best available source for high currents of highly spin-polarized electron beams are photo-electron sources using GaAs-based photocathodes \cite{kessler1985}. The type III-V compound semiconductor material is usually p-doped, hence the notation p-GaAs. The spin of electrons in the conduction band of a nonmagnetic solid like GaAs can be polarized by optical pumping with circularly polarized light, the angular momentum of which sets the direction of the electron spin in the excitation process. The maximum spin-polarization is determined by the selection rules for transitions possible at the energy of the incident circularly-polarized photons. At the valence-band maximum and conduction-band minimum, GaAs has a $p$ band that is sixfold degenerate and split through spin-orbit coupling into a higher fourfold degenerate $P_{3/2}$ level and a lower twofold degenerate $P_{1/2}$ level. By excitation at photon energies close to the band gap, such that only states of the $P_{3/2}$ level are excited, a maximum spin-polarization of $\mathcal{P} =$~\num{0.5} (or \qty{50}{\percent}) is possible for bulk-GaAs \cite{pierce+meier1976}. \\
An increase in spin-polarization can be achieved by constructing the photocathode from alternating layers of two substrates with different lattice constants and different band gaps, creating a so-called strained superlattice~(SL). This splits the $P_{3/2}$ level into two twofold degenerate sub-levels that can be selectively excited by choosing the corresponding photon energy or wavelength, hence enabling a maximum theoretical spin-polarization of up to \qty{100}{\percent}. GaAs/GaAsP strained superlattice photocathodes have so far shown the best performance, providing spin-polarization of up to \qty{92}{\percent} at an incident photon wavelength between \qty{776}{\nano\meter} and \qty{785}{\nano\meter} \cite{maruyama2004, nishitani2005, jin2014, liu2016}. \\
Efficient photoemission at photon energies close to the band gap is not possible in GaAs due to an energy barrier, also called electron affinity, at the surface that excited electrons need to overcome in order to be emitted. This problem can be addressed by applying a thin layer, consisting of Cs in combination with an oxidant such as $\mathrm{O}_2$ or $\mathrm{NF}_3$, to the surface of the material in order to introduce a band-bending effect. Since the Fermi level of the surface layer is aligned to the Fermi level of GaAs, the energy levels at the surface line up such that the vacuum level of the layer is lowered below the conduction-band minimum of GaAs, hence reducing the work function and creating negative electron affinity (NEA) that allows electrons from the conduction-band minimum to be emitted from the material \cite{sonnenberg1969a, spicer+bell1972}. The surface layer therefore greatly influences the quantum efficiency $\eta$ of the photo-emission from the cathode, which is proportional to the current density $J_\mathrm{e}$ of the emitted electrons divided by the intensity $\mathcal{I}_\gamma$ and wavelength $\lambda$ of the incident photons \cite{jensen2006}
\begin{equation}
    \eta(\lambda) = \frac{hc}{e} \cdot \frac{J_\mathrm{e}}{\mathcal{I}_\gamma \cdot \lambda} \quad .
    \label{eq:qe_lambda}
\end{equation}
In practice, the average current of emitted electrons $I_\text{p}$ and average power of incident light $P_\gamma$ are measured when working with photo-electron sources. By canceling out the areas in $J_\text{e}$ and $I_\gamma$ and writing $\frac{hc}{e}$ in practical units, the equation can be expressed as
\begin{equation}
    \eta \approx \qty{123.98}{\nano\meter\milli\watt\percent\per\micro\ampere} \cdot \frac{I_\text{p}}{\lambda\cdot P_\gamma} \quad .
\end{equation}
However, the NEA surface layer is prone to degrading effects during storage and operation, such as gas desorption \cite{uebbing1970, chanlek2014} and ion back-bombardment (IBB \cite{aulenbacher1997, siggins2001}). Due to these effects, $\eta$ drops over time. This behavior can be approximated by a single-exponential decay that is characterized by a decay constant $\tau$, also called the lifetime of the photocathode \cite{ciccacci+chiaia1991}
\begin{equation}
    \eta(t) \approx \eta(0) \cdot e^{-\frac{t}{\tau}} \quad .
\end{equation}
For operation in the high-current beam regime that is of major interest for advanced accelerator applications, cathode lifetime is severely limited by IBB \cite{dunham2013} and heat-induced desorption \cite{aulenbacher2003}. \\
A multitude of measures have been implemented to increase the lifetime of the photocathode. The fabrication of strained superlattice photocathodes with an integrated Distributed Bragg Reflector (DBR) can mitigate the significant decrease of $\eta$ due to the SL structure, yielding quantum efficiencies comparable to bulk-GaAs and hence prolonging effective cathode usability by reducing the required laserpower and the associated heat deposition on the photocathode \cite{liu2016}. Unwanted electron emission outside of the area illuminated by the laser beam can be eliminated by anodization of the photocathode \cite{schwartz1976, grames2005, hernandez-garcia2005, grames2011} and use of a mask during activation \cite{aulenbacher2003, grames2011} in order to reduce the active area of the photocathode. Elimination of field emission from the parts of the gun on high-voltage potential is achieved by gun designs with low field-gradients \cite{palacios-serrano2018, foerster2022} and high-voltage conditiong during gun comissioning \cite{siggins2001, hernandez-garcia2009}. Improving the vacuum conditions inside the gun chamber reduces gas desorption and IBB, hence prolonging lifetime \cite{sinclair2007}. Since ions are accelerated towards the electrostatic center, moving the laser spot away from it will reduce the amount of ions that are accelerated directly towards the point of emission, hence reducing damage to the surface layer from IBB there \cite{grames2011}. Increasing the laser spot size on the cathode distributes the back-accelerated ions over a wider area, reducing the damage caused to the surface layer \cite{grames2011}. IBB can be further reduced by applying a positive bias to the anode \cite{pozdeyev2007, grames2008, yoskowitz2020, yoskowitz2024}. \\
While all of the listed methods have been successfully employed during low-current operations, the overall lifetime increment is not sufficient to achieve prolonged operation at high beam currents \cite{dunham2013} and some methods suffer from trade-offs such as reduced emittance \cite{grames2011}. Hence, additional methods to increase photocathode lifetime are needed. Introducing a geometric offset of the anode has been proposed to reduce IBB \cite{rahman2019a}. Also, an implementation of active photocathode cooling has been demonstrated to reduce thermal desorption at high laser power, although no lifetime experiments have been conducted with this setup as yet \cite{wang2022}. \\
Another way of prolonging photocathode lifetime independent of environmental and operational conditions is to increase the robustness of the NEA layer itself. It has been demonstrated that this can be achieved by introducing an additional agent, such as Sb \cite{cultrera2020}, Te \cite{bae2018}, or Li \cite{mulhollan+bierman2008}, during the activation process. The effects of adding Sb or Te to a Cs layer on a p-GaAs surface were first investigated in 1969, showing an increase in both quantum efficiency and stability when compared to a pure Cs layer \cite{hagino+nishida1969}. The increase in quantum efficiency was attributed to the creation of NEA conditions through the same band-bending effect caused by a Cs-O$_2$ surface layer \cite{sonnenberg1969b, sugiyama2011}. Further experimental studies confirmed this for both Cs-Sb \cite{zhao1993} and Cs-Te \cite{uchida2014, kuriki2015}, and in comparison to Cs-O$_2$ an increase in $\tau$ without significant change in spin-polarization, albeit at the expense of reduced $\eta$, has been reported for both surface layers \cite{bae2018, bae2019}. Recent studies have focused on the addition of O$_2$ to form Cs-O$_2$-Sb \cite{bae2019, cultrera2020, bae2020} and Cs-O$_2$-Te \cite{rahman2019b, biswas2021} layers, i.e. using Sb and Te as enhancement agents for Cs-O$_2$ surface layers. This attenuates the reduction in $\eta$ while still increasing $\tau$, without significant reduction of spin-polarization, although a dependence of $\mathcal{P}$ on the surface layer thickness has been established. \\
It has been discovered that the addition of a second alkali metal to Cs-$\mathrm{NF}_3$ activation has a similar beneficial effect of increasing $\tau$, with Li performing best \cite{mulhollan+bierman2008} without significant impact on spin-polarization \cite{mulhollan2010}. Li-enhancement has also been successfully established for Cs-O$_2$, yielding comparable results \cite{mulhollan2010, kurichiyanil2019}. Similar to enhancement with Sb or Te, an influence on $\eta$ appears to exist for Cs-O$_2$-Li and Cs-$\mathrm{NF}_3$-Li. However, reported results vary from significant increase \cite{mulhollan+bierman2008, herbert2020} to significant decrease \cite{mulhollan2010, kurichiyanil2019} in quantum efficiency. A successful combination of different enhancement agents has also been demonstrated, with a Cs-K-Te surface layer showing an impressive increase in photocathode lifetime, albeit at the cost of greatly reduced $\eta$ \cite{kuriki+masaki2019}. \\
Recent studies have successfully tested Cs-$\mathrm{NF}_3$-Li \cite{herbert2020} and Cs-O$_2$-Sb \cite{bae2022a} surface layers on bulk-GaAs inside DC high-voltage photo-electron guns, demonstrating that enhanced surface layers can be used under operational conditions. However, more research is needed in order to establish this method as reliable way of increasing GaAs photocathode lifetime during high-current operation. The theory behind the increased robustness of enhanced surface layers is not well understood. Also, the lifetime increase must be carefully studied in relation to reduction in $\eta$ to find the right balance between the two effects, else the required increase in laser power would negate the enhanced robustness of the NEA layer. \\
In this work, we present a comparative study on Cs-O$_2$ and Cs-O$_2$-Li surface layers carried out at TU Darmstadt's setup for Photo-Cathode Activation, Testing, and Cleaning using atomic Hydrogen (\mbox{Photo-CATCH}). The manuscript reviews theoretical considerations from literature and is aimed at establishing a well-defined, reproducible procedure for Li-enhanced activation. 

\section{Photoemission from NEA Gallium Arsenide} \label{sec:theory}

\subsection{Three-step model}
The photoemission process from a GaAs semiconductor can be described using the phenomenological three-step model \cite{spicer1958}. It depicts the process as consisting of three independent steps \cite{spicer+herrera-gomez1993}:
\begin{itemize}
    \item[1.] \textbf{Photo-excitation}\\
    This step describes the excitation of electrons within the material from the valence band into the conduction band. The associated probability $P_\text{ex}$ of electron excitation to an energy above the threshold energy $E_\text{th}$ required for emission depends on the reflectance $R(\lambda)$ and the photon absorption length $l_\text{a}(\lambda)$ of the material, with the latter being defined as the depth at which the incident intensity has been reduced to a fraction of $1/e$ of its unreflected value at the surface. Both parameters depend on the wavelength $\lambda$ of the incident light.
    \item[2.] \textbf{Transport}\\
    The energy of excited electrons is reduced during transport to the surface, mainly by scattering processes such as electron-phonon scattering, electron-electron scattering and scattering due to plasmon creation. The probability $P_\text{tr}$ associated with this energy loss during transport is characterized by the escape depth or average electron-electron scattering length $\overline{l}_{e-e}(\lambda)$, defined as the distance after which an exited electron retains at least a fraction of $1/e$ of $E_\text{th}$.
    \item[3.] \textbf{Emission}\\
    An excited electron that reaches the surface with sufficient energy is emitted by tunneling through the surface barrier. This tunneling process is characterized by a finite probability $P_\text{em}$ depending on the wavelength of the incident light. $P_\text{em}$ increases with decreasing wavelength and typically does not exceed \qty{50}{\percent}.
\end{itemize}
Hence, the quantum efficiency of the process is proportional to the product of probabilities:
\begin{equation}
    \eta(\lambda) \propto P_\text{ex}(\lambda) \cdot P_\text{tr}(\lambda) \cdot P_\text{em}(\lambda) \quad .
    \label{eq:three-step}
\end{equation}

\subsection{Photoemission conditions}
The spin-polarization $\mathcal{P}$ of an ensemble of photo-emitted electrons is given as \cite{pierce+meier1976}
\begin{equation}
    \mathcal{P} = \frac{N_\uparrow - N_\downarrow}{N_\uparrow + N_\downarrow} \quad ,
\end{equation}
where $N_\uparrow$ and $N_\downarrow$ are the numbers of electrons with spin parallel and anti-parallel to a preferred orientation. This corresponds to a spin projection $s_z$ of $+ \frac{1}{2}$ and $- \frac{1}{2}$, respectively. Using optical pumping with circularly-polarized light \cite{lampel1968, parsons1969}, electrons in the valence band are photo-excited and transition into the conduction band, with the orientation of the magnetic moment depending on the final state of the electron after transition. 
In order to ensure a pure $P_{3/2} \rightarrow S_{1/2}$ transition for polarized-electron emission,
\begin{equation}
    E_\gamma \leq E_\mathrm{g} + \Delta E_\mathrm{so} = \qty{1.76}{\electronvolt} \quad \text{or} \quad \lambda \geq \qty{705}{\nano\meter}
    \label{eq:spin-pol-cond-1}
\end{equation}
must hold with the energy of incident photons $E_\gamma$, the GaAs band gap energy at room temperature $E_\mathrm{g} =$~\qty{1.42}{\electronvolt}, and the spin-orbit split $\Delta E_\mathrm{so} =$~\qty{0.34}{\electronvolt} \cite{cardona1988} between the $P_{3/2}$ and $P_{1/2}$ levels. \\
The SL structure of GaAs/GaAsP (or similar compounds) separates the four-fold degenerate $P_{3/2}$ level into two two-fold degenerate sub-levels with an energy split $\Delta E_\mathrm{sl}$, also increasing the band gap. For GaAs/GaAsP, values of $E_\mathrm{g, sl} =$~\qty{1.58}{\electronvolt} and $\Delta E_\mathrm{sl} =$~\qty{82}{\milli\electronvolt} have been reported \cite{maruyama2004}. Hence, one chooses
\begin{equation}
    E_\gamma \leq E_\mathrm{g, sl} + \Delta E_\mathrm{sl} = \qty{1.66}{\electronvolt} \quad \text{or} \quad \lambda \geq \qty{750}{\nano\meter} \quad .
    \label{eq:spin-pol-cond-2}
\end{equation}\\
After excitation, the electron must still possess enough energy to be transported to the surface and be emitted into the vacuum level $E_\text{vac}$ \cite{spicer+herrera-gomez1993}. Neglecting energy losses from transport, the threshold energy $E_\mathrm{th}$ for effective photoemission can then be given as \cite{soboleva1974}
\begin{equation}
    E_\mathrm{th} = \phi + E_\mathrm{fv} = E_\mathrm{a} + E_\mathrm{g} \quad ,
    \label{eq:thresh1}
\end{equation}
with the work function $\phi$, the gap $E_\mathrm{fv}$ between Fermi level $E_\text{F}$ and valence band maximum $E_\text{vbm}$, and the electron affinity $E_\mathrm{a}$ which is the difference between $E_\text{vac}$ and the conduction band minimum $E_\text{cbm}$, i.e. the height of the energy barrier at the surface. For bulk-GaAs with $E_\mathrm{g} =$~\qty{1.42}{\electronvolt} and $E_\mathrm{a} =$~\qty{4.1}{\electronvolt}, this gives the condition
\begin{equation}
    E_\gamma \geq E_\mathrm{th} = \qty{5.52}{\electronvolt} \quad \text{or} \quad \lambda \leq \qty{225}{\nano\meter} \quad ,
\end{equation}
which obviously contradicts the condition for spin-polarized emission (Eq.\ (\ref{eq:spin-pol-cond-1})). It is therefore necessary to reduce $E_\mathrm{a}$, which is equivalent to lowering the work function $\phi$. This can be done by p-doping the GaAs crystal and by adsorbing a material as surface layer such that $\phi$ is reduced sufficiently to enable photoemission of spin-polarized electrons.

\subsection{GaAs p-doping}
By heavily p-doping the GaAs crystal, $E_\mathrm{F}$ is shifted from its mid-gap position towards the valence band energy $E_\text{vbm}$ within the bulk of the crystal. Defects at the surface create surface states that exchange electrons with impurities within the bulk, creating a positively charged surface and causing the formation of a space-charge region close to the surface \cite{chubenko2021}. This causes downward bending of both valence and conduction band, pinning the Fermi level at the surface to a different value relative to the valence band maximum \cite{scheer+laar1969}. Hence, $E_\mathrm{fv}$ has different values within the bulk and at the surface, $E_\mathrm{fv,b}$ and $E_\mathrm{fv,s}$, with $E_\mathrm{fv,b} \approx 0$. The magnitude of the band bending caused by p-doping is then \mbox{$\Delta E_\text{bb,p} = E_\mathrm{fv,s} - E_\mathrm{fv,b} \approx E_\mathrm{fv,s}$}. \\
The reduced work function for p-doped semiconductors $\phi_p$ can be calculated by \cite{kampen1998}
\begin{equation}
    \phi_\text{p} = E_\mathrm{ion} - \Delta E_\mathrm{bb,p} - E_\mathrm{fv,b} \quad ,
\end{equation}
with the ionization energy $E_\mathrm{ion} = E_\text{vac} - E_\text{vbm} = E_\text{g} + E_\text{a}$. Using $E_\text{fv,b} \approx 0$ leads to
\begin{equation}
    \phi_\text{p} = E_\text{ion} - \Delta E_\text{bb,p} \quad .
    \label{eq:wf_p-GaAs}
\end{equation}
For clean p-GaAs surfaces, $\Delta E_\mathrm{bb,p} =$~\qty{0.3}{\electronvolt} has been reported \cite{alperovich1995}\footnote{In our convention, we use a positive sign for band bending that reduces the band energies in order to meet the sign convention of Eq.\ (\ref{eq:wf_p-GaAs}).}. Together with $E_\text{ion,p} = \qty{5.52}{\electronvolt}$, this gives $\phi_\text{p} = \qty{5.22}{\electronvolt}$, in good agreement with the reported experimental value of $\phi = \qty{5.2(1)}{\electronvolt}$ for Zn-doped p-type GaAs(110) with a carrier concentration of \qty{3e19}{\per\cubic\centi\meter} \cite{madey+yates1971}. While p-doping reduces $\phi$, it is on its own not sufficient to enable polarized electron emission. 

\subsection{Surface-layer adsorption}
If two materials form an interface, their band structures will align at the boundary surface such that the Fermi levels $E_\mathrm{F}$ are the same \cite{milton+baer1971}. Using \mbox{$E_\text{fv,s} \approx E_\text{vbm,s} - E_\text{vbm,b}$}, the work function of highly doped GaAs with a thin surface layer can be written as
\begin{equation}
    \phi = E_\text{cbm,b} - E_\text{vbm,s} - E_\text{fv,s} \approx E_\text{g} \quad .
    \label{eq:workfunc}
\end{equation}
By choosing the surface layer such that $E_\text{vac}$ lies below $E_\text{cbm,b}$, the difference between these two values, also called effective electron affinity $E_\text{a,eff}$, becomes negative. The created condition is therefore called negative electron affinity. For an interface between GaAs as substrate and a surface layer material meeting this criterion, the constraint for NEA can then be formulated as
\begin{equation}
    \phi = E_\text{g} + E_\text{a,eff} \leq E_\text{g}
    \label{eq:nea-cond}
\end{equation}
with $E_\text{a,eff} \leq 0$. Hence, the NEA condition allows excitation of photo-electrons with $E_\gamma \approx E_\mathrm{g}$ for direct band-gap semiconductors with an appropriate surface layer material, fulfilling the condition for polarized electron emission. \\
One may assume that materials may be chosen as possible NEA surface-layer candidates by evaluating their work functions with Eq.\ (\ref{eq:nea-cond}). However, this is not applicable in practice because the work function of a surface layer on GaAs (or semiconductors in general) differs from the bulk work function of the surface layer material, as has been observed for Cs \cite{burt+heine1978}. This has been attributed to the confinement of the surface layer to a thickness of only one or a few atoms. Also, for compounds involving two or more elements, the structure of the layer formed upon adsorption on the surface differs from the bulk structure of the compound. For example, Cs and O$_2$ form a layer of dipoles instead of adsorbing as Cs$_2$O, which would be the expected bulk structure \cite{biswas2020}. Instead, one needs to consider the final work function of the activated cathode. Hence, the constraint for the final work function of a GaAs photocathode covered with an NEA layer is
\begin{equation}
    \phi \leq E_\mathrm{g} =
    \begin{cases}
    \qty{1.42}{\electronvolt} \quad \text{bulk} \\
    \qty{1.58}{\electronvolt} \quad \text{SL}
    \end{cases}
    \quad .
    \label{eq:nea-crit-1}
\end{equation}
This criterion for NEA surface layers on GaAs has been previously stated by H. Sonnenberg for bulk-GaAs \cite{sonnenberg1970}. \\
The change in work function $\Delta\phi_\text{layer}$ induced by adsorption of a surface layer can be calculated from the change in ionization energy $\Delta E_\mathrm{ion}$ and the band bending at the interface \cite{kampen1998}
\begin{equation}
    \Delta\phi_\text{layer} = \Delta E_\mathrm{ion} - \Delta E_\mathrm{bb} \quad .
\end{equation}
Using the dipole model, $\Delta E_\mathrm{ion}$ can be calculated for dipoles adsorbed on a surface using the model of J. Topping \cite{topping+chapman1927}. Adapting this approach, $\Delta E_\mathrm{ion}$ can be considered as a change in work function $\Delta\phi_\mathrm{dip}$ introduced by the adsorbed dipoles \cite{clemens1978}
\begin{equation}
    \Delta E_\mathrm{ion} = \Delta\phi_\mathrm{dip} = \pm\frac{e}{\epsilon_0}\cdot\frac{\mu\cos{(\vartheta)}\sigma\delta}{1+\kappa\alpha(\sigma\delta)^{(3/2)}} \quad ,
    \label{eq:e_ion_change_1}
\end{equation}
with the bond dipole moment $\mu$, the angle between dipole and surface normal $\vartheta$, the area density of surface dipoles $\sigma$, the relative layer thickness $\delta$, the geometry constant $\kappa$, and the adatom polarizability $\alpha$. \\
The bond dipole moment is given as $\mu = \Delta q \cdot e \cdot d$, with the dipole length equal to the sum of covalent radii of atoms involved in the dipole $d = \sum_i r_{\mathrm{cov},i}$ \cite{kampen+moench1992}, and the ionic character of the bond $\Delta q$, given as \cite{hannay1946}
\begin{equation}
    \Delta q = 0.16 \cdot \left| \chi_1 - \chi_2 \right| + 0.035 \cdot \left| \chi_1 - \chi_2 \right|^2 \quad .
\end{equation}
Here, $\chi_1$ and $\chi_2$ are the electronegativities on the Pauling scale of the substrate and the adatom, respectively. The sign of Eq.\ (\ref{eq:e_ion_change_1}) depends on the relation of $\chi_1$ to $\chi_2$: positive for $\chi_1 < \chi_2$ and negative for $\chi_1 > \chi_2$. This represents the orientation of the adatom-induced dipoles. \\
For adsorption on the GaAs(110) surface, $\vartheta$ is approximated as the angle of the dangling bonds of GaAs with respect to the surface normal: $\vartheta = \qty{90}{\degree} - \frac{\varphi_\text{t}}{2} = \qty{35.26}{\degree}$, where $\varphi_\text{t} = \qty{109.47}{\degree}$ is the tetrahedal angle \cite{kampen+moench1992}. However, this value may change for different surface-layer compositions due to geometrical constraints. A relative thickness of $\delta = 1$ corresponds to the complete coverage of the substrate with one monolayer of adatoms. Depending on the geometry of the surface layer, $\kappa$ ranges from \num{9.03} for a simple square to \num{8.89} for an equitriangular layout. \\
For adatoms with small polarizability, depolarizing interactions between surface dipoles may be neglected, thus reducing Eq.\ (\ref{eq:e_ion_change_1}) to \cite{kampen+moench1992}
\begin{equation}
    \Delta\phi_\text{dip} = \pm\frac{e}{\epsilon_0}\cdot\mu\cos{(\vartheta)}\sigma\delta \quad .
    \label{eq:e_ion_change_2}
\end{equation}
Hence, the final work function of a p-GaAs photocathode with a surface layer can be calculated as
\begin{equation}
    \phi = \phi_\text{p} + \Delta\phi_\text{layer} = \phi_\text{p} + \Delta\phi_\text{dip} - \Delta E_\mathrm{bb} \quad .
\end{equation}
In order to meet NEA requirements, i.e. $\phi \leq E_\text{g}$, any surface layer must therefore fulfill the criterion
\begin{equation}
    \begin{split}
        \Delta\phi_\text{layer} &\leq E_\text{g} - \phi_\text{p} \\
        \text{or} \quad \Delta\phi_\text{layer} &\leq
        \begin{cases}
            -\qty{3.80}{\electronvolt} \quad \text{(bulk)} \\
            -\qty{3.64}{\electronvolt} \quad \text{(SL)}
        \end{cases}
        \quad .
    \end{split}
    \label{eq:nea-crit-2}
\end{equation}
For Cs on p-GaAs(110) with a carrier concentration of \qty{3e19}{\per\cubic\centi\meter}, a final work function of $\phi = \qty{1.2(1)}{\electronvolt}$ has been reported \cite{madey+yates1971}, corresponding to a work function reduction of $\Delta\phi_\text{Cs} = -\qty{4.0(2)}{\electronvolt}$. Hence, a Cs surface layer on p-GaAs already creates an NEA condition, albeit with a low quantum efficiency. \\
Combining Eq.\ (\ref{eq:nea-crit-1}) with the constraints for spin-polarized emission, Eqs.\ (\ref{eq:spin-pol-cond-1}) and (\ref{eq:spin-pol-cond-2}), the ranges for optimal excitation of spin-polarized electrons are given by
\begin{equation}
    \begin{split}
        \qty{705}{\nano\meter} &\leq \lambda \leq \qty{873}{\nano\meter} \quad \text{(bulk)}\\
        \text{and} \quad \qty{750}{\nano\meter} &\leq \lambda \leq \qty{785}{\nano\meter} \quad \text{(SL)} \quad .
    \end{split}
    \label{eq:lambda-ranges}
\end{equation}
These ranges have been confirmed experimentally. For bulk-GaAs, a significant drop in quantum efficiency occurs above \qty{870}{\nano\meter} (cf. e.g. \cite[Fig.\ 4]{cultrera2020}), while $\mathcal{P}$ displays a significant increase above \qty{700}{\nano\meter} (cf. e.g. \cite[Fig.\ 5]{cultrera2020}). A significant rise in $\mathcal{P}$ can be observed for GaAs/GaAsP starting at \qty{750}{\nano\meter}, while a major drop in quantum efficiency is observed at about \qty{785}{\nano\meter} (cf. e.g. \cite[Fig.\ 7]{bae2020}). The experimental results show that these ranges do not represent hard limits, but are smeared out. However, $\mathcal{P}$ is reduced significantly at lower wavelengths. GaAs/GaAsP photocathodes produce a degree of spin-polarization comparable to bulk-GaAs when operated below \qty{750}{\nano\meter}, with both photocathode types sharing a hard lower limit at about \qty{600}{\nano\meter}. At higher wavelengths, $\eta$ is impacted severely, dropping by about two orders of magnitude between \qty{850}{\nano\meter} and \qty{950}{\nano\meter} for bulk-GaAs \cite{cultrera2020} and almost one order of magnitude between \qty{780}{\nano\meter} and \qty{810}{\nano\meter} for GaAs/GaAsP \cite{bae2020}. Nevertheless, SL photocathodes are sometimes operated at higher wavelengths in order to obtain maximum $\mathcal{P}$. \\

\subsection{Optical and emission properties}
Using Eq.\ (\ref{eq:three-step}) from the three-step model, one can express the quantum efficiency $\eta$ as a function of $\lambda$ and $\phi_\text{eff} = \phi - \phi_\text{Schottky}$ \cite{dowell2006}
\begin{equation}
\begin{split}
    \eta(\lambda, \phi_\text{eff}) &\approx \frac{(1 - R(\lambda))}{\left(1+\frac{l_\text{a}(\lambda)}{\overline{l}_{e-e}(\lambda)}\right)} \cdot \frac{(E_\text{F}+E_\gamma(\lambda))}{2E_\gamma(\lambda)} \\
    &\cdot \left(1-\sqrt{\frac{(E_\text{F}+\phi_\text{eff})}{(E_\text{F}+E_\gamma(\lambda))}}\right)^2 \quad .
    \label{eq:qe-wf}
\end{split}
\end{equation}
The defining parameters $R(\lambda)$, $l_\text{a}(\lambda)$, and $\overline{l}_{e-e}(\lambda)$ are also influenced by the surface layer. Since the adatoms cover the GaAs surface, the reflectance of the photocathode is altered. Ideally, the surface layer should possess a reflectance that is equal to or lower than that of the clean GaAs surface, that is $R \approx \num{0.33}$ in the range of \qtyrange{1.42}{1.76}{\electronvolt} \cite{akinlami2013}. The same value can be assumed for GaAs/GaAs-P since the top layer of the superlattice usually is GaAs \cite{liu2016}. \\
The thickness $d$ of the surface layer should be chosen such that $l_\text{a,layer} \gg d$ and $\overline{l}_{e-e,layer} \gg d$. Else a considerable part of the emission process takes place within the surface layer instead of the substrate, which is undesirable since it would significantly reduce the spin-polarization of the emitted electrons. The same conditions should be observed for the parameters of the substrate, i.e. $l_\text{a,subst} \gg d$ and $\overline{l}_{e-e,subst} \gg d$, to make sure that the photoemission process is not hampered significantly by the surface layer. For this reason, the surface layer is usually no thicker than a few atomic layers, i.e. a few hundred picometers, compared to $l_\text{a}$ and $\overline{l}_{e-e}$ that are in the range of a few hundred nanometers for GaAs, e.g. $l_\text{a} \approx \qty{630}{\nano\meter}$ at $E_\gamma = \qty{1.58}{\electronvolt}$ \cite{chubenko2021} and $\overline{l}_{e-e} = \qty{117.5}{\nano\meter}$ at $E_\gamma = \qty{1.6}{\electronvolt}$ \cite{yang1992}. \\
Equation (\ref{eq:qe-wf}) also shows that it is desirable to reduce the work function below the criterion stated in Eq.\ (\ref{eq:nea-crit-2}) in order to increase the number of electrons that retain sufficient energy to be emitted upon arrival at the photocathode surface at incident photon energies close to $E_\text{g}$, effectively increasing the quantum efficiency. The addition of O$_2$ to form a Cs-O$_2$ surface layer reduces the work function further, with final work functions reported in the range of \qtyrange{0.7}{1.0}{\electronvolt} \cite{uebbing+james1970, su1983}, depending on surface layer thickness, corresponding to $\Delta\phi_\text{Cs-O$_2$}$ in the range of \qtyrange{-4.2}{-4.5}{\electronvolt}.

\subsection{Structural and chemical aspects}
As is evident from Eq.\ (\ref{eq:e_ion_change_1}), the work function reduction induced by the surface layer depends on the alignment of the bonds as well as the chemical characteristics of the dipoles on the surface. Hence, it is important to consider the structural aspects given by the crystal surface and the dipole layer. Also, chemical aspects, most importantly the reactivity of the surface dipoles in regard to residual gas particles, need to be taken into consideration when evaluating the quality of a surface layer. \\
GaAs forms a Zincblende crystal structure, with the GaAs(110) surface containing equal numbers of Ga and As atoms. While each atom within the bulk possesses four bonds, the surface atoms only possess three. This leads to a relaxation of the surface. Since As has a higher electronegativity, charge is transferred from Ga to As, leading to a so-called dangling electron pair on each surface As atom \cite{gregory1974}. This creates a distortion of the crystal structure at the surface and in the layers directly below it: As atoms are raised, while Ga atoms are lowered relative to the plane they would inhibit within the bulk structure \cite{kahn1978}, forming dipoles on the surface that increase the ionization energy and hence the work function of the surface \cite{kampen+moench1992}. This effect is also referred to as buckling. Depending on the adsorbed atom species and the overall chemical composition of the surface layer, this relaxation may be eliminated, further reducing the work function. It has been reported that the As surface states associated with As dangling electrons are removed by adsorption of Cs on the surface \cite{manghi1984}, hinting at a possible elimination of the surface relaxation by a Cs adatom layer. A similar surface relaxation effect has been reported for the more commonly used GaAs(100) surface \cite{schailey+ray1999}, although the description of this surface is more complicated due to the occurrence of extensive surface reconstruction \cite{feenstra+stroscio1993}. \\
The exact nature of the NEA surface layer structure is still subject of ongoing research. Proposed models include a heterojunction between the substrate and a thick surface layer \cite{sonnenberg1969a}, single or multiple layers of dipoles \cite{fisher1972, su1983}, or the formation of clusters on the GaAs surface \cite{burt+heine1978}. Experimental studies of GaAs activated with a Cs-O$_2$ layer report that $\eta$ reaches its maximum for an adatom coverage of about two monolayers, suggesting that the dipole layer model is valid at a coverage of a few monolayers, while the heterojunction model is valid for a coverage above four monolayers of adatoms \cite{bakin2015}. \\
Synchrotron radiation photoemission studies have been conducted for Cs-NF$_3$-Li on Zn-doped p-type GaAs(100), suggesting an increase in chemical resistivity against contamination due to the interaction between Li and NF$_3$ \cite{sun2009}. This interaction is proposed to influence the orientation of the NF$_3$ molecules adsorbed on the surface, with Li adsorbing in the Cs adatom layer next to the Cs atoms, and NF$_3$ forming a layer in between the layers of Cs and Li adatoms. \\
Su et al.\ \cite{su1983} showed that O$_2$ contributes to the work function reduction by adsorbing directly to the GaAs surface, forming O-GaAs bonds, and by forming Cs-O-Cs. Since the covalent radius of O is only \qty{63}{\pico\meter}, it can fit in the gaps between the Cs adatoms with a much larger covalent radius of \qty{232}{\pico\meter}. F most likely adsorbs in a similar manner since its covalent radius of \qty{64}{\pico\meter} is almost identical to that of O. The covalent radii of Li, Sb and Te are \qty{133}{\pico\meter}, \qty{140}{\pico\meter} and \qty{136}{\pico\meter}, about \qty{60}{\percent} of the Cs covalent radius and not small enough to fit into the gaps between Cs. However, these atoms may then adsorb on top of these gaps and hence shield the O or F atoms positioned there from residual gas adsorption and IBB. \\
The chemical resistivity of both Cs-NF$_3$-Li and Cs-O$_2$-Li has been tested using exposure to different gas species, showing an increased robustness against CO$_2$ and O$_2$ exposure \cite{mulhollan2010}. For Cs-Sb and Cs-O$_2$-Sb as well as Cs-Te and Cs-O$_2$-Te, the increased robustness to both IBB and chemical poisoning has been attributed to an increased thickness of the NEA layer and lower chemical reactivity observed in both Cs$_3$Sb and Cs$_2$Te photocathodes \cite{bae2018, cultrera2020, bae2020}. Synchrotron X-ray photoemission studies on Cs-Te and Cs-O$_2$-Te suggest that only a part of the NEA layer consists of Cs$_2$Te, with the additional presence of covalent Te as well as GaAs-Cs, GaAs-O$_2$-Cs and Cs-O dipoles that contribute to the work function reduction \cite{biswas2021}.

\section{Experimental Setup}
The experiments have been performed at the test stand for photocathode activation, testing, and cleaning using atomic hydrogen (\mbox{Photo-CATCH}) at the Institut für Kernphysik of Technische Universität Darmstadt. \\
Besides enabling photo-electron gun as well as photocathode research and development \cite{herbert2018, kurichiyanil2019, herbert2022, herbert2023a, herbert2023b}, it serves the institute's superconducting Darmstadt linear electron accelerator \mbox{S-DALINAC} \cite{richter1996, pietralla2018} by providing cleaned and tested photocathodes for the -\qty{100}{\kilo\volt} DC photo-electron gun, situated in the \mbox{S-DALINAC} polarized injector (SPIn \cite{poltoratska2011}), which is an alternative for the more commonly used -\qty{250}{\kilo\volt} thermionic gun of the accelerator. \\
The \mbox{S-DALINAC} can be operated as single- \cite{arnold2020} and multi-turn ERL \cite{schliessmann2023}. Ongoing projects on e.g. laser Compton backscattering \cite{meier2023} might strongly benefit from ERL operation with pulsed electron beams from a photo-electron source.

\subsection{Photo-CATCH test stand}
Photo-CATCH consists of two sections: the first section is used for photocathode cleaning and activation, consisting of three vacuum chambers placed in line and separated by all-metal gate valves. A load-lock chamber is used to put samples in or out of the vacuum system. It is connected to a chamber dedicated to photocathode surface cleaning, equipped with a hydrogen-atom beam source as well as heating coils for both hydrogen cleaning and heat cleaning. Connected to the cleaning chamber is the activation chamber, used for photocathode activation. The experiments presented in this work were carried out in this chamber, and a more detailed description will be given in the next subsection. Both chambers were baked out at \qty{200}{\celsius} to reach base pressures in the \qty{e-11}{\milli\bar} region. \\
The second section, connected to the activation chamber via an all-metal gate valve, is used for photocathode testing. It consists of a gun chamber with adjacent beamline. A -\qty{60}{\kilo\volt} DC photo-electron gun with inverted insulator geometry is installed in the gun chamber, which is connected to a short vertical section of the beamline. At the time of this study, the gun was not available due to extensive maintenance. A dipole magnet transports the electron beam from the gun into the main horizontal beamline. Here, a differential pumping stage separates the gun vacuum, in the region of \qty{e-11}{\milli\bar}, from the vacuum in the rest of the beamline (\qtyrange{e-9}{e-8}{\milli\bar}). The main beamline houses several devices for beam parameter measurements, such as Faraday cups, BeO fluorescent viewscreens, wirescanners, a chopper-slit combination for time-resolved electron-bunchshape measurements, and a Mott-polarimeter for spin-polarization measurements. A Wien filter is available for spin-polarization manipulation. Along the beamline, several quadrupole triplets, steerer dipoles and solenoids are installed for focusing and positioning of the beam, and Helmholtz coils are used to counteract the influence of the Earth's magnetic field. \\
A laser system is installed on a separate laser table, providing laser beams with a wavelength of \qty{785(2)}{\nano\meter} via fiber-optic patch cable to the \mbox{Photo-CATCH} setup. \\
A more detailed description of the test stand and the laser system can be found in \cite{herbert2022}.

\subsection{Activation chamber}
\begin{figure}[h]
    \centering
    \includegraphics[width=\columnwidth]{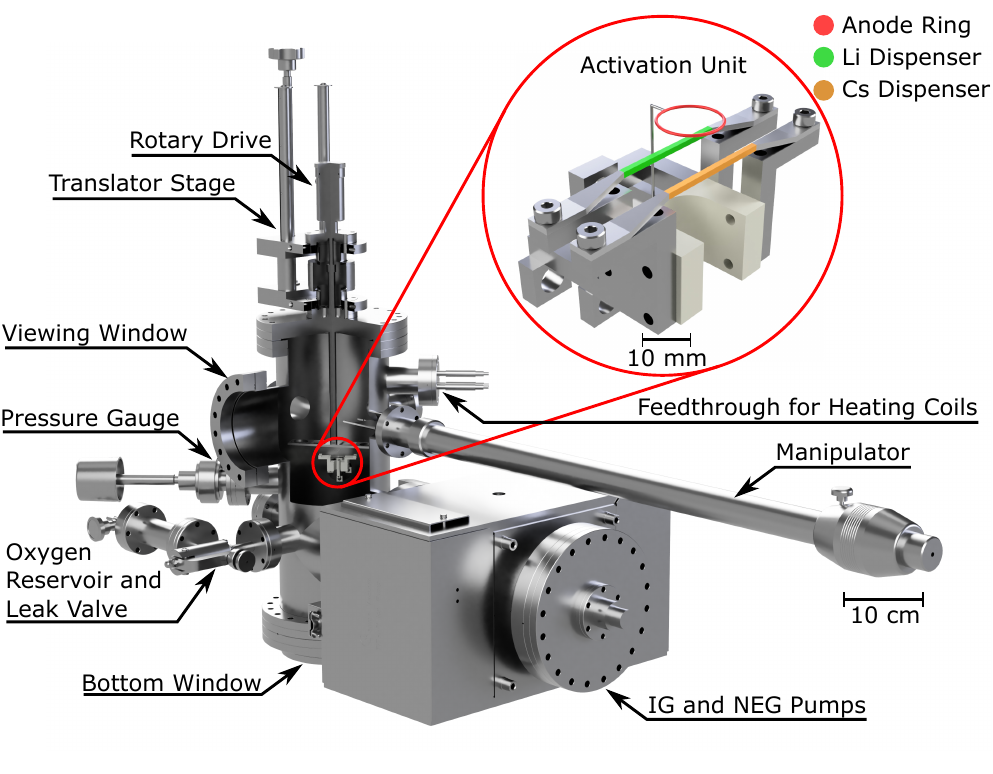}
    \caption{Rendered cut-away view of the activation chamber. The assembly holding the anode ring as well as the Cs and Li dispensers is shown in more detail in a blow-up.}
    \label{fig:act-chamber}
\end{figure}

\begin{figure}[h]
    \centering
    \includegraphics[width=\columnwidth]{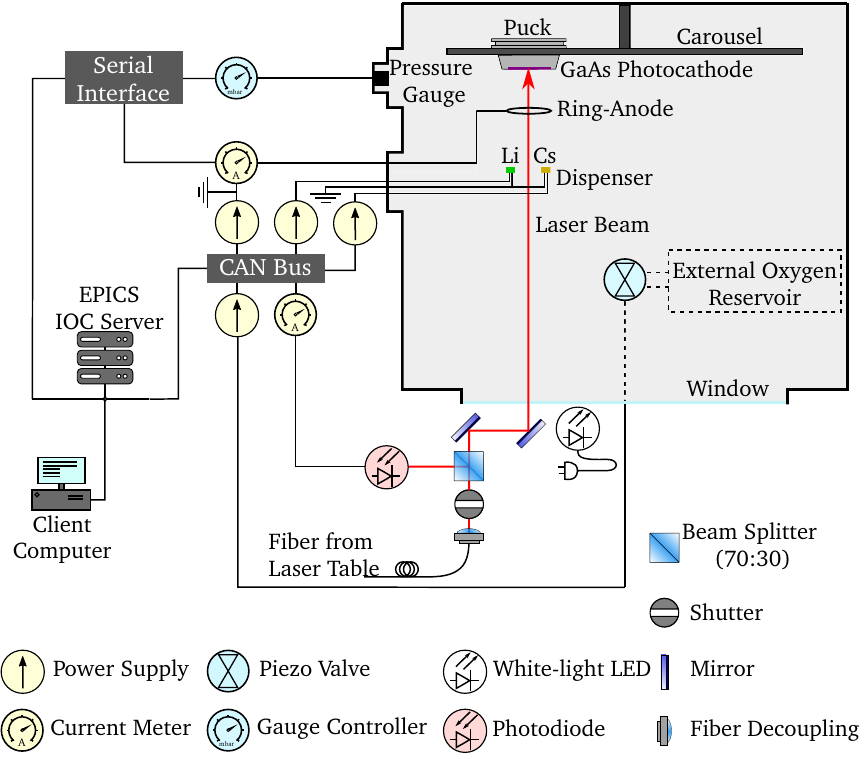}
    \caption{Schematic representation of the relevant components of the activation setup. All components used during the activation process are connected to the EPICS IOC server via CAN bus or serial interface and are remote-controlled from a client computer.}
    \label{fig:act-setup}
\end{figure}

The activation experiments presented in this work have been conducted using the cathode activation chamber (CAC) of the \mbox{Photo-CATCH} test stand. A 3D render of the chamber is shown in Fig.\ \ref{fig:act-chamber}. Within the vacuum chamber, after bakeout a base pressure of about \qty{1e-11}{\milli\bar} is reached using an ion-getter pump and a non-evaporable getter pump. A cold-cathode ionization gauge is used to monitor the pressure. Two tungsten coils mounted on an electric feed-through are used for heat cleaning of samples within the CAC. An assembly with ring anode and two dispensers, one for Cs and one for Li, is mounted on a separate feed-through. During the activation process, the photocathode is placed above this assembly. Vertical and rotational movement of the photocathode, which is placed inside a molybdenum cathode holder, also called puck, is possible using a carousel assembly with translator stage and rotary drive. Oxygen is provided from an external reservoir, connected to the chamber via piezo-electric leak valve. Two large windows, mounted mid-level on the side and at the bottom of the chamber, respectively, allow the observation of the cathode positioning and illumination of the cathode. Below the bottom window, an illumination assembly is mounted, housing a white-light LED array, consisting of 2x6 type 2835 LED chips with a total power of \qty{2.4}{\watt} and a color temperature of \qty{2500}{\kelvin} to \qty{3500}{\kelvin}, as well as a short laser beamline. The laser beam is transported to the assembly from an adjacent laser table via fiber-optic patch cable. For online laser-power monitoring, a 70:30 non-polarizing beam splitter is used to divert a part of the laser beam to a photodiode. A camera is mounted next to the assembly to monitor the laser-spot position. \\
All components of the activation setup in the CAC, shown in Fig.\ \ref{fig:act-setup}, are remote-controlled using the Experimental Physics and Industrial Control System (EPICS \cite{epics2023}). For this purpose, they are connected to an EPICS IOC server, using either Controller Area Network (CAN) bus or serial interface. A Control System Studio (CSS \cite{css2023}) GUI is used to control the connected devices. For further details on the used components, see Ref. \cite{herbert2022}.

\section{Experimental procedures}
For all measurements described here, a single Zn-doped p-type bulk-GaAs sample cut from a wafer\footnote{Wafer Technology Ltd.} polished on one side was used. The orientation of the sample is (100)~$\pm~\qty{0.1}{\degree}$, with a thickness of \qty{500(25)}{\micro\meter} and a carrier concentration in the range of \qtyrange{5e18}{5e19}{\per\centi\meter\cubed}, as per specifications given by the manufacturer. A puck was cleaned, inserted into the load lock chamber and annealed at a temperature of about \qty{500}{\celsius} for several hours, using the tungsten filament installed in the chamber. The annealed puck was then removed and taken to a clean room, where the GaAs sample was freshly cut from the wafer, inserted into the puck and fixed using a tungsten wire. After the installation of the sample, the puck was immediately reintroduced into the load-lock chamber and, after evacuation, transferred to the CAC. There, the sample was heat-cleaned several times in order to remove the oxidized surface layer formed by contact to the air, as well as any other contamination. \\
Before each individual activation, the puck containing the sample was moved directly beneath the tungsten coils, heating the sample to a temperature between \qtyrange{500}{600}{\celsius} for about one hour. During heat-cleaning, the pressure within the CAC rose to the low \qty{e-8}{\milli\bar} range. In order to reduce contamination of the dispensers, both were supplied with a current of \qty{0.3}{\ampere} during operation of the heating coils. After this process, a cooldown period of about four hours was observed to allow the pressure to drop to the low \qty{e-10}{\milli\bar} range before placing the puck approximately \qty{10}{\milli\meter} above the anode ring and starting the activation.

\subsection{Activation procedure}

\subsubsection*{Cs-O$_2$ Co-deposition (Scheme 1)}
\begin{figure}[h]
    \centering
    \includegraphics[width=\columnwidth]{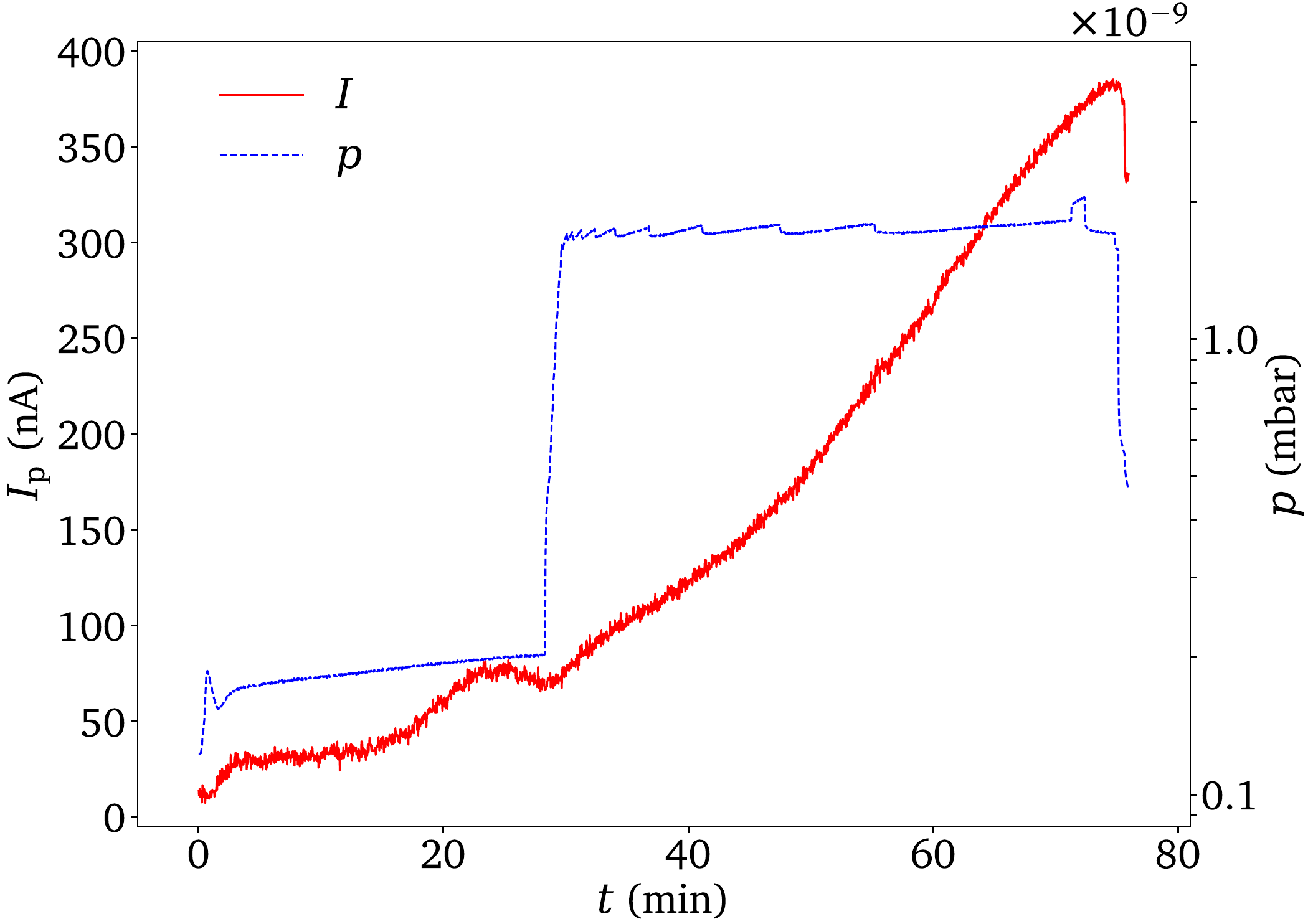}
    \caption{Photocurrent $I_\text{p}$ and pressure $p$ as function of time during a scheme-1 activation process, using white-light illumination. Shown here is scheme-1 activation No.~1 that yielded a final quantum efficiency of \qty{9.06(11)}{\percent} at \qty{785}{\nano\meter} directly after activation.}
    \label{fig:scheme1}
\end{figure}

The so-called co-deposition (Co-De) method was used for all activations conducted in this study. First, the sample is exposed to an optimized rate of Cs by operating the Cs dispenser at a current of \qty{3.1}{\ampere} until a photocurrent is observed and reaches its saturation maximum, also called Cs peak. Cs exposure was continued until the photocurrent dropped to about \qty{75}{\percent} of its maximum value\footnote{The background current from dispenser operation is subtracted from the measured anode current to obtain this value.}. At this point, O$_2$ was introduced by setting the piezo-electric leak valve to an operating voltage $U_\text{ox}$. Upon reaching a specific pressure threshold, $U_\text{ox}$ was then gradually reduced to keep the pressure at the chosen threshold, hence co-depositing Cs and O$_2$ for the remainder of the activation. This causes a rise in photocurrent up to a saturation maximum, at which point both Cs and O$_2$ exposure was stopped. The pressure threshold was chosen to produce a partial pressure ratio of $\frac{p_\text{Cs}}{p_\text{ox}} \approx 0.043$, which was found to yield optimal results for the experimental setup \cite{kurichiyanil2017}. Here, the Cs partial pressure $p_\text{Cs} = p_1 - p_0$ is defined as the difference between the pressure at the beginning of the Cs peak $p_1$ and the initial pressure at the beginning of the activation $p_0$, and the O$_2$ partial pressure $p_\text{ox}$ is calculated from the optimal ratio and $p_\text{Cs}$. Typical trends observed for photocurrent and pressure during the activation process are shown in Fig.\ \ref{fig:scheme1}. During Cs dispenser operation, a background current can be observed, most likely caused by ionization of Cs atoms from the dispenser. This causes a 'step' in photocurrent at the end of the activation process when the Cs dispenser is switched off. \\
Throughout the activation procedure, the sample was illuminated using the white-light LED array because the Cs peak was barely visible when using the laser with a wavelength of \qty{785}{\nano\meter} needed for polarized-electron extraction. Since $\eta$ cannot be measured using the white-light LED, it was determined separately after the activation is finished, using the laser. \\
The ring anode was biased at \qty{102(0.1)}{\volt} throughout the activation process. 

\subsubsection*{Cs-O$_2$-Li Co-deposition (Scheme 2)}
\begin{figure}[h]
    \centering
    \includegraphics[width=\columnwidth]{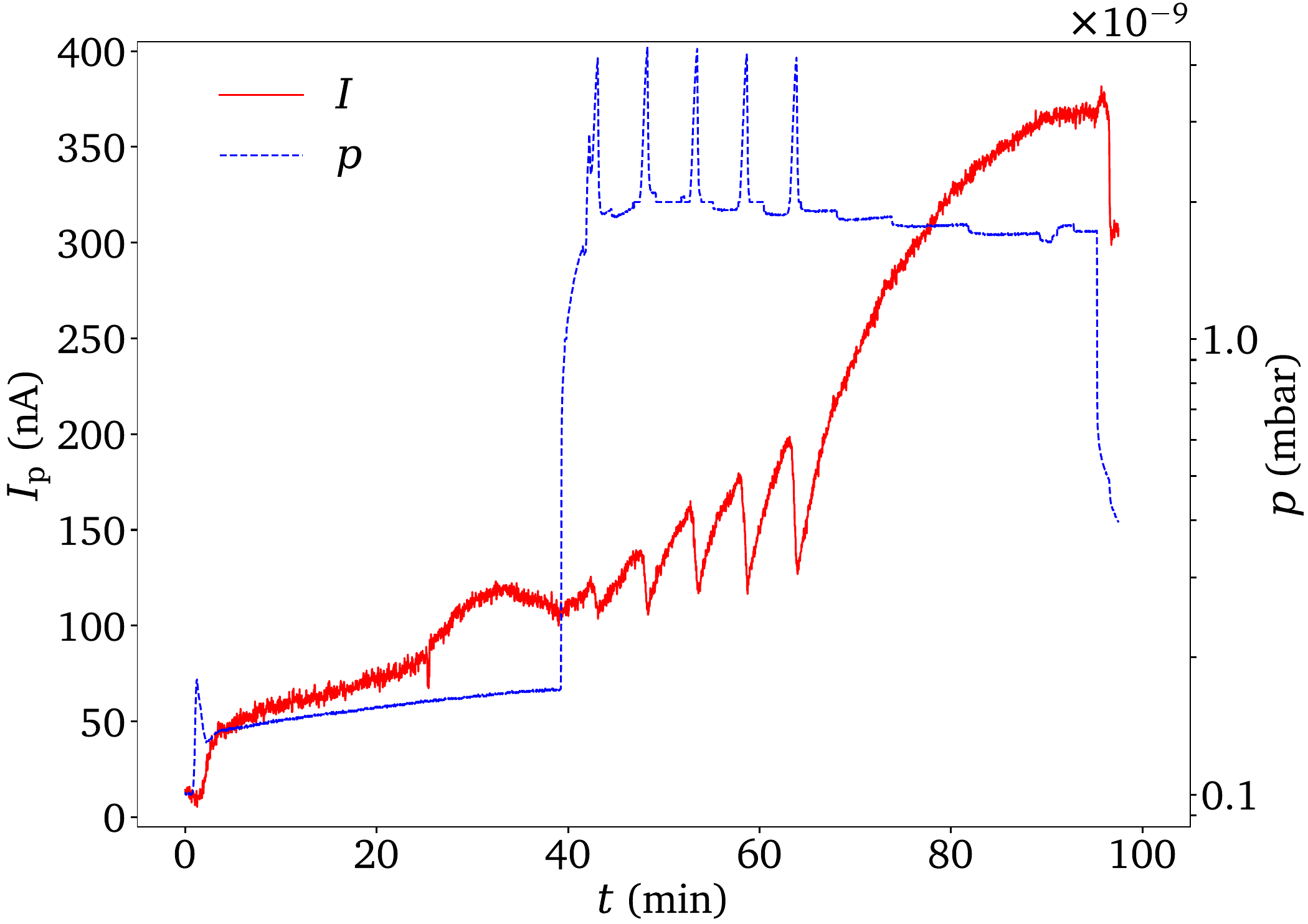}
    \caption{Photocurrent $I_\text{p}$ and pressure $p$ as function of time during a scheme-2a activation process, using white-light illumination. Shown here is scheme-2a activation No.~2 that yielded a final quantum efficiency of \qty{9.66(4)}{\percent} at \qty{785}{\nano\meter} directly after activation.}
    \label{fig:scheme2a}
\end{figure}

\begin{figure}[h]
    \centering
    \includegraphics[width=\columnwidth]{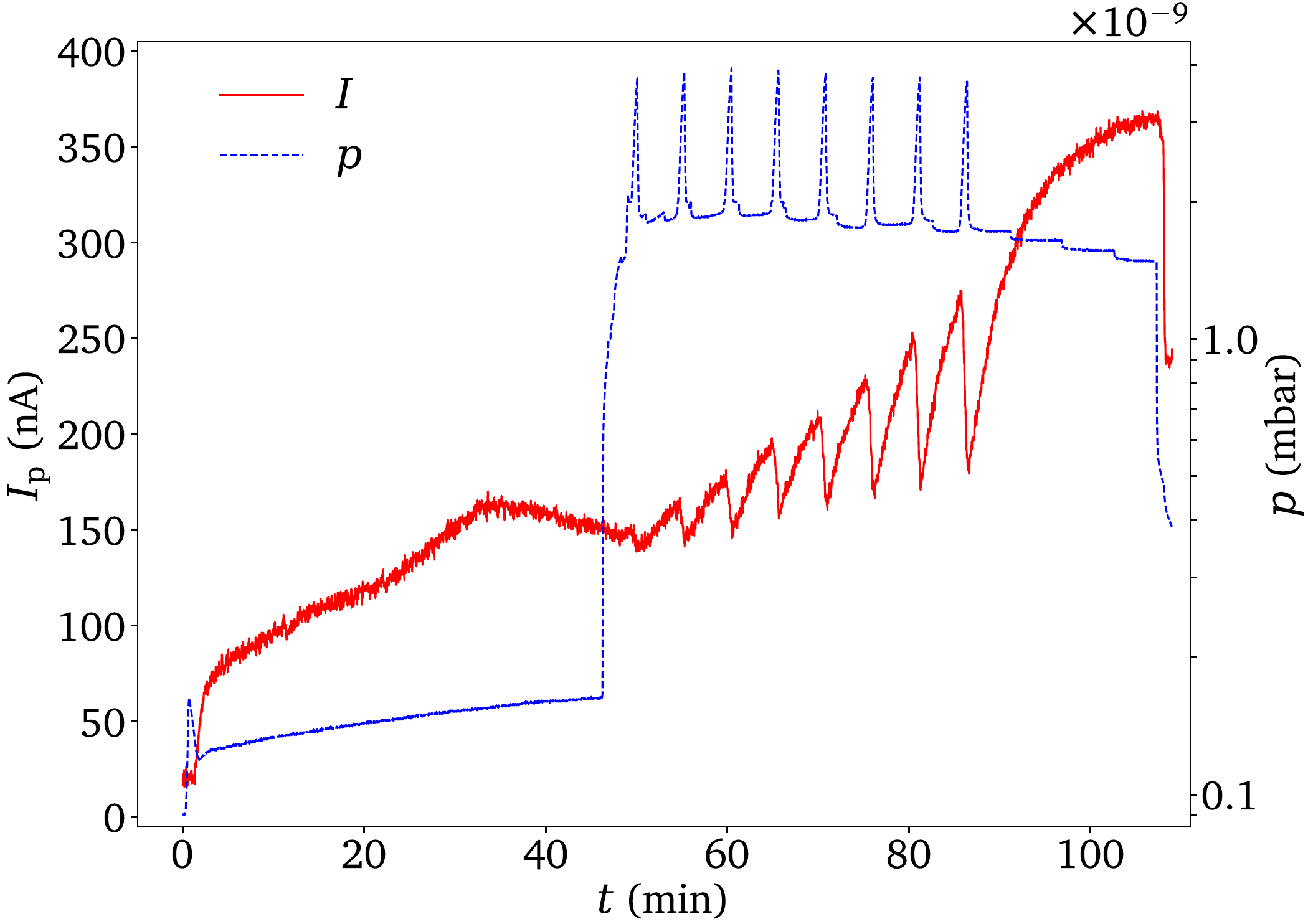}
    \caption{Photocurrent $I_\text{p}$ and pressure $p$ as function of time during a scheme-2b activation process, using white-light illumination. Shown here is scheme-2b activation No.~1 that yielded a final quantum efficiency of \qty{8.15(3)}{\percent} at \qty{785}{\nano\meter} directly after activation.}
    \label{fig:scheme2b}
\end{figure}

For enhancement with Li, scheme 1 was modified by introducing Li in short bursts during O$_2$ exposure, based on the scheme used in Ref.~\cite{kurichiyanil2019}. To increase reproducibility, a python script embedded into the CSS GUI was used to precisely time the Li bursts. During each burst, the Li dispenser was operated at a current of \qty{4.1}{\ampere} for a duration of \qty{75}{\second}. The first pulse was initiated with a delay of \qty{150}{\second} after O$_2$ exposure had begun, and each subsequent pulse started \qty{225}{\second} after the previous pulse ended. This scheme is referred to as scheme 2. Two sub-schemes with five and eight consecutive pulses were used for the experiments, furthermore called scheme~2a and scheme~2b, respectively. Typical trends observed for photocurrent and pressure during scheme~2a and scheme~2b are shown in Figs.\ \ref{fig:scheme2a} and \ref{fig:scheme2b}, respectively. \\
A difference in duration of the activation process was observed for several consecutive activations of a single sample, irrespective of the used activation scheme. For the sample used in our experiments, the duration of the activation process increased during the course of 28 consecutive activations from about \qty{60}{\minute} to almost \qty{120}{\minute} because the value of $p_\text{Cs}$ induced by a fixed operating current $I_\text{Cs}$ of the cesium dispenser declined over the period of the experiments \cite{herbert2023b}. Hence, the scheme-2 activations shown in Figs. \ref{fig:scheme2a} and \ref{fig:scheme2b} are longer than the scheme-1 activation shown in Fig.\ \ref{fig:scheme1} since they were conducted at a later time. The change in duration of the activation procedure, and hence the change in the amount of Cs, had no discernible effect on $\eta$ or $\tau$.

\subsection{Ingredient exposure determination}
\label{subsec:ingredient-calc}
After the activation process is finished, the pressure curve can be separated into three parts to obtain the partial pressures for Cs, Li and O$_2$:
\begin{itemize}
    \item \textbf{Part 1: Cs exposure}\\
    During this part, only Cs is introduced into the chamber. Here, $p_\text{cs}$ is calculated using \mbox{$p_\text{cs} = p(t) - p(0)$} and fitted with a linear function.
    \item \textbf{Part 2: Cs-O$_2$ co-deposition}\\
    During this part, Cs and O$_2$ are introduced simultaneously. Here, $p_\text{cs}$ is approximated using the linear fit from part~1 and $p_\text{ox}$ is calculated using $p_\text{ox} = p(t) - p_\text{cs} - p(0)$.
    \item \textbf{Part 3: Li exposure}\\
    This part represents a special case since Li exposure occurs during part~2. Here, $p_\text{cs}$ is approximated using the linear fit from part~1 and $p_\text{ox}$ is approximated with a linear fit to the data from part~2 prior to Li exposure, allowing the calculation of $p_\text{li}$ using $p_\text{li} = p(t) - p_\text{ox} - p_\text{cs} - p(0)$.
\end{itemize}
The extracted partial pressures are then integrated over the duration of the entire activation procedure to obtain an approximation of the total exposure $D$ for the respective material.

\subsection{Quantum efficiency measurement}
After the activation procedure was finished, the white-light LED array was switched off and the shutter was opened to introduce \qty{785(2)}{\nano\meter} laser light onto the sample, allowing the measurement of $\eta$. The laser had a power\footnote{Since the laser power was not stabilized, its value differed slightly for the individual measurements, ranging from \qty{48.9(1.1)}{\micro\watt} to \qty{51.4(1.2)}{\micro\watt}.} of \qty{50}{\micro\watt} and a $1/e^2$ spotsize of \qty{445(5)}{\micro\meter} on the sample surface. During laser illumination of the sample, the ring anode was biased at \qty{100(0.1)}{\volt}. \\
In order to obtain the maximum quantum efficiency, the position of the laser spot on the photocathode surface was adjusted\footnote{An example of the manual adjustment process can be found in Ref. \cite[Fig.\ 4.2]{herbert2022}}. The position which yielded the highest value of $\eta$ was then used for all subsequent measurements. Two distinctive values for the quantum efficiency were determined: the maximum quantum efficiency $\eta_\text{max}$ measured during laser-spot adjustment, representing the maximum value achieved with the activation process, and the initial quantum efficiency $\eta_0$ at the beginning of the lifetime measurement. Between these two measurements, a period of between \num{5} and \num{10} minutes passed\footnote{After activation, $\eta$ was measured as a function of anode bias voltage. These measurements are not presented here and will be subject of a future publication.}. Hence, $\eta_\text{max}$ and $\eta_0$ were chosen to consider any effects on the surface layer between the end of activation and the beginning of the lifetime measurement. \\
For operation within a photo-electron gun, the photocathode needs to be transferred to the gun chamber directly after activation. Since the duration of the transfer is comparable to the time between the measurement of $\eta_\text{max}$ and $\eta_0$, the latter value can be taken as an approximation for the initial quantum efficiency during gun operation\footnote{This approximation does not include the increase in quantum efficiency in a high-voltage DC gun due to the Schottky effect, cf. \cite{herbert2020}.}, although the measurements in the CAC may be influenced by residual gases following the activation.

\subsection{Lifetime measurement}
Once the laser spot had been adjusted for maximum $\eta$, the illumination of the sample was continued in order to observe the decrease of $\eta$ over time. A laser power of \qty{50}{\micro\watt} was set and monitored\footnote{The laser power was not stabilized and fluctuated during the measurements, ranging from \qty{42(1)}{\micro\watt} to \qty{57.7(1.2)}{\micro\watt}.}, allowing an online calculation of $\eta$ including time-dependent drifts of the laser power. Ideally, the measurement was kept running until $\eta$ had fallen to $1/e$ of its initial value or below, although this could not always be guaranteed due to scheduling or technical issues. All lifetime data presented in this paper were obtained at an anode-bias voltage of \qty{100(0.1)}{\volt}. \\

\section{Results and discussion}
In total, eight activations with subsequent quantum-efficiency and lifetime measurements were performed, with four activations each using scheme~1 and scheme~2. Of the scheme-2 activation process, two each were carried out using scheme~2a and 2b. \\
The eight activations can be classified into five different types:
\begin{itemize}
    \item \textbf{Type i:}
    Scheme~1 activation carried out prior to any activation with scheme~2, e.g.\ prior to introduction of Li into the chamber. Two activations of this type were executed during this study.
    \item \textbf{Type ii:}
    Scheme~1 activation carried out subsequent to a scheme~2a activation. Two activations of this type were executed during this study.
    \item \textbf{Type iii:}
    Scheme~2a activation carried out subsequent to a scheme~1 activation. Two activations of this type were executed during this study.
    \item \textbf{Type iv:}
    Scheme~2b activation carried out subsequent to a scheme~1 activation. One activation of this type was executed during this study.
    \item \textbf{Type v:}
    Scheme~2b activation carried out subsequent to a scheme~2b activation. One activation of this type was executed during this study.
\end{itemize}
It was observed that type-ii activations resulted in a significantly different performance compered to type-i activations, hinting at an influence of Li traces remaining in the activation system after a scheme-2 activation process. This finding suggests that the heat cleaning of the sample prior to activation did not completely remove Li from the sample surface. Another possibility is that during activation, lifetime measurement or heat cleaning, Li may adsorb on the anode ring or Cs dispenser, subsequently desorbing during operation and hence being introduced into the activation process. Regardless of the cause, this effect needs to be taken into consideration when conducting studies on enhanced activation methods. For example, during a previous study that tested a GaAs photocathode activated with \mbox{Cs-NF$_3$-Li} in the gun of the upgraded injector test facility (UITF) at JLab, the Li-enhanced activation was used first, followed by a standard activation \cite{herbert2020}. This may have introduced a systematic error, reducing the observed improvement factor from Li due to the standard activation also being influenced by the positive effect of residual Li. \\
For quantum efficiency, lifetime, and extracted charge, a corresponding factor of change $r$ was calculated by comparing the mean results of type-i activations to those of the respective type $x$:
\begin{equation}
    r_\text{parameter} = \frac{\text{mean parameter (type~} x\text{)}}{\text{mean parameter (type i)}} \quad .
\end{equation}
Table~\ref{tab:ratios} lists the factors of relative change for the individual parameters. The results will be discussed in the following subsections, with an overview given in Tab.\ \ref{tab:results}.

\begingroup
\setlength{\tabcolsep}{15pt}
\renewcommand{\arraystretch}{2.0}
\begin{table}[t]
    \caption{Overview of factors describing the relative change in maximum quantum efficiency $\eta_\text{max}$ after activation, initial quantum efficiency $\eta_0$ and photocurrent $I_0$ at the beginning of the lifetime measurement, lifetime $\tau$, and lifetime charge $Q(\tau)$.}
    \begin{tabular}{cc}
    \hline
    \hline
    Factor & Ratio \\
    \hline
    $r_{\eta_\text{max}}$ & $\frac{\overline{\eta}_\text{max}~\text{(type}~x\text{)}}{\overline{\eta}_\text{max}~\text{(type i)}}$ \\
    $r_{\eta_0}$ & $\frac{\overline{\eta}_0~\text{(type}~x\text{)}}{\overline{\eta}_0~\text{(type i)}}$ \\
    $r_{I_0}$ & $\frac{\overline{I}_0~\text{(type}~x\text{)}}{\overline{I}_0~\text{(type i)}}$ \\
    $r_\tau$ & $\frac{\overline{\tau}~\text{(type}~x\text{)}}{\overline{\tau}~\text{(type i)}}$ \\
    $r_{Q(\tau)}$ & $\frac{\overline{Q(\tau)}~\text{(type}~x\text{)}}{\overline{Q(\tau)}~\text{(type i)}}$ \\
    \hline
    \hline
    \end{tabular}
    \label{tab:ratios}
\end{table}
\endgroup

\begin{table*}[t]
    \caption{Mean photocathode parameters for the different activation schemes measured at $\lambda = \qty{785(2)}{\nano\meter}$ and $P_\gamma = \qty{50}{\micro\watt}$. The relative change compared to type~i activation is given as a factor $r$ for each parameter.}
    \begin{tabular}{c|ccccc}
    \hline
    \hline
    Type & i & ii & iii & iv & v \\
    & & & & & \\
    Scheme & 1 & 1 & 2a & 2b & 2b \\
    & & & & & \\
    $\overline{\eta}_\text{max}$ in \qty{}{\percent} & \num{8.6(5)} & \num{9.7(4)} & \num{9.6(4)} & \num{8.2(3)} & \num{8.3(3)} \\
    $r_{\eta_\text{max}}$ & - & \num{1.13(8)} & \num{1.12(8)} & \num{0.95(7)} & \num{0.97(7)} \\
    & & & & & \\
    $\overline{\eta}_0$ in \qty{}{\percent} & \num{8.3(4)} & \num{9.1(4)} & \num{8.7(2)} & \num{7.5(1)} & \num{7.1(1)} \\
    $r_{\eta_0}$ & - & \num{1.12(0.11)} & \num{1.04(9)} & \num{0.90(5)} & \num{0.86(4)} \\
    & & & & & \\
    $\overline{I}_0$ in \qty{}{\micro\ampere} & \num{2.6(2)} & \num{2.9(2)} & \num{2.7(1)} & \num{2.41(2)} & \num{2.27(3)} \\
    $r_{I_0}$ & - & \num{1.12(0.11)} & \num{1.04(9)} & \num{0.92(6)} & \num{0.87(6)} \\
    & & & & & \\
    $\overline{t}_{\text{off}}$ in \qty{}{\hour} & - & \num{5(1.4)} & \num{22(2)} & \num{48(1)} & \num{120(2)} \\
    & & & & & \\
    $\overline{\eta}_0^*$ in \qty{}{\percent} & \num{8.3(4)} & \num{8.3(5)} & \num{6.9(5)} & \num{6.8(2)} & \num{4.3(1)} \\
    $r_{\eta_0^*}$ & - & \num{1.00(8)} & \num{0.83(7)} & \num{0.82(5)} & \num{0.52(3)} \\
    & & & & & \\
    $\overline{\tau^*}$ in \qty{}{\hour} & \num{10.5(8)} & \num{34(2.5)} & \num{34(2.7)} & \num{103(13)} & \num{79(9)} \\
    $r_{\tau^*}$ & - & \num{3.2(3)} & \num{3.2(4)} & \num{9.8(1.4)} & \num{7.5(1.0)} \\
    & & & & & \\
    $\overline{Q(\tau^*)}$ in \qty{}{\milli\coulomb} & \num{63(9)} & \num{206(11)} & \num{166(23)} & \num{488(63)} & \num{250(29)} \\
    $r_{Q(\tau^*)}$ & - & \num{3.3(5)} & \num{2.6(5)} & \num{7.7(1.5)} & \num{4.0(7)} \\
    & & & & & \\
    $\overline{Q(\tau^*)}_\text{est}$ in \qty{}{\milli\coulomb} & \num{63(9)} & \num{224(23)} & \num{209(23)} & \num{565(71)} & \num{408(47)} \\
    $r_{Q(\tau^*)_\text{est}}$ & - & \num{3.6(6)} & \num{3.3(6)} & \num{9.0(1.7)} & \num{6.5(1.2)} \\
    & & & & & \\
    $\overline{\tau}$ in \qty{}{\hour} & \num{10.5(8)} & \num{38.9(1.9)} & \num{56(2.7)} & \num{151(14)} & \num{199(11)} \\
    $r_{\tau}$ & - & \num{3.7(3)} & \num{5.3(5)} & \num{14.4(1.7)} & \num{19(2)} \\
    & & & & & \\
    $\overline{Q}_{\text{off}}$ in \qty{}{\milli\coulomb} & - & \num{52(18)} & \num{212(14)} & \num{388(7)} & \num{788(10)} \\
    & & & & & \\
    $\overline{Q(\tau)}$ in \qty{}{\milli\coulomb} & \num{63(9)} & \num{258(22)} & \num{378(19)} & \num{876(70)} & \num{1038(39)} \\
    $r_{Q(\tau)}$ & - & \num{4.1(7)} & \num{6(1)} & \num{13.9(2.3)} & \num{16.5(2.4)} \\
    \hline
    \hline
    \end{tabular}
    \label{tab:results}
\end{table*}

\subsection{Ingredient exposure}
Table~\ref{tab:dosage} shows the average dosages of the ingredients for each activation scheme. The uncertainties have been estimated based on the accuracy of the calculation method. As was mentioned before, an elongation of the activation process was observed in the course of multiple consecutive activations \cite{herbert2023b}, hence the dosages of type~i activations are lower because they were performed earlier in the study. However, the ratio $\frac{\overline{D}_\text{cs}}{\overline{D}_\text{ox}}$ was identical within its uncertainties for all types.\\
Since the number of Li-pulses was 5 and 8 for scheme~2a and scheme~2b, respectively, one would expect an increase in Li dosage by a factor of about \num{1.6}. However, a lower increase by a factor of \num{1.3(2)} was observed. Since both $p_\text{cs}$ and $p_\text{ox}$ are approximated in order to calculate $p_\text{li}$, it is possible that the approximations are not accurate enough to properly reproduce the Li partial pressure. Also, the deviation of $p_\text{li}$ produced by the Li-dispenser for the same current setting has not been included in the uncertainties given for $\overline{D}_\text{li}$.

\begingroup
\renewcommand{\arraystretch}{2.0}
\begin{table}[t]
    \caption{Average dosages of ingredients introduced over the duration of the activation process.}
    \begin{tabular}{ccccc}
    \hline
    \hline
    & \multicolumn{3}{c}{Average dosages in \qty{e-7}{\milli\bar\second}} & \\
    Type & $\overline{D}_\text{cs}$ & $\overline{D}_\text{ox}$ & $\overline{D}_\text{li}$ & $\frac{\overline{D}_\text{cs}}{\overline{D}_\text{ox}}$ \\
    \hline
    i & \num{3.9(1)} & \num{39.9(5)} & - & \num{0.098(3)} \\
    ii & \num{5.0(1)} & \num{50.4(5)} & - & \num{0.099(3)} \\
    iii & \num{4.7(2)} & \num{52(3)} & \num{3.3(2)} & \num{0.09(1)} \\
    iv & \num{5.3(1)} & \num{56.2(2)} & \num{4.6(2)} & \num{0.094(2)} \\
    v & \num{5.5(1)} & \num{58.6(3)} & \num{4.3(3)} & \num{0.094(2)} \\
    \hline
    \hline
    \end{tabular}
    \label{tab:dosage}
\end{table}
\endgroup

\subsection{Quantum efficiency}
All values of $\overline{\eta}_\text{max}$ and, hence, $r_{\eta_\text{max}}$ are listed in \mbox{Tab.\ \ref{tab:results}}. As is evident from the data, the maximum quantum efficiency directly after activation is slightly increased by both residual and low amounts of Li, and does not change significantly for higher amounts of Li. A comparable result of \num{1.0(1)} was obtained by a previous study at \mbox{Photo-CATCH} using \qty{405}{\nano\meter} incident laser-light and comparing standard to Li-enhanced co-deposition \cite{kurichiyanil2019}. \\
The effect of Li is more pronounced when comparing the quantum efficiency $\eta_0$ at the beginning of the lifetime measurement, see Tab.\ \ref{tab:results}. While low amounts of Li show no significant change in quantum efficiency, a slight reduction is observed for higher amounts of Li. This is most likely caused by an initial short decrease in $\eta$ observed for Li-enhanced activations, as discussed in the following subsection.  \\
It has been observed for Cs-O$_2$-Sb that an increase in Sb thickness reduces the quantum efficiency significantly \cite{cultrera2020, bae2020}. A similar behavior has been observed for \mbox{Cs-NF$_3$-Li} at \qty{780}{\nano\meter}, with $r_\eta$ of \num{1.2}, \num{0.9} and \num{0.07} for low, medium, and high Li dosage, respectively, as estimated from the data presented in \cite{mulhollan2010}. The impact of Li on the quantum efficiency observed for scheme~2b activations that used the highest amount of Li is comparable to that observed in \cite{mulhollan2010} for medium Li dosage.

\subsection{Lifetime and extracted charge}
\begin{figure}[h]
    \centering
    \includegraphics[width=\columnwidth]{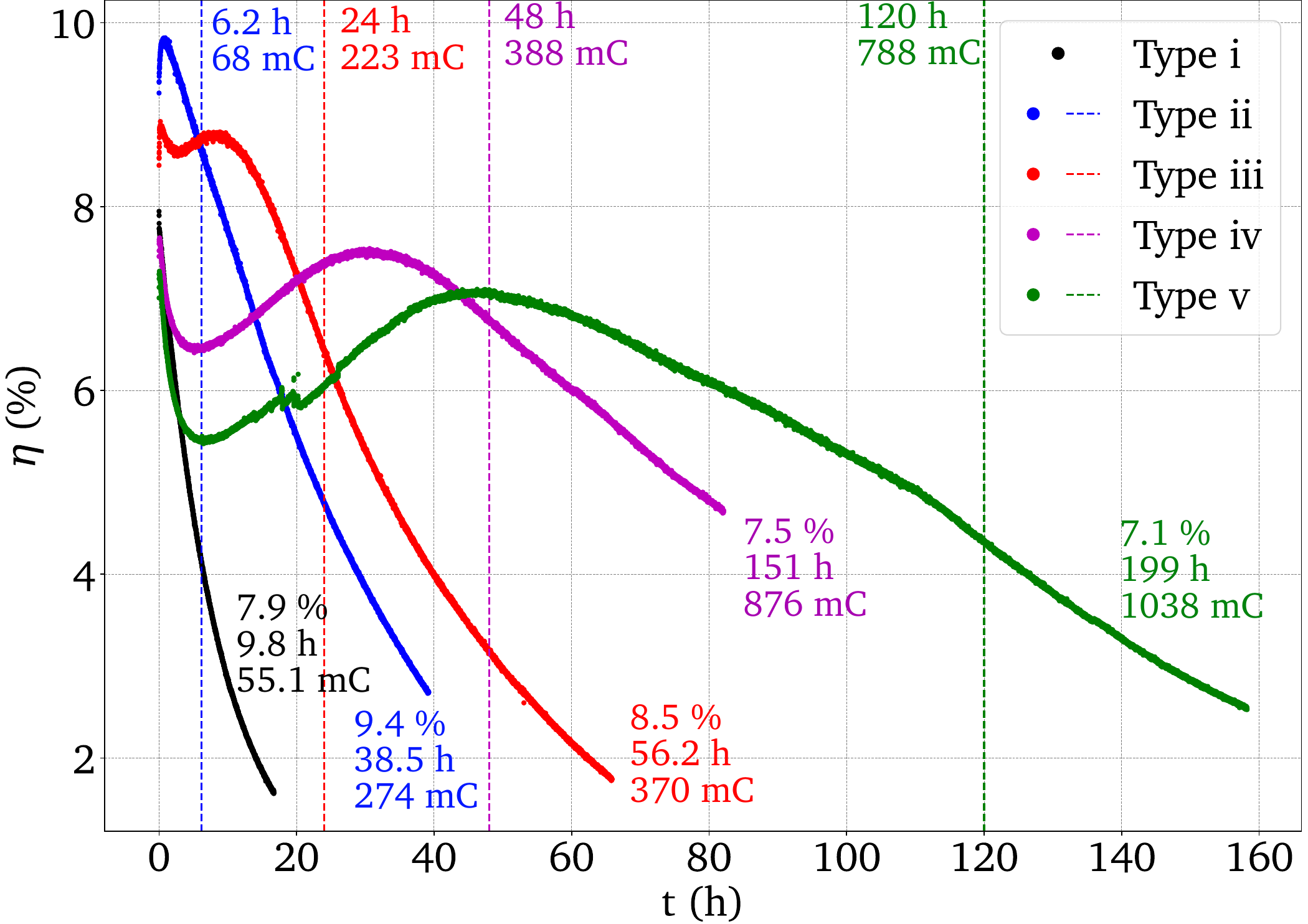}
    \caption{Quantum efficiency at \qty{785}{\nano\meter} as a function of time for selected measurements of the different activation schemes. Each curve is labeled with its parameters $\eta_0$, $\tau$, and $Q(\tau)$ in the bottom half of the figure. The end of the offset period is marked by a dashed vertical line for each curve and labeled with the offset parameters $t_\text{off}$ and $Q_\text{off}$ at the top of the figure.}
    \label{fig:decay_comp}
\end{figure}

The lifetime data was fitted with a single-exponential decay function of the form
\begin{equation}
    \eta(t) = \eta(t=0) \cdot e^{-\frac{t}{\tau}} = \eta_0 \cdot e^{-\frac{t}{\tau}}
    \label{eq:qe_t}
\end{equation}
to obtain the lifetime $\tau$. The extracted charge $Q(t)$ was calculated by integrating the measured photocurrent. To obtain the charge $Q(\tau)$ extracted during the period $\tau$, furthermore referred to as lifetime charge, the calculated charge was fitted with the function
\begin{equation}
    Q(t) = \int_{t=0}^{t} I(\Tilde{t}) \text{d}\Tilde{t} = I_0 \cdot \tau \cdot \left(1 - e^{-\frac{t}{\tau}}\right) \quad ,
    \label{eq:charge_t}
\end{equation}
yielding
\begin{equation}
    Q(\tau) = I_0 \cdot \tau \cdot \left(1 - \frac{1}{e}\right) \approx \num{0.63}\cdot I_0 \cdot \tau \quad ,
    \label{eq:charge_tau}
\end{equation}
with $I(t=0) = I_0$. The lifetimes obtained from Eqs.\ (\ref{eq:qe_t}) and (\ref{eq:charge_t}) were compared to assess the quality of the fits, showing good agreement\footnote{One should note that the lifetime charges presented here were measured at relatively low extracted currents, with the initial current $I_0$ ranging from \qtyrange{2.3}{2.9}{\micro\ampere} and the final current at the end of the lifetime measurement ranging from \qtyrange{0.5}{1.0}{\micro\ampere}.}. \\
It should also be possible to obtain $Q(\tau)$ by fitting the quantum efficiency as function of extracted charge with\footnote{Sometimes, $Q(\tau)$ is obtained from Eq.\ (\ref{eq:qe_charge}) and called $\tau_\text{c}$ or "charge lifetime", given in units of C. However, for the sake of consistency, we use $\tau$ only for lifetimes in units of time, cf. e.g. \cite{aulenbacher2005}.}
\begin{equation}
    \eta(Q(t)) = \eta_0 \cdot e^{-\frac{Q(t)}{Q(\tau)}} \quad ,
    \label{eq:qe_charge}
\end{equation}
with $\eta(Q(t=0)) = \eta_0$. However, Eq.\ (\ref{eq:charge_t}) proved to be the more reliable fit to the data presented here since $\eta(Q(t))$ did not show the assumed exponential decay behavior. \\
During the lifetime measurements, a different behavior in the decay of the quantum efficiency was observed: instead of the decay starting immediately after activation, the quantum efficiency increased for a period of time before the exponential decay started. For Cs-O$_2$-Li activations, this increase was preceded by a short exponential decay. Figure \ref{fig:decay_comp} shows a comparison of curves for $\eta(t)$ observed for the different activation types. This kind of behavior was previously observed at \mbox{Photo-CATCH} and attributed to Cs background in the chamber. However, a closer analysis of the effect was not carried out \cite{kurichiyanil2019}. Similar behaviors have also been observed for Cs-O$_2$-Te activated samples \cite{rahman2019b, bae2022b}, but have not been addressed in these publications. \\
The offset period appears to be influenced significantly by the dosage of Li. While the effect was negligible for type~i, it was perceivable for type~ii and steadily increased with the amount of Li, being strongest for type~v. For this reason, Eqs.\ (\ref{eq:qe_t}) and (\ref{eq:charge_t}) did not yield good fits to data with large offset. \\
It is possible that the NEA layer does not reach full chemical equilibrium during the activation and requires a period of rearrangement. Since the lifetime measurements presented here have been conducted in the activation chamber, the pressure kept falling towards its base level during the measurements. However, we did not observe any clear direct relation between the trends of pressure and quantum efficiency. Nevertheless, since residual gas within the chamber is the dominant cause of quantum efficiency decay, the composition of the residual gas may play a role in distorting the decay behavior of the surface layer. \\
Hence, the initial fast decay observed after Cs-O$_2$-Li activations may be caused by the presence of excessive Cs and, or Li due to overexposure during the activation process. Another possibility is additional IBB from Li residue remaining in the chamber after the activation process. Previous experiments have shown that GaAs surfaces can be cleaned by means of sputtering with Ar ions at ion energies as low as \qty{50}{\electronvolt} \cite{rabinzohn1984}. Hence, damage to the NEA surface layer from IBB is to be expected for the anode potential of \qty{100}{V} used here. \\
For an electron energy of \qty{100}{\electronvolt}, Li has an ionization cross section between \qty{1.0e-16}{\centi\meter\squared} and \qty{2.2e-16}{\centi\meter\squared} \cite{jalin1973, mcfarland1965}, compared to about \qty{0.9e-16}{\centi\meter\squared} for H$_2$ \cite{kim+rudd1994}, \qty{1.7e-16}{\centi\meter\squared} for O$_2$ \cite{maerk1975}, \qty{1.6e-16}{\centi\meter\squared} for N$_2$ \cite{maerk1975}, and \qty{7.0e-16}{\centi\meter\squared} for Cs \cite{mcfarland1965}. Hence, while residual Cs is expected to contribute the most to IBB during operation directly after the activation process, residual Li can also be expected to contribute significantly to IBB during this stage since its ionization cross section is about equal or up to \num{2.4} times larger than that of H$_2$ and comparable to the cross sections of O$_2$ and N$_2$. \\
On the contrary, the adsorption of remaining Cs and Li atoms onto the photocathode surface likely has positive effects on the lifetime of the cathode. The observed rise in quantum efficiency after the initial fast decay appears to be consistent with the previously observed rise during re-cesiation of a GaAs-cathode with decayed surface layer \cite{rodway+allenson1986}. Since this effect was more pronounced for higher amounts of Li used during the activation and Li was also observed to significantly inhibit quantum efficiency decay, one may assume that the adsorption of additional Li leads to a slow change in surface structure that is observable as a rise in work function or quantum efficiency, until the amount of residual Li in the chamber becomes too low to sustain this effect and the exponential decay of the NEA layer becomes dominant. Hence, we separated the observed trend in quantum efficiency into three different regimes for further analysis:
\begin{itemize}
    \item[1.] Surface layer poisoning from excessive activation agents and, or IBB from residual activation agents are dominant, causing a fast exponential decay of $\eta$.
    \item[2.] Adsorption of residual activation agents becomes dominant, instigating a partial reformation of the surface layer, observable as a recovery of $\eta$.
    \item[3.] IBB and adsorption of residual gas atoms other than the activation agents become dominant, leading to an exponential decay of $\eta$.
\end{itemize}
For type-i, type-ii and type-iii activations, the first regime was either not observed or much shorter and less pronounced. \\
In order to test if this behavior also occurs if no Li is present during and after the activation process, we conducted additional lifetime measurements of a different bulk-GaAs sample activated with Cs-O$_2$ within the CAC. Preliminary results show that the predicted behavior is indeed observed in our setup for Cs-O$_2$ layers, with the second regime starting after about \qty{20}{\hour}. Since the lifetime measurements for both type-i activations presented here did not last longer than \qty{17}{\hour}, it is possible that the predicted trend was not observed during those measurements because they were ended earlier than those for the Cs-O$_2$-Li layer and therefore before the second regime was reached. However, all lifetime measurements of type~ii and type~iii lasted at least \qty{39}{\hour} and a short second regime was observed. Considering that the beginning of the second regime was observed after about \qty{6}{\hour} for both type~iv and type~v, the assumption made for type~i cannot be made for type~ii and type~iii. \\
After one of the additional lifetime measurement, at which point $\eta$ had declined below $1/e$ of its initial value and the pressure within the chamber had stabilized at about \qty{2e-11}{\milli\bar}, the laser spot was re-positioned to the location of the remaining maximum value of $\eta$, which was close to $\eta(0)$ at the initial laser spot position. During the subsequent lifetime measurement, all three regimes could be observed, although the first two were much less pronounced than during the preceding lifetime measurement. Hence, the behavior is also present if no residual Cs and O$_2$ from the activation process is present. This indicates that the mechanisms behind the first and second regime are mainly caused by the photo-electron emission process and leads to several assumptions:
\begin{itemize}
    \item
    The three regimes arise from an evolution of the surface-layer composition. Upon starting photo-electron emission, loosely bound parts of the surface layer that are more susceptible to adsorption and IBB are removed, quickly reducing $\eta$. Simultaneously, the surface-layer structure is optimized by desorption of loosely bound parts and formation of additional firmly bound components by re-adsorption of desorbed activation agents. This leads to the first two regimes: initially, the desorption of loosely bound surface-layer components and IBB from residual and desorbed atoms has a much greater impact on $\eta$ than the surface layer reconstruction, leading to the exponential decay observed during the first regime. As the amount of loosely bound components is drastically reduced, the impact of surface reconstruction becomes greater, leading to the increase in $\eta$ observed during the second regime until the surface layer reaches an equilibrium state, marked by a maximum in $\eta$. Finally, the surface layer deterioration due to IBB and adsorption from residual gas particles other than the activation ingredients becomes the dominant influence on the trend of $\eta$ because most ingredient particles that either remained after the activation or were desorbed due to IBB have been either pumped away or adsorbed on the surface. Hence, the shape of all three regimes depends on the chemical composition of the surface layer after the activation process and should be reproducible for a standardized activation scheme.
    \item
    Since Li inhibits the decay of the surface layer, it compensates the effect of additional IBB from Li atoms. Hence, if the enhancement of the surface-layer stability has a greater impact on the decay than the additional IBB, the first regime will be less pronounced and shorter. Also, the length of the second regime as well as the maximum of $\eta$ reached within it will increase with increasing Li dosage since more Li is available to contribute to the rearrangement process. Hence, the addition of Li will change the shape of all three regimes in two ways: by changing the composition of the surface layer and by contributing to the mechanisms that change the layer during the first and second regime. This, in combination with the previous prediction, likely caused the different quantum efficiency trends observed during this study.
    \item
    For lifetime measurements of the surface layer in an environment without residual agents from the activation process, e.g. during operation in a DC photo-electron gun at high voltage and high current, the first and second regimes are caused by activation agents desorbed from the surface through IBB, albeit in a less distinct form due to the lower amount of residual atoms. Hence, the third regime is dominant in this case. This may explain the observation of a double-exponential decay for both Cs-NF$_3$ and Cs-NF$_3$-Li during operation in a photo-electron gun \cite{herbert2020}.
    \item
    The electron-impact ionization cross sections for Cs, O$_2$ and N$_2$ is maximal at electron energies around or below \qty{100}{\electronvolt} and decreases with increasing electron energy. Hence, the decay observed in the first and third regime should be less severe for higher bias voltages, e.g. operation in a DC photo-electron gun at high voltage.
    \item
    As IBB is strongest in the electrostatic centre of the photocathode \cite{grames2011}, the decay during the first and third regimes will be stronger the closer the laser spot is positioned to the electrostatic centre. The influence of this effect on the observed quantum efficiency trends could not be verified in this study because the available manual adjustment did not allow for a precise determination of the exact position of the laser spot on the photocathode surface.
\end{itemize}
Since the exact cause for this behavior and its impact on the overall decay of the quantum efficiency could not be verified in the course of this study, we decided to analyse the data in two ways: considering the entire decay curve and considering the third regime only. This allows for a direct comparison of the exponential decay as well as taking into account the impact of the observed recovery of $\eta$. In order to separate the third regime, we defined an offset period $t_\text{off}$ determining the end of the first two regimes. This enables the data to be analyzed separately, including and excluding $t_\text{off}$. \\
The offset period $t_\text{off}$ was determined by calculating the first derivative of the smoothed data and then identifying the point at which the data curve begins to show the form of a negative exponential decay after the first zero-crossing of the derivative. Then, the lifetime $\tau^*$ of the decay was extracted by fitting Eqs.\ (\ref{eq:qe_t}) and (\ref{eq:charge_t}) to the data starting at $t = t_\text{off}$ and comparing the results. The value of $Q(\tau^*)$ was then calculated using Eq.\ (\ref{eq:charge_tau}) with $I_0^* = I(t=t_\text{off})$. \\
In order to include the offset period, $\tau$ was determined by fitting Eqs. (\ref{eq:qe_t}) and (\ref{eq:charge_t}) to the data (type~i) or, if this fit did not produce a good match due to the distorted trend of the data (type~ii to v), by using the approximation
\begin{equation}
    \tau \approx \tau^* + t_\text{off} \quad .
\end{equation}
The lifetime charge $Q(\tau)$ was then calculated in two ways: using Eq.\ (\ref{eq:charge_tau}) and using the approximation
\begin{equation}
    Q(\tau) \approx Q(\tau^*) + Q_\text{off} \quad ,
\end{equation}
with the charge extracted during the offset period $Q_\text{off} = Q(t=t_\text{off})$. The resulting values for $Q(\tau)$ from both methods where in good agreement within their respective uncertainties for our data. \\
However, the chosen offset approach displays significant discrepancies considering quantum efficiency and charge when considering the data excluding the offset period, as can be seen when comparing $\eta_0^* = \eta(t_\text{off})$. For type~i, $t_\text{off} = 0$ and, hence, $\eta_0^* = \eta_0$. For all other types, $\eta$ has already partially decayed at $t = t_\text{off}$, ranging from about \qty{90}{\percent} of $\eta_0$ for type~ii to about \qty{60}{\percent} of $\eta_0$ for type~v. The effect of lower $\eta_0^*$ and, hence, $I_0^*$ leads to a lower extracted charge $Q(\tau^*)$. Therefore, $r_{\eta_0}^*$ and $r_{Q(\tau^*)}$ are smaller and the effect of Li is misrepresented in these cases. In order to estimate the lifetime charge $Q(\tau^*)_\text{est}$ that would be extracted during a single-exponential decay until $t=\tau^*$, we assumed $r_{\eta^*} = r_\eta$, i.e. $\eta_0^* = \eta_0$ and $I_0^* = I_0$ for all types and calculated $Q(\tau^*)_\text{est}$ for each type using \mbox{Eq.\ (\ref{eq:charge_tau})}. \\
All resulting mean values for each type are shown in \mbox{Tab.\ \ref{tab:results}}. For each parameter, a factor of relative change or enhancement factor $r$ compared to type~i was calculated. The enhancement factors for lifetime and lifetime charge are related through Eq.\ (\ref{eq:charge_tau}):
\begin{equation}
    r_{Q(\tau)} = r_{I_0} \cdot r_\tau \quad .
    \label{eq:r_charge_tau_1}
\end{equation}
With $P_{\gamma_0} \approx$~const. for all types, $r_{I_0} \approx r_{\eta_0}$, yielding
\begin{equation}
    r_{Q(\tau)} \approx r_{\eta_0} \cdot r_\tau \quad .
    \label{eq:r_charge_tau_2}
\end{equation}
The results are in good agreement with this correlation for all types. Moreover, $r_\tau$ and $r_{Q(\tau)}$ are in agreement for each individual type within their respective ranges of uncertainty. Hence, the reduction of quantum efficiency caused by Li addition can be considered minimal. The only exception is type~v, where $r_{\tau^*}$ and $r_{Q(\tau^*)}$ are not in agreement due to the reasons discussed above. For all types, $r_{\tau^*}$ and $r_{Q(\tau^*)_\text{est}}$ are in agreement within their uncertainties, with $r_{Q(\tau^*)_\text{est}}$ producing values closer to $r_{\tau^*}$. Hence, this approach appears to better represent the extracted charge without offset behavior. For all activations using scheme~2, the results gained for the third regime only are lower than those with included offset period. Therefore, the results obtained for the third regime only can be used as a conservative lower bound for the effect of Li. \\
Types~ii and iii produced surprisingly similar results, considering that Li was not actively added to type-ii activations. The effect of active addition of Li is only visible when including the offset period. Activations using scheme~2b showed by a considerable margin the best results, with type~v performing best when including the offset. It is possible that this was amplified by the fact that this activation was directly subsequent to the type-iv activation, hence benefiting from the same additional enhancement due to residual Li that was observed for type-ii activations. Since no more than two consecutive activations using Li were conducted during this study, we were not able to determine if and in what way this effect is cumulative for further subsequent Li-enhanced activations. It is also of interest for how long the influence of Li lingers in the chamber after one activation with Li was conducted. Additional measurements not presented here suggest that up to seven consecutive activations without Li, conducted over the course of nine days, are necessary to reduce the traces of Li to a level that has no observable effect on the activated photocathode \cite{herbert2022}. \\
Overall, the type-iv and type-v activations using the \mbox{Cs-O$_2$-Li} scheme~2b showed a very promising performance, yielding an increase in lifetime by a factor between \num{7.5(1.0)} and \num{19(2)} as well as an increase in lifetime charge between \num{6.5(1.2)} and \num{16.5(2.4)} compared to the common Cs-O$_2$ type~i, with only minimal reduction of the quantum efficiency by a factor between \num{0.90(5)} and \num{0.86(4)}. The observed increase is also significantly larger than that achieved by more conventional methods. For example, an increase in lifetime charge by a factor of about \num{3} has been reported for masked activation \cite{grames2011}, while an increase in lifetime charge by a factor of up to \num{2.43(0.20)} has been shown for biasing the anode at a voltage of \qty{1}{\kilo\volt} \cite{yoskowitz2020, yoskowitz2024}. \\
While it is not clear to which extent the observed offset period would play a role when operating a photocathode with Li-enhanced surface layer within a photogun and, hence, with a stable and low base pressure, we expect the effect on lifetime and lifetime charge to be somewhere in the range between our results with and without considering the offset, corresponding to an increase by about one order of magnitude. \\

\begin{table*}[t]
    \caption{Figure-of-merit enhancement ratios $\mathcal{F}$ for different activation schemes as calculated with Eqs.\ (\ref{eq:fom_2}) for $\mathcal{F}(r_\eta,r_\tau)$ and (\ref{eq:fom_3}) for $\mathcal{F}(r_\gamma,r_{Q(\tau)})$, with and without considering the offset period.}
    \begin{tabular}{c|cccc}
    \hline
    \hline
    Type & ii & iii & iv & v \\
    & & & & \\
    $\frac{r_{\eta_0^*}}{r_{I_0^*}}$ & \num{0.96(0.11)} & \num{1.02(0.12)} & \num{1.04(0.10)} & \num{1.00(0.10)} \\
    & & & & \\
    $\frac{r_{\eta_0}}{r_{I_0}}$ & \num{0.98(0.11)} & \num{1.01(0.10)} & \num{0.98(8)} & \num{0.99(8)} \\
    & & & & \\
    $\mathcal{F}(r_{\eta_0},r_{\tau^*})$ & \num{3.6(5)} & \num{3.3(5)} & \num{8.8(1.4)} & \num{6.5(9)} \\
    $\mathcal{F}(r_\gamma,r_{Q(\tau*)_\text{est}})$ & \num{3.5(7)} & \num{3.3(7)} & \num{8.8(1.8)} & \num{6.4(1.3)} \\
    & & & & \\
    $\mathcal{F}(r_{\eta_0},r_\tau)$ & \num{4.1(4)} & \num{5.6(6)} & \num{13.0(1.7)} & \num{16.3(1.9)} \\
    $\mathcal{F}(r_{\gamma_0},r_{Q(\tau)})$ & \num{4.0(8)} & \num{6.1(1.2)} & \num{13.5(2.6)} & \num{16.3(2.7)} \\
    \hline
    \hline
    \end{tabular}
    \label{tab:results_fom}
\end{table*}

\subsection{Figure of merit}
\begin{figure}[h]
    \centering
    \includegraphics[width=\columnwidth]{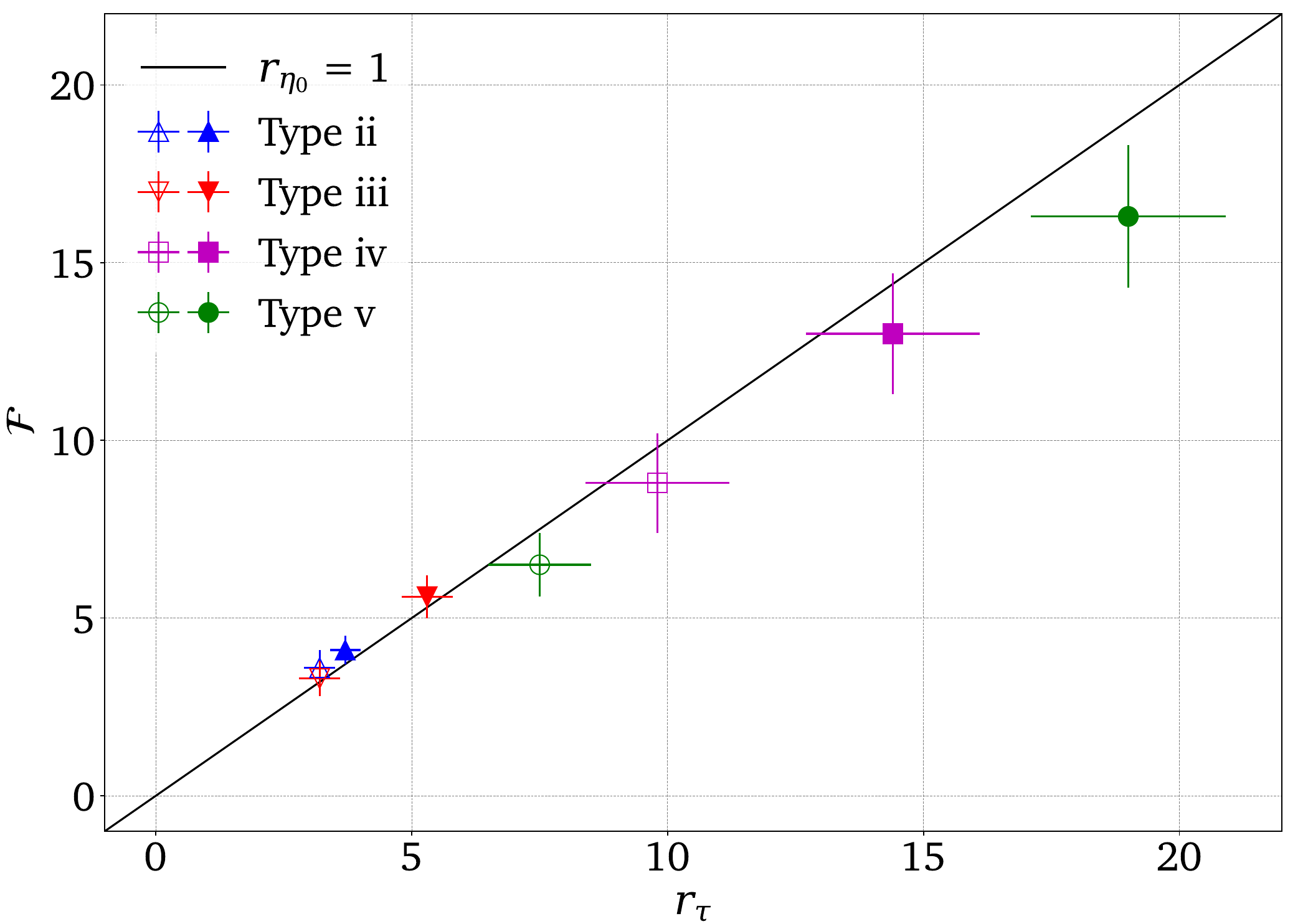}
    \caption{Figure-of-merit ratio $\mathcal{F}$ as a function of $r_\tau$. The values for the different activation types agree with $r_{\eta_0} = 1$ within their respective uncertainties, with the exception of type~v.}
    \label{fig:fom_comp_types}
\end{figure}

While most studies on surface-layer enhancement focus upon the effect on lifetime and extracted charge, the impact on quantum efficiency is equally important for high-current applications since a reduction of $\eta$ necessitates a higher laser power to reach the same current. This increases the heat load on the photocathode surface and, hence, reduces the lifetime due to heat desorption, mitigating the overall lifetime gain. \\
In order to better assess and compare the overall improvement of photocathode performance, one may introduce a figure of merit (FOM). For emission of spin-polarized electrons, a commonly used FOM connects quantum efficiency, lifetime and degree of spin-polarization to gain a measure for the performance of the spin-polarized electron source \cite{cultrera2020}
\begin{equation}
    \text{FOM} = \eta \cdot \tau \cdot \mathcal{P}^2 \quad .
\end{equation}
From this, one can derive a comparative figure of merit $\mathcal{F}$ for the improvement of photocathode performance by calculating the ratio of the FOMs for the enhanced method (index 2) and the standard method (index 1)
\begin{equation}
    \mathcal{F} = \frac{FOM_2}{FOM_1} = \frac{\eta_2}{\eta_1}\cdot\frac{\tau_2}{\tau_1}\cdot\frac{\mathcal{P}_2^2}{\mathcal{P}_1^2} = r_\eta \cdot r_\tau \cdot r_\mathcal{P}^2 \quad .
    \label{eq:fom_1}
\end{equation}
Using Eq.\ (\ref{eq:qe_lambda}), we find
\begin{equation}
    r_\eta = \frac{\eta_2}{\eta_1} = \frac{I_\text{p,2}P_{\gamma,1}\lambda_1}{I_\text{p,1}P_{\gamma,2}\lambda_2} = \frac{r_I}{r_\gamma \cdot r_\lambda} \quad ,
\end{equation}
with $r_\gamma = \frac{P_{\gamma, 2}}{P_{\gamma, 1}}$ and $r_\lambda = \frac{\lambda_2}{\lambda_1}$. When comparing an enhanced method with the standard method, one should use the same laser source for the measurements of both methods in order to exclude any effects from changes in the incident laser light. For typical laser sources used in photo-electron guns, one can assume $\lambda \approx$~const. and hence $r_\lambda \approx 1$. Combined with Eq.\ (\ref{eq:charge_tau}), this yields
\begin{equation}
    \mathcal{F} \approx \frac{1}{r_{\gamma_0}} \cdot r_{I_0} \cdot r_\tau \cdot r_\mathcal{P}^2 = \frac{1}{r_{\gamma_0}} \cdot r_{Q(\tau)} \cdot r_\mathcal{P}^2 \quad ,
\end{equation}
with the ratio of initial laser power $r_{\gamma_0}$ and the ratio of initial current $r_{I_0}$. Assuming\footnote{We make this assumption here since $r_\mathcal{P} \approx 1$ has been reported in all studies on enhanced surface layers using SL GaAs samples so far. Nevertheless, the effect on $\mathcal{P}$ needs to be carefully studied and taken into account to evaluate the overall performance of the surface layer.} $r_\mathcal{P} \approx 1$, this is reduced to
\begin{equation}
    \mathcal{F} \approx r_{\eta_0} \cdot r_\tau = \frac{1}{r_{\gamma_0}} \cdot r_{Q(\tau)} = \frac{r_{\eta_0}}{r_{I_0}} \cdot r_{Q(\tau)} \quad .
    \label{eq:fom_2}
\end{equation}
For $P_{\gamma,1} \approx P_{\gamma,2}$, $r_{\gamma_0} \approx 1$ and thus $r_{I_0} \approx r_{\eta_0}$, leading to
\begin{equation}
    \mathcal{F} = r_{\eta_0} \cdot r_\tau \approx r_{Q(\tau)} \quad ,
    \label{eq:fom_3}
\end{equation}
which contains the correlation between $r_\tau$ and $r_{Q(\tau)}$ from Eq.\ (\ref{eq:r_charge_tau_2}). \\
During operation in a gun, the beam current is usually kept at a pre-defined value depending on the experimental application the beam is used for. $I_1 \approx I_2$ or $r_{I_0} \approx 1$ and thus $r_{\gamma_0} \approx \frac{1}{r_{\eta_0}}$ gives
\begin{equation}
    \mathcal{F} = r_{\eta_0} \cdot r_\tau \approx r_{\eta_0} \cdot r_{Q(\tau)} \quad ,
    \label{eq:fom_4}
\end{equation}
containing the correlation between $r_\tau$ and $r_{Q(\tau)}$ from \mbox{Eq.\ (\ref{eq:r_charge_tau_1})} for the case $r_{I_0} \approx 1$. Hence, $r_{\eta_0}$ needs to be taken into consideration when evaluating the performance of an enhanced activation scheme, except when comparing the extracted charge for equal initial laser power. \\
In order to verify this model, we used Eq.\ (\ref{eq:fom_2}) to calculate both $\mathcal{F}(r_{\eta_0},r_\tau) = r_{\eta_0} \cdot r_\tau$ and $\mathcal{F}(r_{\gamma_0},r_{Q(\tau)}) = \frac{r_{\eta_0}}{r_{I_0}} \cdot r_{Q(\tau)}$. The results for the measurements presented here are shown in Tab.\ \ref{tab:results_fom}. The ratio $\frac{r_{\eta_0}}{r_{I_0}}$ is approximately 1 for all types, confirming that the initial laser power used during our experiments was approximately equal. Within the given uncertainties, the values of $\mathcal{F}(r_{\eta_0},r_\tau)$ and $\mathcal{F}(r_{\gamma_0},r_{Q(\tau)})$ are in good agreement, confirming the validity of Eq.\ (\ref{eq:charge_t}) for the data presented here. Since $\mathcal{F}(r_{\eta_0},r_\tau)$ produces lower uncertainties, we assume that this method of calculating $\mathcal{F}$ yields higher precision for the data presented here and use the resulting values for comparison with other methods. \\
Figure \ref{fig:fom_comp_types} shows $\mathcal{F}$ as a function of $r_\tau$ for the different activation types used in this work. For higher dosage of Li, the data points lie below, but close to the diagonal line representing $r_{\eta_0} = 1$, meaning that $\eta$ is slightly reduced. Values for ideal enhancement would lie either on the line or above it, i.e. $r_{\eta_0} \geq 1$. Hence, this type of plot can be used as benchmark plot for surface layer enhancement methods.
 \\

\begin{table*}[t]
    \caption{Comparison of enhancement factors $r$ at different wavelengths for different methods, as available in literature. The mode of operation used in the experiment is given with 'a' for $r_{\gamma,0} \approx$~1 and 'b' for $r_{I,0} \approx$~1. If only $r_\eta$ is given, the value is derived from the maximum quantum efficiency measured directly after activation. Bold values have been stated directly in the corresponding source. The uncertainties of values marked with $^\dagger$ were not given and have been estimated. All other values for $r$ were not stated explicitly and have been extracted from the data shown in the corresponding reference. Values of $\mathcal{F}$ were calculated using Eq.\ (\ref{eq:fom_2}). Values of $r$ and $\mathcal{F}$ marked with $^\ddagger$ have been estimated using either Eq.\ (\ref{eq:fom_3}) or Eq.\ (\ref{eq:fom_4}). All studies presented here used GaAs(100) samples, either as bulk or superlattice (SL).}
    \begin{tabular}{p{1.5cm}|ccccccccc}
    \hline
    \hline
    Agents & Sample & $\lambda$ in nm & $P_\gamma$ in \qty{}{\micro\watt} & Mode & $r_{\eta,0}$ & $r_\tau$ & $r_{Q(\tau)}$ & $\mathcal{F}$ & Ref. \\
    \hline
    \multirow{9}{5em}{Cs-O$_2$-Li} & \multirow{3}{2em}{bulk} & \multirow{3}{3em}{\num{785(2)}} & \multirow{3}{4em}{\num{40} to \num{60}} & \multirow{3}{0.5em}{a} & \textbf{\num{0.95(7)}} & - & - & - & \multirow{3}{2em}{this work} \\
    &  &  &  &  & \textbf{\num{0.86(4)}} & \textbf{\num{7.5(1.0)}} & \textbf{\num{6.5(1.2)}} & \num{6.5(9)} &  \\
    &  &  &  &  & \textbf{\num{0.86(4)}} & \textbf{\num{19(2)}} & \textbf{\num{16.5(2.4)}} & \num{16.3(1.9)} &  \\
    & & & & & & & & & \\
    & \multirow{2}{2em}{bulk} & \num{850} & ? & \multirow{2}{0.5em}{a} & \num{1.0(1)} & \num{2.7(1)} & \num{2.7(3)}$^\ddagger$ & \num{2.7(3)} & \multirow{2}{1.5em}{\cite{mulhollan2010}} \\
    &  & \num{633} & ? & & \num{1.0(1)} & - & - & - &  \\
    & & & & & & & & & \\
    & \multirow{2}{2em}{bulk} & \multirow{2}{1.5em}{\num{405}} & \multirow{2}{2em}{\num{5(3)}} & \multirow{2}{0.5em}{a} & \num{1.0(1)} & - & - & - & \multirow{2}{1.5em}{\cite{kurichiyanil2019}} \\
    &  &  &  &  & \num{0.78(8)} & \num{3.5(5)} & \num{2.7(5)}$^\ddagger$ & \num{2.7(5)} &  \\
    \hline
    \multirow{11}{5em}{\mbox{Cs-NF$_3$-Li}} & \multirow{2}{2em}{bulk} & \num{850} & ? & a & \num{1.38(9)} & - & - & - & \multirow{2}{1.5em}{\cite{mulhollan+bierman2008}} \\
    & & \num{780} & ? & a & \num{1.1(1)} & - & - & - &  \\
    & & & & & & & & & \\
    & SL & \num{780} & ? & a & \num{0.9(1)} & \num{5.7(1)} & \num{5.1(7)}$^\ddagger$ & \num{5.1(7)} & \cite{mulhollan2010} \\
    & & & & & & & & & \\
    & \multirow{2}{2em}{bulk} & \multirow{2}{3em}{\num{780(5)}} & \num{26(1)} & a & \textbf{\num{1.40(0.05)}} & - & - & - & \multirow{2}{1.5em}{\cite{herbert2022}} \\
    & & & \num{3800(500)} & a & \textbf{\num{1.13(2)}} & \textbf{\num{1.7(6)}} & \num{1.6(6)} & \num{1.9(7)} &  \\
    & & & & & & & & \\
    & bulk & \num{773(5)} & \num{50(2)} & a & \textbf{\num{1.2(4)}} & \textbf{\num{1.9(6)}} & \num{2.3(1.0)}$^\ddagger$ & \num{2.3(1.0)} & \cite{herbert2022} \\
    & & & & & & & & \\
    & bulk & \num{633} & ? & a & \num{1.49(4)} & \num{1.8(2)} & \num{2.7(3)}$^\ddagger$ & \num{2.7(3)} & \cite{mulhollan+bierman2008} \\
    \hline
    \multirow{5}{5em}{Cs-Sb} & \multirow{5}{2em}{bulk} & \multirow{2}{1.5em}{\num{780}} & \multirow{2}{1em}{\num{10}} & \multirow{2}{0.5em}{a} & \num{0.023(2)} & - & - & - & \cite{bae2019} \\
    & & &  &  & \num{0.27(7)} & - & - & - & \cite{bae2020} \\
    & & & & & & & & \\
    &  & \multirow{2}{1.5em}{\num{505}} & \multirow{2}{1em}{\num{20}} & \multirow{2}{0.5em}{a} & \num{0.35(3)} & \num{161(14)}$^\ddagger$ & \textbf{\num{56.4(1)}}$^\dagger$ & \num{56.4(1)}$^\ddagger$ & \cite{bae2019} \\
    & & &  &  & \num{0.27(12)}$^\dagger$ & \textbf{\num{26.9(1)}}$^\dagger$ & \num{7.3(3.2)}$^\ddagger$ & \num{7.3(3.2)} & \cite{bae2020} \\
    \hline
    \multirow{12}{5em}{Cs-O$_2$-Sb} & \multirow{3}{2em}{bulk} & \multirow{5}{1.5em}{\num{780}} & \num{20} & a & \num{0.32(3)} & - & - & - & \cite{bae2019} \\
    &  &  & ? & a & \num{0.32(6)} & \num{32(7)}$^\ddagger$ & \textbf{\num{10.0(2)}}$^\dagger$ & \num{10.0(2)}$^\ddagger$ & \cite{cultrera2020} \\
    &  &  & \num{10} & a & \num{0.30(4)} & \textbf{\num{6.9(1)}}$^\dagger$ & \num{2.1(3)}$^\ddagger$ & \num{2.1(3)} & \multirow{2}{1.5em}{\cite{bae2020}} \\
    & SL &  & \num{10} & a & \num{0.36(1)} & \textbf{\num{7.2(1.1)}}$^\dagger$ & \num{2.6(8)}$^\ddagger$ & \num{2.6(8)} &  \\
    & bulk &  & ? & b & \num{0.08(4)} & \num{1.5(4)}$^\ddagger$ & \num{1.5(4)} & \num{0.12(7)}$^\ddagger$ & \cite{bae2022a} \\
    & & & & & & & & & \\
    & \multirow{4}{2em}{bulk} & \multirow{4}{1.5em}{\num{505}} & \num{20} & a & \num{0.57(4)} & \num{31.4(2.3)}$^\ddagger$ & \textbf{\num{17.9(3)}}$^\dagger$ & \num{17.9(3)}$^\ddagger$ & \cite{bae2019} \\
    &  &  & \multirow{2}{3.5em}{\num{5} to \num{30}} & \multirow{2}{0.5em}{b} & \num{0.23(1)} & \textbf{\num{12.2(4)}}$^\dagger$ & \num{12.2(4)}$^\ddagger$ & \num{2.8(2)} & \multirow{2}{1.5em}{\cite{cultrera2020}} \\
    & & &  &  & \num{0.23(1)} & \num{24(9)} & \num{24(11)} & \num{5.5(2)} &  \\
    & & & \num{20} & a & \num{0.56(13)} & \textbf{\num{9.9(1)}}$^\dagger$ & \num{5.5(1.3)}$^\ddagger$ & \num{5.5(1.3)} & \cite{bae2020} \\
    & & & & & & & & & \\
    & bulk & \num{488} & ? & b & \num{0.4(3)} & \num{0.8(4)}$^\ddagger$ & \num{0.8(4)} & \num{0.32(0.29)} & \cite{bae2022a} \\
    \hline
    \multirow{3}{5em}{Cs-Te} & \multirow{3}{2em}{bulk} & \num{780} & \num{40} & a & \num{0.12(5)} & \num{0.7(1)} & \num{0.08(4)}$^\ddagger$ & \num{0.08(4)} & \cite{bae2022b} \\
    & & & & & & & & & \\
    & & \num{532} & \num{50} & a & \num{1.8(3)} & \num{2.8(1)}$^\ddagger$ & \textbf{\num{5.0(3)}}$^\dagger$ & \num{5.0(3)}$^\ddagger$ & \cite{bae2018} \\
    \hline
    \multirow{3}{5em}{Cs-O$_2$-Te} & \multirow{3}{2em}{bulk} & \num{780} & \num{40} & a & \num{1.8(3)} & \num{0.9(1)} & \num{1.6(3)}$^\ddagger$ & \num{1.6(3)} & \cite{bae2022b} \\
    & & & & & & & & & \\
    & & \num{532} & \num{50} & a & \num{1.2(2)} & \num{4.6(9)}$^\ddagger$ & \num{5.5(5)} & \num{5.5(5)}$^\ddagger$ & \cite{rahman2019b} \\
    \hline
    Cs-Te-K & bulk & \num{602} & ? & a & - & \textbf{\num{19(3)}} & - & - & \cite{kuriki+masaki2019} \\ 
    \hline\hline
    \end{tabular}
    \label{tab:comparison}
\end{table*}

\subsection{Comparison with other enhancement methods}
Many studies on Li, Sb, and Te as enhancement agents for GaAs photocathodes with Cs and O$_2$ or NF$_3$ have been conducted. Table \ref{tab:comparison} lists the results with values for $\mathcal{F}$ as calculated in this work. If enhancement factors were not given explicitly, they were estimated by extracting the data from the respective reference and calculating the ratios from the extracted data. Some studies were conducted for several different dosages of the involved enhancement agent, in which case we picked the dosage with best performance as shown in the corresponding paper. Substantial work has been conducted on Cs-O$_2$-Sb, followed by Cs-NF$_3$-Li, Cs-O$_2$-Li, and Cs-Sb. Cs-Te, Cs-O$_2$-Te, and Cs-Te-K have so far only received minor attention.  \\

\subsubsection{Cs-O$_2$-Li}
Only limited research has been carried out on \mbox{Cs-O$_2$-Li} prior to this work. At \mbox{Photo-CATCH}, Kurichiyanil et al. previously studied this surface layer at \qty{405}{\nano\meter} \cite{kurichiyanil2019}. Quantum efficiency was not significantly impacted when adding Li to a Co-De scheme. However, the lifetime performance of the layer produced by this Li-enhanced scheme was not investigated. The impact of Li on a two-stage activation process \cite{stocker1975} was also analyzed, comparing a Nagoya+Nagoya\footnote{The Nagoya activation procedure has been named after the Nagoya university where it was first developed, cf.\ Ref.\ \cite{togawa1998}.} with a Nagoya+Co-De(Li) procedure. This yielded a quantum efficiency reduction by a factor of \num{0.78(8)}, but a lifetime enhancement factor of \num{3.5(5)}, corresponding to $r_{Q(\tau)} = \num{2.7(5)}$. \\
G. A. Mulhollan conducted research on the Cs-O$_2$-Li layer at both \qty{633}{\nano\meter} and \qty{850}{\nano\meter} \cite{mulhollan2010}. The observed impact on quantum efficiency was negligible. A lifetime increase in regard to CO$_2$ exposure by a factor of \num{2.7(1)} was observed at \qty{850}{\nano\meter}, identical to the one observed in \cite{kurichiyanil2019}. \\
The impact of quantum efficiency reported in \cite{mulhollan2010} and \cite{kurichiyanil2019} is comparable to our results obtained for type~ii and iii, i.e. lower dosages of Li. Hence, we assume that both previous studies used lower amounts of Li. The smaller values for $r_\tau$ and $r_{Q(\tau)}$ are likely caused by a combination of lower Li dosage and differences in the activation procedure. \\
Aside from our work, no measurements at wavelengths in the range between \qty{750}{\nano\meter} and \qty{785}{\nano\meter} that is of interest for emission of spin-polarized electrons from GaAs/GaAsP photocathodes have been conducted so far.

\subsubsection{Cs-NF$_3$-Li}
Beside Cs-O$_2$-Sb, \mbox{Cs-NF$_3$-Li} has received the most attention so far. Several studies have been carried out at or close to \qty{780}{\nano\meter} \cite{mulhollan+bierman2008, mulhollan2010, herbert2020, herbert2022}, yielding values of $r_{\eta_0}$ between \num{0.9} and \num{1.4}, as well as $r_\tau$ between \num{1.7} and \num{5.7}, corresponding to $r_{Q(\tau)}$ between \num{1.6} and \num{5.1}. The best performance was reported by G. A. Mulhollan for an SL sample \cite{mulhollan2010}, with $r_{\eta_0} = \num{0.9(1)}$, $r_\tau = \num{5.7(1)}$ and $r_{Q(\tau)} = \num{5.1(7)}$. Activating the same sample with a lower dosage showed $r_{\eta_0} = \num{1.2(1)}$, $r_\tau = \num{2.5(1)}$ and $r_{Q(\tau)} = \num{3.0(3)}$, which is comparable to the other reported results. Hence, the addition of Li appears to increase lifetime and lifetime charge at the expense of quantum efficiency, with higher dosage resulting in a higher impact. Further data presented in \cite{mulhollan2010} suggest an upper limit of Li dosage after which a further increase in dosage has negative effects on the photocathode performance, with an observed reduction in quantum efficiency down to a factor of \num{0.07(1)} for higher Li dosages. Hence, the amount of used Li, as well as the exact procedure of its addition to the activation process, needs to be carefully optimized to obtain the best possible photocathode performance. \\
G. A. Mulhollan \cite{mulhollan2010} also investigated the impact of the altered surface layer on the degree of spin-polarization, comparing $\mathcal{P}(\lambda)$ of Cs-NF$_3$ and Cs-NF$_3$ for both bulk and SL samples. This yielded $r_\mathcal{P} \approx 1$ for both samples, with $\lambda$ in the range of \qtyrange{660}{890}{\nano\meter} and \qtyrange{650}{800}{\nano\meter} for bulk and SL, respectively. In both cases, a high dosage of Li was used, resulting in a high reduction of $\eta$ with $r_\eta \approx \num{0.38}$ at $\lambda = \qty{780}{\nano\meter}$. Since even the addition of significant amounts of Li does not decrease $\mathcal{P}$ significantly, the Li dosage for this layer can be adjusted freely in order to maximize $\mathcal{F}$ by balancing $r_\eta$ and $r_\tau$. \\ 
Cs-NF$_3$-Li has been tested at Jefferson National Lab in a DC photo-gun at -\qty{144}{\kilo\volt} and \qty{100(5)}{\micro\ampere}, yielding only a small increase in lifetime \cite{herbert2020, herbert2022}. However, this result needs to be treated with caution for several reasons. First, the Li-enhanced cathode was initially run at \qty{20}{\micro\ampere} for \qty{18}{\hour} before switching to \qty{100}{\micro\ampere}, and the lifetime comparison was done for the data at \qty{100}{\micro\ampere} only. Second, the Li-enhanced activation was conducted first, followed by a conventional Cs-NF$_3$ activation. As discussed above, the effect of residual Li from the previous activation might have influenced the photocathode performance. Surface quantum efficiency scans conducted both before and after beam extraction showed that while no significant quantum efficiency could be observed across the entire surface after beam extraction from the Cs-NF$_3$ coated sample, the Li-enhanced surface layer had only been depleted at and close to the position of beam extraction, with plenty of quantum efficiency remaining across the rest of the photocathode surface \cite{herbert2022}. Hence, even if the direct increase in lifetime was low, the time of operation could have been increased significantly by simply re-positioning the laser spot on the surface for further beam extraction. \\
The best result reported for Cs-NF$_3$-Li, yielding $\mathcal{F} = \num{5.1(7)}$, is comparable to our results excluding the offset period. Hence, it is safe to assume that our method for Cs-O$_2$-Li performs equal or better to those used for Cs-NF$_3$-Li so far. This may originate from different amounts of Li used during the respective activation procedures. Further studies are required to determine the optimal Li dosage for the best possible performance enhancement. Also, the differences, if any, between \mbox{Cs-NF$_3$-Li} and \mbox{Cs-O$_2$-Li} need to be investigated.

\subsubsection{Cs-Sb}
Cs-Sb has so far shown a promising performance at a wavelength of \qty{505}{\nano\meter}. While it reduces the quantum efficiency by a factor between \num{0.27} and \num{0.35}, both lifetime and lifetime charge are increased significantly. A lifetime charge enhancement factor of \num{56.4(1)}, corresponding to an increase in lifetime by a factor of \num{161(14)} for the shown reduction in quantum efficiency by a factor of \num{0.35(3)}, has been observed by Bae et al.\ \cite{bae2019}. However, this result was not consistently reproduced by a subsequent publication of the same authors, which showed a lifetime enhancement factor of \num{26.9(1)}, corresponding to a lifetime charge enhancement factor of \num{7.3(3.2)} for the observed quantum efficiency reduction by a factor of \num{0.27(12)} \cite{bae2020}. The results yield diverging values for $\mathcal{F}$ between \num{7.3(3.2)} and \num{56.4(1)}, making an accurate assessment of the surface layer performance difficult. \\
At \qty{780}{\nano\meter} or similar wavelengths, the performance of Cs-Sb compared to standard surface layers has not been studied in detail. Bae et al. \cite{bae2020} measured $\mathcal{P}$ as a function of $\lambda$ for a bulk-GaAs sample activated with Cs-Sb and compared it to a Cs-O$_2$ activated sample, yielding $r_\mathcal{P} = \num{0.82(9)}$ in the range of \qtyrange{760}{840}{\nano\meter}. The behavior in terms of quantum efficiency has been studied by the same authors, resulting in conflicting factors of \num{0.023(2)} and \num{0.27(7)} \cite{bae2019, bae2020}. It remains to be seen if the high increase in lifetime and lifetime charge can be consistently reproduced in future studies and is confirmed at higher wavelengths. If so, the high enhancement factors could mitigate the observed reduction in quantum efficiency, resulting in a performance similar or better than our results for Cs-O$_2$-Li and hence making Cs-Sb a good surface layer candidate for high-current applications.

\subsubsection{Cs-O$_2$-Sb}
At \qty{505}{\nano\meter}, Bae et al.\ observed a quantum efficiency reduction by a factor of \num{0.57(4)} and a charge lifetime increase by a factor of \num{17.9(3)}, corresponding to an increase in lifetime by a factor of \num{31.4(2.3)} \cite{bae2019}. While a subsequent publication of the same authors was able to reproduce the reduction in quantum efficiency at the same wavelength, observing a factor of \num{0.56(0.13)}, the reported increase in lifetime was significantly lower, showing only a factor of \num{9.9(1)} \cite{bae2020}. Cultrera et al. reported a reduction in quantum efficiency by a factor of \num{0.23(1)} and enhancement factors for dark lifetime and charge lifetime of up to \num{12.2} and \num{58}, respectively \cite{cultrera2020}. However, the latter two factors are not directly comparable to the enhancement factors of lifetime and lifetime charge presented here, since the relation
\begin{equation}
    \eta(t, Q) = \eta_0 \cdot e^{-t/\tau_\text{d}} \cdot e^{-Q(t)/\tau_\text{c}}
    \label{eq:qe_t_q}
\end{equation}
with dark lifetime $\tau_\text{d}$ in units of time and charge lifetime $\tau_\text{c}$ in units of charge was used. The former was obtained from measurements with \qty{2}{\percent} duty cycle and then used to fit the above function to \qty{100}{\percent} duty-cycle measurements in order to obtain $\tau_\text{c}$. Hence, two cases can be examined: (a) $\tau_\text{d} = \tau$ for the \qty{2}{\percent} duty cycle and (b) $\tau_\text{d} \neq \tau$ and $\tau_\text{c} \neq Q(\tau)$ for the \qty{100}{\percent} duty cycle. Case (a) yields an enhancement factor of \num{12.2(4)} for both lifetime and lifetime charge. In order to compare the data from case~(b) with those from this work and other publications, $\tau$ and $Q(\tau)$ were extracted. By equating Eqs.\ (\ref{eq:qe_t}) and (\ref{eq:qe_t_q}), inserting $t=\tau$ and using Eq.\ (\ref{eq:charge_tau}), one obtains
\begin{equation}
    \frac{1}{\tau} = \frac{1}{\tau_\text{d}} + \frac{(1-\frac{1}{e})\cdot I_0}{\tau_\text{c}} \quad .
\end{equation}
Using this relation and Eq.\ (\ref{eq:charge_tau}) with $I_0 = \qty{250(50)}{\nano\ampere}$ yields $r_\tau = \num{24(9)}$ and $r_{Q(\tau)} = \num{24(11)}$. Interestingly, this is almost double the result obtained from the \qty{2}{\percent} duty-cycle data. It is possible that this is caused by combining positive effects of Sb addition on the surface layer robustness against residual gas adsorption and IBB. While the former is the dominant process for low-duty-cycle measurements, the latter becomes the main influence in full-duty-cycle measurements. A similar behavior can be observed for Cs-O$_2$-Li when comparing low-duty-cycle data from \cite{kurichiyanil2019} with the data obtained in this work at full duty cycle. Within the given uncertainties, the results of case~(b) are well comparable with the results for Cs-O$_2$-Li presented in this work. \\
The values of $\mathcal{F}$ obtained at \qty{505}{\nano\meter} range from \num{2.8(2)} to \num{17.9(3)}. Interestingly, the values for case (b) of \cite{cultrera2020} and for \cite{bae2020} are in good agreement, despite twice the dosage of Sb being used in the former publication. It appears that within a certain range of Sb dosage, the quantum efficiency is reduced in an approximately linear way when adding more Sb, as evident when comparing $r_{\eta_0}$. If this holds true, the amount of Sb could be adjusted to minimize quantum efficiency reduction while retaining the optimal performance increase. \\
At \qty{780}{\nano\meter}, $r_{\eta_0} \approx \num{0.3}$ has been reported for comparable Sb dosages. The only exception is \cite{bae2022a}, where Bae et al.\ studied the decay of Cs-O$_2$-Sb in a -\qty{200}{\kilo\volt} photo-electron gun during constant extraction of \qty{1}{\milli\ampere} beam. However, the reduction in $\eta$ by a factor of \num{0.08(4)} was measured after transferring the samples from the activation chamber to the gun chamber with a vacuum suitcase, with the transport procedure taking about one day. Hence, this value is not well comparable since it most likely contains unknown effects of the transfer process on the surface layer. \\
Cultrera et al.\ reported $r_\eta = \num{0.32(6)}$ and $r_{Q(\tau)} = \num{10.0(2)}$ for a Sb thickness of \qty{0.25}{\nano\meter}, corresponding to $r_\tau = \num{43(2)}$ \cite{cultrera2020}.  Bae et al.\ reported $r_\eta = \num{0.30(4)}$ and $r_\tau = \num{6.9(1)}$, corresponding to $r_{Q(\tau)} = \num{2.1(3)}$ \cite{bae2020}. The same authors reported an estimated enhancement factor of \num{1.5(4)} for both lifetime and charge during \qty{1}{\milli\ampere} operation in a -\qty{200}{\kilo\volt} photo-electron gun \cite{bae2022a}. However, all of these values are estimates taken from quantum-efficiency measurements at \qty{780}{\nano\meter} before and after lifetime measurements at the respective wavelengths (\qty{505}{\nano\meter} for \cite{cultrera2020} and \cite{bae2020}, and \qty{488}{\nano\meter} for \cite{bae2022a}) and should therefore be treated with caution. The higher decrease in quantum efficiency as well as the higher increase in both lifetime and lifetime charge reported in \cite{cultrera2020} result from double the dosage of Sb being used in comparison to the other studies. \\
The impact on spin-polarization was investigated by Cultrera et al. \cite{cultrera2020}, applying Cs-Sb layers with varying thickness of Sb onto a bulk-GaAs. This yielded a clear dependency of $\mathcal{P}$ on the applied Sb thickness, yielding ratios ranging from $r_\mathcal{P} \approx \num{1.0}$ for \qty{0.12}{\nano\meter} Sb thickness to $r_\mathcal{P} \approx \num{0.6}$ for \qty{0.5}{\nano\meter} Sb thickness within the wavelength range of \qtyrange{750}{850}{\nano\meter}. For an Sb thickness of \qty{0.25}{\nano\meter}, which yielded the best balance between reduction of quantum efficiency and increase in lifetime, a spin-polarization ratio of about $r_\mathcal{P} \approx \num{0.9}$ was observed in the same range of $\lambda$. This value was confirmed by Bae et al. \cite{bae2020}, yielding $r_\mathcal{P} = \num{0.88(6)}$ in the wavelength range of \qtyrange{760}{840}{\nano\meter} for a Cs-O$_2$-Sb layer with Sb thickness of \qty{0.25}{\nano\meter} on a bulk-GaAs sample. \\
Bae et al.\ \cite{bae2020} also investigated the effect of Cs-O$_2$-Sb on an SL sample at \qty{780}{\nano\meter}, yielding $r_{\eta_0} = \num{0.36(1)}$, $r_\tau = \num{7.2(1.1)}$ and $r_{Q(\tau)} = \num{2.6(8)}$, in good agreement to the values obtained for a bulk sample. No significant effect on the degree of spin-polarization was observed, with $r_\mathcal{P} \approx \num{1.0}$ in the wavelength range of \qtyrange{750}{800}{\nano\meter}. \\
The resulting values for $\mathcal{F}$, excluding those from \cite{bae2022a} for reasons stated above, are between \num{2.1} and \num{17.9}. This is well comparable to the results obtained with \mbox{Cs-O$_2$-Li} so far. In the wavelength range that is of interest for SL photoacthodes, Cs-O$_2$-Li appears to perform better than Cs-O$_2$-Sb, with the highest reported value for $\mathcal{F}$ being \num{10(2)} at \qty{780}{\nano\meter}. \\
The large increase in lifetime and lifetime charge observed for Cs-O$_2$-Sb is diminished by a significant reduction of quantum efficiency. However, the results reported at \qty{505}{\nano\meter} incident wavelength suggest that Cs-O$_2$-Sb has not been optimized to its full potential yet. Hence, further studies on the performance of Cs-O$_2$-Sb are required in order to reliably compare the performance of \mbox{Cs-O$_2$-Li} and Cs-O$_2$-Sb. Also, Cs-NF$_3$-Sb has not been investigated so far and may represent another potential enhanced surface candidate.

\subsubsection{Cs-Te, Cs-O$_2$-Te, and Cs-Te-K}
The available data on Cs-Te, Cs-O$_2$-Te, and Cs-Te-K are sparse. For Cs-Te, the results are inconclusive. While a significant increase in both $\eta$ and $Q(\tau)$ by respective factors of \num{1.8(3)} and \num{5.0(3)} has been observed at \qty{532}{\nano\meter} \cite{bae2018}, a significant decrease in $\eta$ and $Q(\tau)$ by respective factors of \num{0.12(5)} and \num{0.08(4)} was shown at \qty{780}{\nano\meter} \cite{bae2022b}. For Cs-Te with \qty{1.2}{\nano\meter} thickness on a bulk-GaAs sample, Bae et al. \cite{bae2018} reported no significant change in $\mathcal{P}$ compared to a Cs-O$_2$ layer, with $r_\mathcal{P} \approx \num{1.0}$ for $\lambda$ in the range of \qtyrange{620}{880}{\nano\meter}. Hence, the enhancing properties of Cs-Te need to be studied further. \\
Combining Te with O$_2$ to form Cs-O$_2$-Te appears to have a positive effect on the quantum efficiency, with mixed results for lifetime and lifetime charge and varying performance dependent on laser wavelength. At \qty{780}{\nano\meter}, $\eta$ is increased significantly by a factor of \num{1.8(3)}, while no significant effect on the lifetime was observed \cite{bae2022b}. At \qty{532}{\nano\meter}, the increase in quantum efficiency is less significant with an observed factor of \num{1.2(2)}. The charge lifetime, however, is increased significantly by a factor of \num{5.5(5)}, which is in the same range as the conservative lower estimate of \num{6.5(1.2)} observed in our work, albeit at a different wavelength \cite{rahman2019b}. Hence, further work on Cs-O$_2$-Te is needed to allow a proper evaluation of its performance. An evaluation of Cs-NF$_3$-Te has so far not been conducted and would also be of interest. \\
Cs-Te-K showed a significant increase in lifetime by a factor of \num{19(3)} \cite{kuriki+masaki2019}, comparable to the lifetime increase presented in this work. However, while no direct comparison between Cs-Te-K and Cs-O$_2$ in terms of quantum efficiency was presented, the reported value of about \qty{0.04}{\percent} for Cs-Te-K is two orders of magnitude lower than commonly reported values for both Cs-O$_2$ and Cs-NF$_3$ activation of bulk-GaAs. Hence, the overall performance of Cs-Te-K is likely lower than that of standard surface layers. \\
Further research on these types of enhanced surface layers could be of interest. Cs-O$_2$-Te has not received as much attention as Cs-O$_2$-Li, Cs-NF$_3$-Li, or Cs-O$_2$-Sb, and Cs-NF$_3$-Te has not been investigated at all. While the more exotic Cs-Te-K layer yielded low quantum efficiencies that would hinder its use, the addition of O$_2$ in order to form Cs-O$_2$-Te-K could have a beneficial effect on quantum efficiency, similar to the addition of O$_2$ to Cs-Te. Further blind spots of systematic evaluations comprise the agents Cs-O$_2$-Li-Te, Cs-NF$_3$-Li-Te, \mbox{Cs-O$_2$-Li-Sb}, and Cs-NF$_3$-Li-Sb.

\subsubsection{Comparison for \num{750}-\qty{785}{\nano\meter} incident light}
\begin{figure}[h]
    \centering
    \includegraphics[width=\columnwidth]{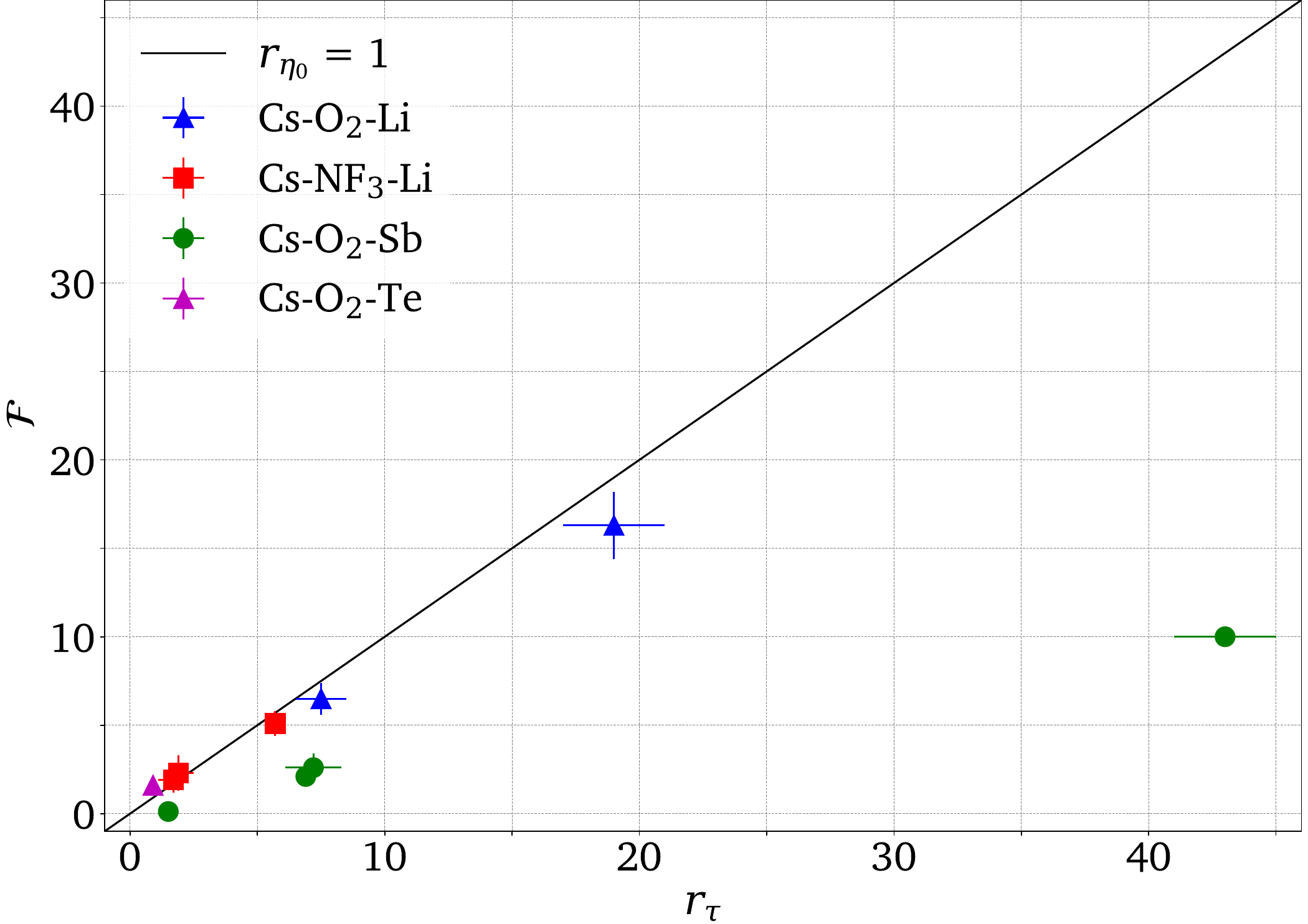}
    \caption{Figure-of-merit ratio $\mathcal{F}$ as a function of $r_\tau$ for different activation agents in the wavelength range of \qtyrange{750}{785}{\nano\meter}. The values have been taken from Tab.\ \ref{tab:comparison}. Both Cs-O$_2$-Li and Cs-NF$_3$ have so far yielded values close to $r_{\eta_0} = 1$. While Cs-O$_2$-Sb has yielded the highest value for $r_\tau$, it is also furthest away from $r_{\eta_0} = 1$, and, hence, yields lower values for $\mathcal{F}$. The only available value for Cs-O$_2$-Te lies above $r_{\eta_0} = 1$, but is the lowest in comparison with the results of the other agents since no significant increase in lifetime has been reported for that layer.}
    \label{fig:fom_comp}
\end{figure}

Overall, comparing our values including the offset period to the reported and calculated performance of other methods in the range of \qty{750}{\nano\meter} to \qty{785}{\nano\meter} shows that the method tested in this work performs best in regard to lifetime charge. A better value for $r_\tau$ has only been reported for Cs-O$_2$-Sb, although this result has been approximated from an estimated value of $r_{Q(\tau)}$. Also, this impressive increase in lifetime is greatly diminished by the significant reduction of quantum efficiency by almost \qty{70}{\percent}, as is apparent when comparing $\mathcal{F}$ where our results yield the highest value. \\
When considering our results without offset period, the performance is comparable to the best reported for \mbox{Cs-NF$_3$-Li}, with the values of $\mathcal{F}$ agreeing within their respective uncertainties. In this case, Cs-O$_2$-Sb shows the best performance even when considering the impact of decreased quantum efficiency. \\
Figure \ref{fig:fom_comp} shows $\mathcal{F}$ as a function of $r_\tau$ for the different enhancement methods. The values for Cs-O$_2$-Li, Cs-NF$_3$-Li, and Cs-O$_2$-Te are close to $r_{\eta_0} = 1$, while the values for Cs-O$_2$-Sb deviate significantly. Cs-O$_2$-Li shows the highest value of $\mathcal{F}$ close to $r_{\eta_0} = 1$. 
 \\
In summary, our results for Cs-O$_2$-Li show that it is a very promising surface layer, yielding a performance enhancement at least as good as that of other enhancement methods for NEA surface layers on GaAs photocathodes.

\FloatBarrier

\subsection{Predicted impact}
For the LHeC, a beam current of \qty{20}{\milli\ampere} is projected \cite{agostini2021}, equivalent to an extracted charge of \qty{70}{\coulomb} per hour. Continuous beam-time should be possible for at least one week and up to one month or more, roughly corresponding to photocathode lifetimes between \qty{100}{\hour} and \qty{1000}{\hour} and extracted charges between \qty{7.2}{\kilo\coulomb} and \qty{72}{\kilo\coulomb}. So far, a lifetime of about \qty{1.7}{\hour} and a lifetime charge of about \qty{200}{\coulomb} has been demonstrated for operating a GaAs photocathode at a high current of $I_0 = \qty{52}{\milli\ampere}$ and $P_\gamma = \qty{5}{\watt}$ \cite{dunham2013}. \\
The photocathode used in \cite{dunham2013} was activated without mask or anodization. Also, while the used photogun features an anode \cite{dunham2007}, it was not biased. Hence, assuming enhancement factors of \num{2.5} for masked activation \cite{aulenbacher2003} and \num{2.4} for anode biasing \cite{yoskowitz2020, yoskowitz2024}, an increase to a lifetime of \qty{10.2}{\hour} and a lifetime charge of \qty{1.2}{\kilo\coulomb} can be expected by implementing this method. Considering that all other established methods have been used, this falls well short of the LHeC operational requirements. \\
Using an enhancement of Cs-O$_2$ with Li could be expected to increase $\tau$ to between \qty{76.5}{\hour} and \qty{193.9}{\hour} and $Q(\tau)$ to between \qty{8.2}{\kilo\coulomb} and \qty{20.5}{\kilo\coulomb}. This significant increase would enable photo-gun performances satisfying LHeC operational requirements and enabling experiments with high-current spin-polarized electron beams, with a further increase of operational lifetime to be expected by changing the laser spot position. \\
Furthermore, an increment of charge lifetime by a factor of \num{6} to \num{11} by offsetting the anode \cite{rahman2019a} has been proposed. Considering operation at a fixed beam current and hence $r_\tau = r_{Q(\tau)}$, this increases the predicted values to between \qty{459}{\hour} and \qty{2133}{\hour} and between \qty{49.4}{\kilo\coulomb} and \qty{224.7}{\kilo\coulomb}. This estimate exceeds the range required for LHeC, especially when considering that the beam current used in Ref. \cite{dunham2007} was a factor of \num{2.6} higher than the beam current required for the LHeC. \\
Another proposed upgrade to further improve photogun performance is the application of an active cooling to the photocathode. At $P_\gamma = \qty{10}{\watt}$, a decrease of cathode temperature from \qty{60}{\celsius} to \qty{15}{\celsius} has been demonstrated \cite{wang2022}. However, it is not clear how this affects the photocathode lifetime. While a recovery of quantum efficiency, i.e. a perceived increase in lifetime, has been shown when heating the photocathode to \qty{150}{\celsius} \cite{schade1972}, a clear reduction in lifetime was observed when operating an activated GaAs photocathode at temperatures as high as \qty{84.5}{\celsius} \cite{iijima2010}. Hence, the effect and optimal application of active cooling on photocathode lifetime needs to be studied further. If it indeed has a positive effect, it could not only further increase photocathode performance, but also mitigate the effect of a reduced quantum efficiency since the impact of higher laser power on the photocathode lifetime would be greatly reduced. \\
Therefore, the results obtained in this study and those available in literature suggest that enhanced activation with Li or Sb will be able to provide a significant contribution to reliable high-current operation of GaAs photocathodes for prolonged beam-times, enabling a multitude of accelerator applications with high-current spin-polarized electron beams using otherwise already existing technology.

\section{Conclusion and outlook}
Activating bulk-GaAs photocathodes with a Cs-O$_2$-Li NEA surface layer at Photo-CATCH yielded a significant increase in lifetime and lifetime charge by factors of up to \num{19(2)} and \num{16.5(2.4)}, compared to traditional NEA activation with Cs-O$_2$. This effect increased with rising dosage of Li applied in the activation process. While the maximum quantum efficiency measured directly after activation displayed a slight increase for small amounts of Li, it yielded no significant change for the highest Li dosage that produced the best increase in lifetime and lifetime charge. Residual Li from previous activations showed a beneficial effect on photocathode performance. \\
During quantum efficiency decay, a combination of fast initial decay followed by a slow rise to a saturation point was observed, delaying the typical single-exponential decay by a time offset. This behavior got more pronounced for higher dosages of Li used during the activation process, resulting in longer offsets. While the exact cause for this behavior could not yet be determined, we find evidence for a combination of increased IBB and adsorption of residual Cs and Li. The fast initial decay observed for the \mbox{Cs-O$_2$-Li} surface layer yielded a reduction in quantum efficiency when compared to the conventional Cs-O$_2$ layer at the beginning of the lifetime measurement several minutes after activation, corresponding to a time interval commonly required to transfer a photocathode from an activation chamber to a gun chamber for beam extraction. \\
However, no more than two activations of each type were performed, and lifetime measurements were conducted at low anode voltage in the activation chamber. Therefore, further studies are required to verify the reproducibility of our results, especially during operation in a DC photo-electron gun at high beam currents. Also, the impact of further increase in Li dosage and its influence on spin-polarization have not been investigated. Hence, the performance presented here may not represent the peak performance of the surface layer. \\
Comparison of our results and other data available for Cs-O$_2$-Li to those of other enhanced layers such as Cs-NF$_3$-Li and Cs-O$_2$-Sb with incident light in the wavelength range of \qtyrange{750}{785}{\nano\meter} shows an equal or better performance. The resulting values for $\mathcal{F}$, ranging from \num{2.7} to \num{16.3}, are comparable with the range of results reported for Cs-NF$_3$-Li and Cs-O$_2$-Sb. Hence, \mbox{Cs-O$_2$-Li} shows promising performance as enhanced NEA surface layer for GaAs photocathodes.

\section*{Acknowledgments}
The authors acknowledge support by the Deutsche
Forschungsgemeinschaft (DFG) - Projektnummer 264883531 (GRK 2128 ``AccelencE'') and the German BMBF (05H18RDRB1).

\bibliography{biblio}

\begin{thebibliography}{119}%
\makeatletter
\providecommand \@ifxundefined [1]{%
 \@ifx{#1\undefined}
}%
\providecommand \@ifnum [1]{%
 \ifnum #1\expandafter \@firstoftwo
 \else \expandafter \@secondoftwo
 \fi
}%
\providecommand \@ifx [1]{%
 \ifx #1\expandafter \@firstoftwo
 \else \expandafter \@secondoftwo
 \fi
}%
\providecommand \natexlab [1]{#1}%
\providecommand \enquote  [1]{``#1''}%
\providecommand \bibnamefont  [1]{#1}%
\providecommand \bibfnamefont [1]{#1}%
\providecommand \citenamefont [1]{#1}%
\providecommand \href@noop [0]{\@secondoftwo}%
\providecommand \href [0]{\begingroup \@sanitize@url \@href}%
\providecommand \@href[1]{\@@startlink{#1}\@@href}%
\providecommand \@@href[1]{\endgroup#1\@@endlink}%
\providecommand \@sanitize@url [0]{\catcode `\\12\catcode `\$12\catcode
  `\&12\catcode `\#12\catcode `\^12\catcode `\_12\catcode `\%12\relax}%
\providecommand \@@startlink[1]{}%
\providecommand \@@endlink[0]{}%
\providecommand \url  [0]{\begingroup\@sanitize@url \@url }%
\providecommand \@url [1]{\endgroup\@href {#1}{\urlprefix }}%
\providecommand \urlprefix  [0]{URL }%
\providecommand \Eprint [0]{\href }%
\providecommand \doibase [0]{https://doi.org/}%
\providecommand \selectlanguage [0]{\@gobble}%
\providecommand \bibinfo  [0]{\@secondoftwo}%
\providecommand \bibfield  [0]{\@secondoftwo}%
\providecommand \translation [1]{[#1]}%
\providecommand \BibitemOpen [0]{}%
\providecommand \bibitemStop [0]{}%
\providecommand \bibitemNoStop [0]{.\EOS\space}%
\providecommand \EOS [0]{\spacefactor3000\relax}%
\providecommand \BibitemShut  [1]{\csname bibitem#1\endcsname}%
\let\auto@bib@innerbib\@empty
\bibitem [{\citenamefont {Sinclair}(2006)}]{sinclair2006}%
  \BibitemOpen
  \bibfield  {author} {\bibinfo {author} {\bibfnamefont {C.~K.}\ \bibnamefont
  {Sinclair}},\ }\bibfield  {title} {\bibinfo {title} {{DC} photoemission
  electron guns as {ERL} sources},\ }\bibfield  {journal} {\bibinfo  {journal}
  {Nucl. Instrum. Methods, Sect. A}\ }\textbf {\bibinfo {volume} {557}},\ \href
  {https://doi.org/https://doi.org/10.1016/j.nima.2005.10.053}
  {https://doi.org/10.1016/j.nima.2005.10.053} (\bibinfo {year}
  {2006})\BibitemShut {NoStop}%
\bibitem [{\citenamefont {Rao}\ \emph {et~al.}(2006)\citenamefont {Rao} \emph
  {et~al.}}]{rao2006}%
  \BibitemOpen
  \bibfield  {author} {\bibinfo {author} {\bibfnamefont {T.}~\bibnamefont
  {Rao}} \emph {et~al.},\ }\bibfield  {title} {\bibinfo {title} {Photocathodes
  for the energy recovery linacs},\ }\bibfield  {journal} {\bibinfo  {journal}
  {Nucl. Instrum. Methods, Sect. A}\ }\textbf {\bibinfo {volume} {557}},\ \href
  {https://doi.org/https://doi.org/10.1016/j.nima.2005.10.112}
  {https://doi.org/10.1016/j.nima.2005.10.112} (\bibinfo {year}
  {2006})\BibitemShut {NoStop}%
\bibitem [{\citenamefont {Heine}(2021)}]{heine2021}%
  \BibitemOpen
  \bibfield  {author} {\bibinfo {author} {\bibfnamefont {R.}~\bibnamefont
  {Heine}},\ }\bibfield  {title} {\bibinfo {title} {Preaccelerator concepts for
  an energy-recovering superconducting accelerator},\ }\href
  {https://doi.org/10.1103/PhysRevAccelBeams.24.011602} {\bibfield  {journal}
  {\bibinfo  {journal} {Phys. Rev. Accel. Beams}\ }\textbf {\bibinfo {volume}
  {24}},\ \bibinfo {pages} {011602} (\bibinfo {year} {2021})}\BibitemShut
  {NoStop}%
\bibitem [{\citenamefont {Brachmann}\ \emph {et~al.}(2007)\citenamefont
  {Brachmann}, \citenamefont {Clendenin}, \citenamefont {Garwin}, \citenamefont
  {Ioakeimidi}, \citenamefont {Kirby}, \citenamefont {Maruyama}, \citenamefont
  {Prescott}, \citenamefont {Sheppard}, \citenamefont {Turner},\ and\
  \citenamefont {Zhou}}]{brachmann2007}%
  \BibitemOpen
  \bibfield  {author} {\bibinfo {author} {\bibfnamefont {A.}~\bibnamefont
  {Brachmann}}, \bibinfo {author} {\bibfnamefont {J.~E.}\ \bibnamefont
  {Clendenin}}, \bibinfo {author} {\bibfnamefont {E.~L.}\ \bibnamefont
  {Garwin}}, \bibinfo {author} {\bibfnamefont {K.}~\bibnamefont {Ioakeimidi}},
  \bibinfo {author} {\bibfnamefont {R.~E.}\ \bibnamefont {Kirby}}, \bibinfo
  {author} {\bibfnamefont {T.}~\bibnamefont {Maruyama}}, \bibinfo {author}
  {\bibfnamefont {C.~Y.}\ \bibnamefont {Prescott}}, \bibinfo {author}
  {\bibfnamefont {J.}~\bibnamefont {Sheppard}}, \bibinfo {author}
  {\bibfnamefont {J.}~\bibnamefont {Turner}},\ and\ \bibinfo {author}
  {\bibfnamefont {F.}~\bibnamefont {Zhou}},\ }\bibfield  {title} {\bibinfo
  {title} {The polarized electron source for the international collider ({ILC})
  project},\ }\href {https://doi.org/10.1063/1.2750959} {\bibfield  {journal}
  {\bibinfo  {journal} {AIP Conf. Proc.}\ }\textbf {\bibinfo {volume} {915}},\
  \bibinfo {pages} {1091} (\bibinfo {year} {2007})}\BibitemShut {NoStop}%
\bibitem [{\citenamefont {Moortgat-Pick}\ \emph {et~al.}(2008)\citenamefont
  {Moortgat-Pick} \emph {et~al.}}]{moortgat-pick2008}%
  \BibitemOpen
  \bibfield  {author} {\bibinfo {author} {\bibfnamefont {G.}~\bibnamefont
  {Moortgat-Pick}} \emph {et~al.},\ }\bibfield  {title} {\bibinfo {title}
  {Polarized positrons and electrons at the linear collider},\ }\href
  {https://doi.org/10.1016/j.physrep.2007.12.003} {\bibfield  {journal}
  {\bibinfo  {journal} {Physics Reports}\ }\textbf {\bibinfo {volume} {460}},\
  \bibinfo {pages} {131} (\bibinfo {year} {2008})}\BibitemShut {NoStop}%
\bibitem [{\citenamefont {Skaritka}\ \emph {et~al.}(2018)\citenamefont
  {Skaritka}, \citenamefont {Wang}, \citenamefont {Willeke}, \citenamefont
  {Lambiase}, \citenamefont {Lui}, \citenamefont {Ptitsyn},\ and\ \citenamefont
  {Rahman}}]{skaritka2018}%
  \BibitemOpen
  \bibfield  {author} {\bibinfo {author} {\bibfnamefont {J.}~\bibnamefont
  {Skaritka}}, \bibinfo {author} {\bibfnamefont {E.}~\bibnamefont {Wang}},
  \bibinfo {author} {\bibfnamefont {F.}~\bibnamefont {Willeke}}, \bibinfo
  {author} {\bibfnamefont {R.}~\bibnamefont {Lambiase}}, \bibinfo {author}
  {\bibfnamefont {W.}~\bibnamefont {Lui}}, \bibinfo {author} {\bibfnamefont
  {V.}~\bibnamefont {Ptitsyn}},\ and\ \bibinfo {author} {\bibfnamefont
  {O.}~\bibnamefont {Rahman}},\ }\bibfield  {title} {\bibinfo {title}
  {{Conceptual design of a Polarized Electron Ion Collider at Brookhaven
  National Laboratory}},\ }in\ \href {https://doi.org/10.22323/1.324.0015}
  {\emph {\bibinfo {booktitle} {Proc. XVII International Workshop on Polarized
  Sources, Targets \& Polarimetry {\textemdash} PoS(PSTP2017)}}},\ Vol.\
  \bibinfo {volume} {324}\ (\bibinfo {year} {2018})\ p.\ \bibinfo {pages}
  {015}\BibitemShut {NoStop}%
\bibitem [{\citenamefont {Abbott}\ \emph {et~al.}(2016)\citenamefont {Abbott}
  \emph {et~al.}}]{abbott2016}%
  \BibitemOpen
  \bibfield  {author} {\bibinfo {author} {\bibfnamefont {D.}~\bibnamefont
  {Abbott}} \emph {et~al.} (\bibinfo {collaboration} {PEPPo Collaboration}),\
  }\bibfield  {title} {\bibinfo {title} {Production of highly polarized
  positrons using polarized electrons at {MeV} energies},\ }\href
  {https://doi.org/10.1103/PhysRevLett.116.214801} {\bibfield  {journal}
  {\bibinfo  {journal} {Phys. Rev. Lett.}\ }\textbf {\bibinfo {volume} {116}},\
  \bibinfo {pages} {214801} (\bibinfo {year} {2016})}\BibitemShut {NoStop}%
\bibitem [{\citenamefont {Cardman}(2018)}]{cardman2018}%
  \BibitemOpen
  \bibfield  {author} {\bibinfo {author} {\bibfnamefont {L.~S.}\ \bibnamefont
  {Cardman}},\ }\bibfield  {title} {\bibinfo {title} {The {PEPPo} method for
  polarized positrons and {PEPPo II}},\ }\href
  {https://doi.org/10.1063/1.5040220} {\bibfield  {journal} {\bibinfo
  {journal} {AIP Conf. Proc.}\ }\textbf {\bibinfo {volume} {1970}},\ \bibinfo
  {pages} {050001} (\bibinfo {year} {2018})}\BibitemShut {NoStop}%
\bibitem [{\citenamefont {Orlov}\ \emph {et~al.}(2009)\citenamefont {Orlov},
  \citenamefont {Krantz}, \citenamefont {Shornikov}, \citenamefont {Lestinsky},
  \citenamefont {Hoffmann}, \citenamefont {Jaroshevich}, \citenamefont
  {Kosolobov}, \citenamefont {Terekhov},\ and\ \citenamefont
  {Wolf}}]{orlov2009}%
  \BibitemOpen
  \bibfield  {author} {\bibinfo {author} {\bibfnamefont {D.~A.}\ \bibnamefont
  {Orlov}}, \bibinfo {author} {\bibfnamefont {C.}~\bibnamefont {Krantz}},
  \bibinfo {author} {\bibfnamefont {A.}~\bibnamefont {Shornikov}}, \bibinfo
  {author} {\bibfnamefont {M.}~\bibnamefont {Lestinsky}}, \bibinfo {author}
  {\bibfnamefont {J.}~\bibnamefont {Hoffmann}}, \bibinfo {author}
  {\bibfnamefont {A.~S.}\ \bibnamefont {Jaroshevich}}, \bibinfo {author}
  {\bibfnamefont {S.~N.}\ \bibnamefont {Kosolobov}}, \bibinfo {author}
  {\bibfnamefont {A.~S.}\ \bibnamefont {Terekhov}},\ and\ \bibinfo {author}
  {\bibfnamefont {A.}~\bibnamefont {Wolf}},\ }\bibfield  {title} {\bibinfo
  {title} {Ultra cold photoelectron beams for ion storage rings},\ }\href
  {https://doi.org/10.1063/1.3215583} {\bibfield  {journal} {\bibinfo
  {journal} {AIP Conf. Proc.}\ }\textbf {\bibinfo {volume} {1149}},\ \bibinfo
  {pages} {1007} (\bibinfo {year} {2009})}\BibitemShut {NoStop}%
\bibitem [{\citenamefont {Vollmer}\ \emph {et~al.}(2003)\citenamefont
  {Vollmer}, \citenamefont {Etzkorn}, \citenamefont {{P. S. Anil Kumar}},
  \citenamefont {Ibach},\ and\ \citenamefont {Kirschner}}]{vollmer2003}%
  \BibitemOpen
  \bibfield  {author} {\bibinfo {author} {\bibfnamefont {R.}~\bibnamefont
  {Vollmer}}, \bibinfo {author} {\bibfnamefont {M.}~\bibnamefont {Etzkorn}},
  \bibinfo {author} {\bibnamefont {{P. S. Anil Kumar}}}, \bibinfo {author}
  {\bibfnamefont {H.}~\bibnamefont {Ibach}},\ and\ \bibinfo {author}
  {\bibfnamefont {J.}~\bibnamefont {Kirschner}},\ }\bibfield  {title} {\bibinfo
  {title} {Spin-polarized electron energy loss spectroscopy of high energy,
  large wave vector spin waves in ultrathin fcc {Co} films on {Cu}(001)},\
  }\href {https://doi.org/10.1103/PhysRevLett.91.147201} {\bibfield  {journal}
  {\bibinfo  {journal} {Phys. Rev. Lett.}\ }\textbf {\bibinfo {volume} {91}},\
  \bibinfo {pages} {147201} (\bibinfo {year} {2003})}\BibitemShut {NoStop}%
\bibitem [{\citenamefont {Suzuki}\ \emph {et~al.}(2010)\citenamefont {Suzuki}
  \emph {et~al.}}]{suzuki2010}%
  \BibitemOpen
  \bibfield  {author} {\bibinfo {author} {\bibfnamefont {M.}~\bibnamefont
  {Suzuki}} \emph {et~al.},\ }\bibfield  {title} {\bibinfo {title} {Real time
  magnetic imaging by spin-polarized low energy electron microscopy with highly
  spin-polarized and high brightness electron gun},\ }\href
  {https://doi.org/10.1143/apex.3.026601} {\bibfield  {journal} {\bibinfo
  {journal} {Appl. Phys. Express}\ }\textbf {\bibinfo {volume} {3}},\ \bibinfo
  {pages} {026601} (\bibinfo {year} {2010})}\BibitemShut {NoStop}%
\bibitem [{\citenamefont {Kuwahara}\ \emph {et~al.}(2012)\citenamefont
  {Kuwahara}, \citenamefont {Kusunoki}, \citenamefont {Jin}, \citenamefont
  {Nakanishi}, \citenamefont {Takeda}, \citenamefont {Saitoh}, \citenamefont
  {Ujihara}, \citenamefont {Asano},\ and\ \citenamefont
  {Tanaka}}]{kuwahara2012}%
  \BibitemOpen
  \bibfield  {author} {\bibinfo {author} {\bibfnamefont {M.}~\bibnamefont
  {Kuwahara}}, \bibinfo {author} {\bibfnamefont {S.}~\bibnamefont {Kusunoki}},
  \bibinfo {author} {\bibfnamefont {X.~G.}\ \bibnamefont {Jin}}, \bibinfo
  {author} {\bibfnamefont {T.}~\bibnamefont {Nakanishi}}, \bibinfo {author}
  {\bibfnamefont {Y.}~\bibnamefont {Takeda}}, \bibinfo {author} {\bibfnamefont
  {K.}~\bibnamefont {Saitoh}}, \bibinfo {author} {\bibfnamefont
  {T.}~\bibnamefont {Ujihara}}, \bibinfo {author} {\bibfnamefont
  {H.}~\bibnamefont {Asano}},\ and\ \bibinfo {author} {\bibfnamefont
  {N.}~\bibnamefont {Tanaka}},\ }\bibfield  {title} {\bibinfo {title} {30-{kV}
  spin-polarized transmission electron microscope with {GaAs–GaAsP} strained
  superlattice photocathode},\ }\href {https://doi.org/10.1063/1.4737177}
  {\bibfield  {journal} {\bibinfo  {journal} {Appl. Phys. Lett.}\ }\textbf
  {\bibinfo {volume} {101}},\ \bibinfo {pages} {033102} (\bibinfo {year}
  {2012})}\BibitemShut {NoStop}%
\bibitem [{\citenamefont {Kessler}(1985)}]{kessler1985}%
  \BibitemOpen
  \bibfield  {author} {\bibinfo {author} {\bibfnamefont {J.}~\bibnamefont
  {Kessler}},\ }\href {https://doi.org/10.1007/978-3-662-02434-8} {\emph
  {\bibinfo {title} {Polarized electrons}}},\ Vol.~\bibinfo {volume} {1}\
  (\bibinfo  {publisher} {Springer Berlin Heidelberg},\ \bibinfo {year}
  {1985})\BibitemShut {NoStop}%
\bibitem [{\citenamefont {Pierce}\ and\ \citenamefont
  {Meier}(1976)}]{pierce+meier1976}%
  \BibitemOpen
  \bibfield  {author} {\bibinfo {author} {\bibfnamefont {D.~T.}\ \bibnamefont
  {Pierce}}\ and\ \bibinfo {author} {\bibfnamefont {F.}~\bibnamefont {Meier}},\
  }\bibfield  {title} {\bibinfo {title} {Photoemission of spin-polarized
  electrons from {GaAs}},\ }\href {https://doi.org/10.1103/PhysRevB.13.5484}
  {\bibfield  {journal} {\bibinfo  {journal} {Phys. Rev. B}\ }\textbf {\bibinfo
  {volume} {13}},\ \bibinfo {pages} {5484} (\bibinfo {year}
  {1976})}\BibitemShut {NoStop}%
\bibitem [{\citenamefont {Maruyama}\ \emph {et~al.}(2004)\citenamefont
  {Maruyama} \emph {et~al.}}]{maruyama2004}%
  \BibitemOpen
  \bibfield  {author} {\bibinfo {author} {\bibfnamefont {T.}~\bibnamefont
  {Maruyama}} \emph {et~al.},\ }\bibfield  {title} {\bibinfo {title}
  {Systematic study of polarized electron emission from strained {GaAs-GaAsP}
  superlattice photocathodes},\ }\href {https://doi.org/10.1063/1.1795358}
  {\bibfield  {journal} {\bibinfo  {journal} {Appl. Phys. Lett.}\ }\textbf
  {\bibinfo {volume} {85}},\ \bibinfo {pages} {2640} (\bibinfo {year}
  {2004})}\BibitemShut {NoStop}%
\bibitem [{\citenamefont {Nishitani}\ \emph {et~al.}(2005)\citenamefont
  {Nishitani} \emph {et~al.}}]{nishitani2005}%
  \BibitemOpen
  \bibfield  {author} {\bibinfo {author} {\bibfnamefont {T.}~\bibnamefont
  {Nishitani}} \emph {et~al.},\ }\bibfield  {title} {\bibinfo {title} {Highly
  polarized electrons from {GaAs-GaAsP} and {InGaAs-AlGaAs} strained-layer
  superlattice photocathodes},\ }\href {https://doi.org/10.1063/1.1886888}
  {\bibfield  {journal} {\bibinfo  {journal} {J. Appl. Phys.}\ }\textbf
  {\bibinfo {volume} {97}},\ \bibinfo {pages} {094907} (\bibinfo {year}
  {2005})}\BibitemShut {NoStop}%
\bibitem [{\citenamefont {Jin}\ \emph {et~al.}(2014)\citenamefont {Jin},
  \citenamefont {Ozdol}, \citenamefont {Yamamoto}, \citenamefont {Mano},
  \citenamefont {Yamamoto},\ and\ \citenamefont {Takeda}}]{jin2014}%
  \BibitemOpen
  \bibfield  {author} {\bibinfo {author} {\bibfnamefont {X.}~\bibnamefont
  {Jin}}, \bibinfo {author} {\bibfnamefont {B.}~\bibnamefont {Ozdol}}, \bibinfo
  {author} {\bibfnamefont {M.}~\bibnamefont {Yamamoto}}, \bibinfo {author}
  {\bibfnamefont {A.}~\bibnamefont {Mano}}, \bibinfo {author} {\bibfnamefont
  {N.}~\bibnamefont {Yamamoto}},\ and\ \bibinfo {author} {\bibfnamefont
  {Y.}~\bibnamefont {Takeda}},\ }\bibfield  {title} {\bibinfo {title} {Effect
  of crystal quality on performance of spin-polarized photocathode},\ }\href
  {https://doi.org/10.1063/1.4902337} {\bibfield  {journal} {\bibinfo
  {journal} {Appl. Phys. Lett.}\ }\textbf {\bibinfo {volume} {105}},\ \bibinfo
  {pages} {203509} (\bibinfo {year} {2014})}\BibitemShut {NoStop}%
\bibitem [{\citenamefont {Liu}\ \emph {et~al.}(2016)\citenamefont {Liu},
  \citenamefont {Chen}, \citenamefont {Lu}, \citenamefont {Moy}, \citenamefont
  {Poelker}, \citenamefont {Stutzman},\ and\ \citenamefont {Zhang}}]{liu2016}%
  \BibitemOpen
  \bibfield  {author} {\bibinfo {author} {\bibfnamefont {W.}~\bibnamefont
  {Liu}}, \bibinfo {author} {\bibfnamefont {Y.}~\bibnamefont {Chen}}, \bibinfo
  {author} {\bibfnamefont {W.}~\bibnamefont {Lu}}, \bibinfo {author}
  {\bibfnamefont {A.}~\bibnamefont {Moy}}, \bibinfo {author} {\bibfnamefont
  {M.}~\bibnamefont {Poelker}}, \bibinfo {author} {\bibfnamefont
  {M.}~\bibnamefont {Stutzman}},\ and\ \bibinfo {author} {\bibfnamefont
  {S.}~\bibnamefont {Zhang}},\ }\bibfield  {title} {\bibinfo {title}
  {Record-level quantum efficiency from a high polarization strained
  {GaAs/GaAsP} superlattice photocathode with distributed {Bragg} reflector},\
  }\href {https://doi.org/10.1063/1.4972180} {\bibfield  {journal} {\bibinfo
  {journal} {Appl. Phys. Lett.}\ }\textbf {\bibinfo {volume} {109}},\ \bibinfo
  {pages} {252104} (\bibinfo {year} {2016})}\BibitemShut {NoStop}%
\bibitem [{\citenamefont {Sonnenberg}(1969{\natexlab{a}})}]{sonnenberg1969a}%
  \BibitemOpen
  \bibfield  {author} {\bibinfo {author} {\bibfnamefont {H.}~\bibnamefont
  {Sonnenberg}},\ }\bibfield  {title} {\bibinfo {title} {Low‐work‐function
  surfaces for negative-electron-affinity photoemitters},\ }\href
  {https://doi.org/10.1063/1.1652819} {\bibfield  {journal} {\bibinfo
  {journal} {Appl. Phys. Lett.}\ }\textbf {\bibinfo {volume} {14}},\ \bibinfo
  {pages} {289} (\bibinfo {year} {1969}{\natexlab{a}})}\BibitemShut {NoStop}%
\bibitem [{\citenamefont {Spicer}\ and\ \citenamefont
  {Bell}(1972)}]{spicer+bell1972}%
  \BibitemOpen
  \bibfield  {author} {\bibinfo {author} {\bibfnamefont {W.~E.}\ \bibnamefont
  {Spicer}}\ and\ \bibinfo {author} {\bibfnamefont {R.~L.}\ \bibnamefont
  {Bell}},\ }\bibfield  {title} {\bibinfo {title} {The {III-V} photocathode: A
  major detector development},\ }\href {https://doi.org/10.1086/129256}
  {\bibfield  {journal} {\bibinfo  {journal} {Publ. Astron. Soc. Pac.}\
  }\textbf {\bibinfo {volume} {84}},\ \bibinfo {pages} {110} (\bibinfo {year}
  {1972})}\BibitemShut {NoStop}%
\bibitem [{\citenamefont {Jensen}\ \emph {et~al.}(2006)\citenamefont {Jensen},
  \citenamefont {Feldman}, \citenamefont {Moody},\ and\ \citenamefont
  {O’Shea}}]{jensen2006}%
  \BibitemOpen
  \bibfield  {author} {\bibinfo {author} {\bibfnamefont {K.~L.}\ \bibnamefont
  {Jensen}}, \bibinfo {author} {\bibfnamefont {D.~W.}\ \bibnamefont {Feldman}},
  \bibinfo {author} {\bibfnamefont {N.~A.}\ \bibnamefont {Moody}},\ and\
  \bibinfo {author} {\bibfnamefont {P.~G.}\ \bibnamefont {O’Shea}},\
  }\bibfield  {title} {\bibinfo {title} {A photoemission model for low work
  function coated metal surfaces and its experimental validation},\ }\href
  {https://doi.org/10.1063/1.2203720} {\bibfield  {journal} {\bibinfo
  {journal} {J. Appl. Phys.}\ }\textbf {\bibinfo {volume} {99}},\ \bibinfo
  {pages} {124905} (\bibinfo {year} {2006})}\BibitemShut {NoStop}%
\bibitem [{\citenamefont {Uebbing}(1970)}]{uebbing1970}%
  \BibitemOpen
  \bibfield  {author} {\bibinfo {author} {\bibfnamefont {J.~J.}\ \bibnamefont
  {Uebbing}},\ }\bibfield  {title} {\bibinfo {title} {Use of auger electron
  spectroscopy in determining the effect of carbon and other surface
  contaminants on {GaAs–Cs–O} photocathodes},\ }\href
  {https://doi.org/10.1063/1.1658753} {\bibfield  {journal} {\bibinfo
  {journal} {J. Appl. Phys.}\ }\textbf {\bibinfo {volume} {41}},\ \bibinfo
  {pages} {802} (\bibinfo {year} {1970})}\BibitemShut {NoStop}%
\bibitem [{\citenamefont {Chanlek}\ \emph {et~al.}(2014)\citenamefont
  {Chanlek}, \citenamefont {Herbert}, \citenamefont {Jones}, \citenamefont
  {Jones}, \citenamefont {Middleman},\ and\ \citenamefont
  {Militsyn}}]{chanlek2014}%
  \BibitemOpen
  \bibfield  {author} {\bibinfo {author} {\bibfnamefont {N.}~\bibnamefont
  {Chanlek}}, \bibinfo {author} {\bibfnamefont {J.~D.}\ \bibnamefont
  {Herbert}}, \bibinfo {author} {\bibfnamefont {R.~M.}\ \bibnamefont {Jones}},
  \bibinfo {author} {\bibfnamefont {L.~B.}\ \bibnamefont {Jones}}, \bibinfo
  {author} {\bibfnamefont {K.~J.}\ \bibnamefont {Middleman}},\ and\ \bibinfo
  {author} {\bibfnamefont {B.~L.}\ \bibnamefont {Militsyn}},\ }\bibfield
  {title} {\bibinfo {title} {The degradation of quantum efficiency in negative
  electron affinity {GaAs} photocathodes under gas exposure},\ }\href
  {https://doi.org/10.1088/0022-3727/47/5/055110} {\bibfield  {journal}
  {\bibinfo  {journal} {J. Phys. D}\ }\textbf {\bibinfo {volume} {47}},\
  \bibinfo {pages} {055110} (\bibinfo {year} {2014})}\BibitemShut {NoStop}%
\bibitem [{\citenamefont {Aulenbacher}\ \emph {et~al.}(1997)\citenamefont
  {Aulenbacher} \emph {et~al.}}]{aulenbacher1997}%
  \BibitemOpen
  \bibfield  {author} {\bibinfo {author} {\bibfnamefont {K.}~\bibnamefont
  {Aulenbacher}} \emph {et~al.},\ }\bibfield  {title} {\bibinfo {title} {The
  {MAMI} source of polarized electrons},\ }\href
  {https://doi.org/10.1016/S0168-9002(97)00528-7} {\bibfield  {journal}
  {\bibinfo  {journal} {Nucl. Instrum. Methods, Sect. A}\ }\textbf {\bibinfo
  {volume} {391}},\ \bibinfo {pages} {498} (\bibinfo {year}
  {1997})}\BibitemShut {NoStop}%
\bibitem [{\citenamefont {Siggins}\ \emph {et~al.}(2001)\citenamefont
  {Siggins}, \citenamefont {Sinclair}, \citenamefont {Bohn}, \citenamefont
  {Bullard}, \citenamefont {Douglas}, \citenamefont {Grippo}, \citenamefont
  {Gubeli}, \citenamefont {Krafft},\ and\ \citenamefont {Yunn}}]{siggins2001}%
  \BibitemOpen
  \bibfield  {author} {\bibinfo {author} {\bibfnamefont {T.}~\bibnamefont
  {Siggins}}, \bibinfo {author} {\bibfnamefont {C.}~\bibnamefont {Sinclair}},
  \bibinfo {author} {\bibfnamefont {C.}~\bibnamefont {Bohn}}, \bibinfo {author}
  {\bibfnamefont {D.}~\bibnamefont {Bullard}}, \bibinfo {author} {\bibfnamefont
  {D.}~\bibnamefont {Douglas}}, \bibinfo {author} {\bibfnamefont
  {A.}~\bibnamefont {Grippo}}, \bibinfo {author} {\bibfnamefont
  {J.}~\bibnamefont {Gubeli}}, \bibinfo {author} {\bibfnamefont {G.~A.}\
  \bibnamefont {Krafft}},\ and\ \bibinfo {author} {\bibfnamefont
  {B.}~\bibnamefont {Yunn}},\ }\bibfield  {title} {\bibinfo {title}
  {Performance of a {DC GaAs} photocathode gun for the {Jefferson} lab {FEL}},\
  }\href {https://doi.org/10.1016/S0168-9002(01)01596-0} {\bibfield  {journal}
  {\bibinfo  {journal} {Nucl. Instrum. Methods, Sect. A}\ }\textbf {\bibinfo
  {volume} {475}},\ \bibinfo {pages} {549} (\bibinfo {year} {2001})},\ \bibinfo
  {note} {{FEL}2000: Proc. 22nd Int. Free Electron Laser Conference and 7th FEL
  Users Workshop}\BibitemShut {NoStop}%
\bibitem [{\citenamefont {Ciccacci}\ and\ \citenamefont
  {Chiaia}(1991)}]{ciccacci+chiaia1991}%
  \BibitemOpen
  \bibfield  {author} {\bibinfo {author} {\bibfnamefont {F.}~\bibnamefont
  {Ciccacci}}\ and\ \bibinfo {author} {\bibfnamefont {G.}~\bibnamefont
  {Chiaia}},\ }\bibfield  {title} {\bibinfo {title} {Comparative study of the
  preparation of negative electron affinity {GaAs} photocathodes with {O2} and
  with {NF3}},\ }\href {https://doi.org/10.1116/1.577161} {\bibfield  {journal}
  {\bibinfo  {journal} {J. Vac. Sci. Technol. A}\ }\textbf {\bibinfo {volume}
  {9}},\ \bibinfo {pages} {2991} (\bibinfo {year} {1991})}\BibitemShut
  {NoStop}%
\bibitem [{\citenamefont {Dunham}\ \emph {et~al.}(2013)\citenamefont {Dunham}
  \emph {et~al.}}]{dunham2013}%
  \BibitemOpen
  \bibfield  {author} {\bibinfo {author} {\bibfnamefont {B.}~\bibnamefont
  {Dunham}} \emph {et~al.},\ }\bibfield  {title} {\bibinfo {title} {Record
  high-average current from a high-brightness photoinjector},\ }\href
  {https://doi.org/10.1063/1.4789395} {\bibfield  {journal} {\bibinfo
  {journal} {Appl. Phys. Lett.}\ }\textbf {\bibinfo {volume} {102}},\ \bibinfo
  {pages} {034105} (\bibinfo {year} {2013})}\BibitemShut {NoStop}%
\bibitem [{\citenamefont {Aulenbacher}\ \emph {et~al.}(2003)\citenamefont
  {Aulenbacher}, \citenamefont {Tioukine}, \citenamefont {Wiessner},\ and\
  \citenamefont {Winkler}}]{aulenbacher2003}%
  \BibitemOpen
  \bibfield  {author} {\bibinfo {author} {\bibfnamefont {K.}~\bibnamefont
  {Aulenbacher}}, \bibinfo {author} {\bibfnamefont {V.}~\bibnamefont
  {Tioukine}}, \bibinfo {author} {\bibfnamefont {M.}~\bibnamefont {Wiessner}},\
  and\ \bibinfo {author} {\bibfnamefont {K.}~\bibnamefont {Winkler}},\
  }\bibfield  {title} {\bibinfo {title} {Status of the polarized source at
  {MAMI}},\ }\href {https://doi.org/10.1063/1.1607302} {\bibfield  {journal}
  {\bibinfo  {journal} {AIP Conf. Proc.}\ }\textbf {\bibinfo {volume} {675}},\
  \bibinfo {pages} {1088} (\bibinfo {year} {2003})}\BibitemShut {NoStop}%
\bibitem [{\citenamefont {Schwartz}\ \emph {et~al.}(1976)\citenamefont
  {Schwartz}, \citenamefont {Ermanis},\ and\ \citenamefont
  {Brastad}}]{schwartz1976}%
  \BibitemOpen
  \bibfield  {author} {\bibinfo {author} {\bibfnamefont {B.}~\bibnamefont
  {Schwartz}}, \bibinfo {author} {\bibfnamefont {F.}~\bibnamefont {Ermanis}},\
  and\ \bibinfo {author} {\bibfnamefont {M.~H.}\ \bibnamefont {Brastad}},\
  }\bibfield  {title} {\bibinfo {title} {The anodization of {GaAs} and {GaP} in
  aqueous solutions},\ }\href {https://doi.org/10.1149/1.2133002} {\bibfield
  {journal} {\bibinfo  {journal} {J. Electrochem. Soc.}\ }\textbf {\bibinfo
  {volume} {123}},\ \bibinfo {pages} {1089} (\bibinfo {year}
  {1976})}\BibitemShut {NoStop}%
\bibitem [{\citenamefont {Grames}\ \emph {et~al.}(2005)\citenamefont {Grames},
  \citenamefont {Adderley}, \citenamefont {Brittian}, \citenamefont {Charles},
  \citenamefont {Clark}, \citenamefont {Hansknecht}, \citenamefont {Poelker},
  \citenamefont {Stutzman},\ and\ \citenamefont {Surles-Law}}]{grames2005}%
  \BibitemOpen
  \bibfield  {author} {\bibinfo {author} {\bibfnamefont {J.}~\bibnamefont
  {Grames}}, \bibinfo {author} {\bibfnamefont {P.}~\bibnamefont {Adderley}},
  \bibinfo {author} {\bibfnamefont {J.}~\bibnamefont {Brittian}}, \bibinfo
  {author} {\bibfnamefont {D.}~\bibnamefont {Charles}}, \bibinfo {author}
  {\bibfnamefont {J.}~\bibnamefont {Clark}}, \bibinfo {author} {\bibfnamefont
  {J.}~\bibnamefont {Hansknecht}}, \bibinfo {author} {\bibfnamefont
  {M.}~\bibnamefont {Poelker}}, \bibinfo {author} {\bibfnamefont
  {M.}~\bibnamefont {Stutzman}},\ and\ \bibinfo {author} {\bibfnamefont
  {K.}~\bibnamefont {Surles-Law}},\ }\bibfield  {title} {\bibinfo {title} {Ion
  back-bombardment of {GaAs} photocathodes inside {DC} high voltage electron
  guns},\ }in\ \href {https://doi.org/10.1109/PAC.2005.1591299} {\emph
  {\bibinfo {booktitle} {Proc. 2005 Particle Accelerator Conference
  (PAC'05)}}}\ (\bibinfo {year} {2005})\ pp.\ \bibinfo {pages}
  {2875--2877}\BibitemShut {NoStop}%
\bibitem [{\citenamefont {Hernandez-Garcia}\ \emph {et~al.}(2005)\citenamefont
  {Hernandez-Garcia}, \citenamefont {Siggins}, \citenamefont {Benson},
  \citenamefont {Bullard}, \citenamefont {Dylla}, \citenamefont {Jordan},
  \citenamefont {Murray}, \citenamefont {Neil}, \citenamefont {Shinn},\ and\
  \citenamefont {Walker}}]{hernandez-garcia2005}%
  \BibitemOpen
  \bibfield  {author} {\bibinfo {author} {\bibfnamefont {C.}~\bibnamefont
  {Hernandez-Garcia}}, \bibinfo {author} {\bibfnamefont {T.}~\bibnamefont
  {Siggins}}, \bibinfo {author} {\bibfnamefont {S.}~\bibnamefont {Benson}},
  \bibinfo {author} {\bibfnamefont {D.}~\bibnamefont {Bullard}}, \bibinfo
  {author} {\bibfnamefont {H.}~\bibnamefont {Dylla}}, \bibinfo {author}
  {\bibfnamefont {K.}~\bibnamefont {Jordan}}, \bibinfo {author} {\bibfnamefont
  {C.}~\bibnamefont {Murray}}, \bibinfo {author} {\bibfnamefont
  {G.}~\bibnamefont {Neil}}, \bibinfo {author} {\bibfnamefont {M.}~\bibnamefont
  {Shinn}},\ and\ \bibinfo {author} {\bibfnamefont {R.}~\bibnamefont
  {Walker}},\ }\bibfield  {title} {\bibinfo {title} {A high average current
  {DC} {GaAs} photocathode gun for {ERLs} and {FELs}},\ }in\ \href
  {https://doi.org/10.1109/PAC.2005.1591383} {\emph {\bibinfo {booktitle}
  {Proc. 2005 Particle Accelerator Conference (PAC'05)}}}\ (\bibinfo {year}
  {2005})\ pp.\ \bibinfo {pages} {3117--3119}\BibitemShut {NoStop}%
\bibitem [{\citenamefont {Grames}\ \emph {et~al.}(2011)\citenamefont {Grames},
  \citenamefont {Suleiman}, \citenamefont {Adderley}, \citenamefont {Clark},
  \citenamefont {Hansknecht}, \citenamefont {Machie}, \citenamefont {Poelker},\
  and\ \citenamefont {Stutzman}}]{grames2011}%
  \BibitemOpen
  \bibfield  {author} {\bibinfo {author} {\bibfnamefont {J.}~\bibnamefont
  {Grames}}, \bibinfo {author} {\bibfnamefont {R.}~\bibnamefont {Suleiman}},
  \bibinfo {author} {\bibfnamefont {P.~A.}\ \bibnamefont {Adderley}}, \bibinfo
  {author} {\bibfnamefont {J.}~\bibnamefont {Clark}}, \bibinfo {author}
  {\bibfnamefont {J.}~\bibnamefont {Hansknecht}}, \bibinfo {author}
  {\bibfnamefont {D.}~\bibnamefont {Machie}}, \bibinfo {author} {\bibfnamefont
  {M.}~\bibnamefont {Poelker}},\ and\ \bibinfo {author} {\bibfnamefont {M.~L.}\
  \bibnamefont {Stutzman}},\ }\bibfield  {title} {\bibinfo {title} {Charge and
  fluence lifetime measurements of a {DC} high voltage {GaAs} photogun at high
  average current},\ }\href {https://doi.org/10.1103/PhysRevSTAB.14.043501}
  {\bibfield  {journal} {\bibinfo  {journal} {Phys. Rev. Accel. Beams}\
  }\textbf {\bibinfo {volume} {14}},\ \bibinfo {pages} {043501} (\bibinfo
  {year} {2011})}\BibitemShut {NoStop}%
\bibitem [{\citenamefont {Palacios-Serrano}\ \emph {et~al.}(2018)\citenamefont
  {Palacios-Serrano}, \citenamefont {Hannon}, \citenamefont {Hernandez-Garcia},
  \citenamefont {Poelker},\ and\ \citenamefont
  {Baumgart}}]{palacios-serrano2018}%
  \BibitemOpen
  \bibfield  {author} {\bibinfo {author} {\bibfnamefont {G.}~\bibnamefont
  {Palacios-Serrano}}, \bibinfo {author} {\bibfnamefont {F.}~\bibnamefont
  {Hannon}}, \bibinfo {author} {\bibfnamefont {C.}~\bibnamefont
  {Hernandez-Garcia}}, \bibinfo {author} {\bibfnamefont {M.}~\bibnamefont
  {Poelker}},\ and\ \bibinfo {author} {\bibfnamefont {H.}~\bibnamefont
  {Baumgart}},\ }\bibfield  {title} {\bibinfo {title} {{E}lectrostatic design
  and conditioning of a triple point junction shield for a
  $\ensuremath{-}200\text{ }\mathrm{kV}$ {DC} high voltage photogunn},\ }\href
  {https://doi.org/10.1063/1.5048700} {\bibfield  {journal} {\bibinfo
  {journal} {Rev. Sci. Instrum.}\ }\textbf {\bibinfo {volume} {89}},\ \bibinfo
  {pages} {104703} (\bibinfo {year} {2018})}\BibitemShut {NoStop}%
\bibitem [{\citenamefont {F\"orster}\ \emph {et~al.}(2022)\citenamefont
  {F\"orster}, \citenamefont {Sch\"ops}, \citenamefont {Enders}, \citenamefont
  {Herbert},\ and\ \citenamefont {Simona}}]{foerster2022}%
  \BibitemOpen
  \bibfield  {author} {\bibinfo {author} {\bibfnamefont {P.}~\bibnamefont
  {F\"orster}}, \bibinfo {author} {\bibfnamefont {S.}~\bibnamefont {Sch\"ops}},
  \bibinfo {author} {\bibfnamefont {J.}~\bibnamefont {Enders}}, \bibinfo
  {author} {\bibfnamefont {M.}~\bibnamefont {Herbert}},\ and\ \bibinfo {author}
  {\bibfnamefont {A.}~\bibnamefont {Simona}},\ }\bibfield  {title} {\bibinfo
  {title} {Freeform shape optimization of a compact dc photoelectron gun using
  isogeometric analysis},\ }\href
  {https://doi.org/10.1103/PhysRevAccelBeams.25.034601} {\bibfield  {journal}
  {\bibinfo  {journal} {Phys. Rev. Accel. Beams}\ }\textbf {\bibinfo {volume}
  {25}},\ \bibinfo {pages} {034601} (\bibinfo {year} {2022})}\BibitemShut
  {NoStop}%
\bibitem [{\citenamefont {Hernandez‐Garcia}\ \emph
  {et~al.}(2009)\citenamefont {Hernandez‐Garcia} \emph
  {et~al.}}]{hernandez-garcia2009}%
  \BibitemOpen
  \bibfield  {author} {\bibinfo {author} {\bibfnamefont {C.}~\bibnamefont
  {Hernandez‐Garcia}} \emph {et~al.},\ }\bibfield  {title} {\bibinfo {title}
  {{DC} high voltage conditioning of photoemission guns at {Jefferson} lab
  {FEL}},\ }\href {https://doi.org/10.1063/1.3215595} {\bibfield  {journal}
  {\bibinfo  {journal} {AIP Conf. Proc.}\ }\textbf {\bibinfo {volume} {1149}},\
  \bibinfo {pages} {1071} (\bibinfo {year} {2009})}\BibitemShut {NoStop}%
\bibitem [{\citenamefont {Sinclair}\ \emph {et~al.}(2007)\citenamefont
  {Sinclair} \emph {et~al.}}]{sinclair2007}%
  \BibitemOpen
  \bibfield  {author} {\bibinfo {author} {\bibfnamefont {C.~K.}\ \bibnamefont
  {Sinclair}} \emph {et~al.},\ }\bibfield  {title} {\bibinfo {title}
  {Development of a high average current polarized electron source with long
  cathode operational lifetime},\ }\href
  {https://doi.org/10.1103/PhysRevSTAB.10.023501} {\bibfield  {journal}
  {\bibinfo  {journal} {Phys. Rev. Accel. Beams}\ }\textbf {\bibinfo {volume}
  {10}},\ \bibinfo {pages} {023501} (\bibinfo {year} {2007})}\BibitemShut
  {NoStop}%
\bibitem [{\citenamefont {Pozdeyev}(2007)}]{pozdeyev2007}%
  \BibitemOpen
  \bibfield  {author} {\bibinfo {author} {\bibfnamefont {E.}~\bibnamefont
  {Pozdeyev}},\ }\bibfield  {title} {\bibinfo {title} {Ion trapping and cathode
  bombardment by trapped ions in dc photoguns},\ }\href
  {https://doi.org/10.1103/PhysRevSTAB.10.083501} {\bibfield  {journal}
  {\bibinfo  {journal} {Phys. Rev. Accel. Beams}\ }\textbf {\bibinfo {volume}
  {10}},\ \bibinfo {pages} {083501} (\bibinfo {year} {2007})}\BibitemShut
  {NoStop}%
\bibitem [{\citenamefont {Grames}\ \emph {et~al.}(2008)\citenamefont {Grames},
  \citenamefont {Adderley}, \citenamefont {Brittian}, \citenamefont {Clark},
  \citenamefont {Hansknecht}, \citenamefont {Machie}, \citenamefont {Poelker},
  \citenamefont {Pozdeyev}, \citenamefont {Stutzman},\ and\ \citenamefont
  {Surles‐Law}}]{grames2008}%
  \BibitemOpen
  \bibfield  {author} {\bibinfo {author} {\bibfnamefont {J.}~\bibnamefont
  {Grames}}, \bibinfo {author} {\bibfnamefont {P.}~\bibnamefont {Adderley}},
  \bibinfo {author} {\bibfnamefont {J.}~\bibnamefont {Brittian}}, \bibinfo
  {author} {\bibfnamefont {J.}~\bibnamefont {Clark}}, \bibinfo {author}
  {\bibfnamefont {J.}~\bibnamefont {Hansknecht}}, \bibinfo {author}
  {\bibfnamefont {D.}~\bibnamefont {Machie}}, \bibinfo {author} {\bibfnamefont
  {M.}~\bibnamefont {Poelker}}, \bibinfo {author} {\bibfnamefont
  {E.}~\bibnamefont {Pozdeyev}}, \bibinfo {author} {\bibfnamefont
  {M.}~\bibnamefont {Stutzman}},\ and\ \bibinfo {author} {\bibfnamefont
  {K.}~\bibnamefont {Surles‐Law}},\ }\bibfield  {title} {\bibinfo {title} {A
  biased anode to suppress ion back‐bombardment in a {DC} high voltage
  photoelectron gun},\ }\href {https://doi.org/10.1063/1.2888075} {\bibfield
  {journal} {\bibinfo  {journal} {AIP Conf. Proc.}\ }\textbf {\bibinfo {volume}
  {980}},\ \bibinfo {pages} {110} (\bibinfo {year} {2008})}\BibitemShut
  {NoStop}%
\bibitem [{\citenamefont {Yoskowitz}\ \emph {et~al.}(2020)\citenamefont
  {Yoskowitz}, \citenamefont {Grames}, \citenamefont {Hansknecht},
  \citenamefont {Hernandez-Garcia}, \citenamefont {Krafft}, \citenamefont
  {Poelker}, \citenamefont {Suleiman}, \citenamefont {Palacios-Seranno},
  \citenamefont {Wijethunga},\ and\ \citenamefont {Van
  Der~Geer}}]{yoskowitz2020}%
  \BibitemOpen
  \bibfield  {author} {\bibinfo {author} {\bibfnamefont {J.}~\bibnamefont
  {Yoskowitz}}, \bibinfo {author} {\bibfnamefont {J.}~\bibnamefont {Grames}},
  \bibinfo {author} {\bibfnamefont {J.}~\bibnamefont {Hansknecht}}, \bibinfo
  {author} {\bibfnamefont {C.}~\bibnamefont {Hernandez-Garcia}}, \bibinfo
  {author} {\bibfnamefont {G.}~\bibnamefont {Krafft}}, \bibinfo {author}
  {\bibfnamefont {M.}~\bibnamefont {Poelker}}, \bibinfo {author} {\bibfnamefont
  {R.}~\bibnamefont {Suleiman}}, \bibinfo {author} {\bibfnamefont
  {G.}~\bibnamefont {Palacios-Seranno}}, \bibinfo {author} {\bibfnamefont
  {S.}~\bibnamefont {Wijethunga}},\ and\ \bibinfo {author} {\bibfnamefont
  {B.}~\bibnamefont {Van Der~Geer}},\ }\bibfield  {title} {\bibinfo {title}
  {{New Simulations for Ion-Production and Back-Bombardment in {GaAs}
  Photo-guns}},\ }in\ \href {https://doi.org/10.22323/1.379.0040} {\emph
  {\bibinfo {booktitle} {Proc. 18th International Workshop on Polarized
  Sources, Targets, and Polarimetry {\textemdash} PoS(PSTP2019)}}},\ Vol.\
  \bibinfo {volume} {379}\ (\bibinfo {year} {2020})\ p.\ \bibinfo {pages}
  {040}\BibitemShut {NoStop}%
\bibitem [{\citenamefont {Yoskowitz}\ \emph {et~al.}(2024)\citenamefont
  {Yoskowitz}, \citenamefont {Krafft}, \citenamefont {Palacios-Serrano},
  \citenamefont {Wijethunga}, \citenamefont {Grames}, \citenamefont
  {Hansknecht}, \citenamefont {Hernandez-Garcia}, \citenamefont {Poelker},
  \citenamefont {Stutzman}, \citenamefont {Suleiman}, \citenamefont
  {Valerio-Lizarraga},\ and\ \citenamefont {Van Der~Geer}}]{yoskowitz2024}%
  \BibitemOpen
  \bibfield  {author} {\bibinfo {author} {\bibfnamefont {J.~T.}\ \bibnamefont
  {Yoskowitz}}, \bibinfo {author} {\bibfnamefont {G.~A.}\ \bibnamefont
  {Krafft}}, \bibinfo {author} {\bibfnamefont {G.}~\bibnamefont
  {Palacios-Serrano}}, \bibinfo {author} {\bibfnamefont {S.~A.~K.}\
  \bibnamefont {Wijethunga}}, \bibinfo {author} {\bibfnamefont
  {J.}~\bibnamefont {Grames}}, \bibinfo {author} {\bibfnamefont
  {J.}~\bibnamefont {Hansknecht}}, \bibinfo {author} {\bibfnamefont
  {C.}~\bibnamefont {Hernandez-Garcia}}, \bibinfo {author} {\bibfnamefont
  {M.}~\bibnamefont {Poelker}}, \bibinfo {author} {\bibfnamefont {M.~L.}\
  \bibnamefont {Stutzman}}, \bibinfo {author} {\bibfnamefont {R.}~\bibnamefont
  {Suleiman}}, \bibinfo {author} {\bibfnamefont {C.~A.}\ \bibnamefont
  {Valerio-Lizarraga}},\ and\ \bibinfo {author} {\bibfnamefont {S.~B.}\
  \bibnamefont {Van Der~Geer}},\ }\bibfield  {title} {\bibinfo {title} {Charge
  lifetime improvement of the continuous electron beam accelerator facility
  photogun with a biased anode},\ }\href
  {https://doi.org/10.1103/PhysRevAccelBeams.27.123401} {\bibfield  {journal}
  {\bibinfo  {journal} {Phys. Rev. Accel. Beams}\ }\textbf {\bibinfo {volume}
  {27}},\ \bibinfo {pages} {123401} (\bibinfo {year} {2024})}\BibitemShut
  {NoStop}%
\bibitem [{\citenamefont {Rahman}\ \emph
  {et~al.}(2019{\natexlab{a}})\citenamefont {Rahman}, \citenamefont {Wang},
  \citenamefont {Ben-Zvi}, \citenamefont {Biswas},\ and\ \citenamefont
  {Skaritka}}]{rahman2019a}%
  \BibitemOpen
  \bibfield  {author} {\bibinfo {author} {\bibfnamefont {O.}~\bibnamefont
  {Rahman}}, \bibinfo {author} {\bibfnamefont {E.}~\bibnamefont {Wang}},
  \bibinfo {author} {\bibfnamefont {I.}~\bibnamefont {Ben-Zvi}}, \bibinfo
  {author} {\bibfnamefont {J.}~\bibnamefont {Biswas}},\ and\ \bibinfo {author}
  {\bibfnamefont {J.}~\bibnamefont {Skaritka}},\ }\bibfield  {title} {\bibinfo
  {title} {Increasing charge lifetime in dc polarized electron guns by
  offsetting the anode},\ }\href
  {https://doi.org/10.1103/PhysRevAccelBeams.22.083401} {\bibfield  {journal}
  {\bibinfo  {journal} {Phys. Rev. Accel. Beams}\ }\textbf {\bibinfo {volume}
  {22}},\ \bibinfo {pages} {083401} (\bibinfo {year}
  {2019}{\natexlab{a}})}\BibitemShut {NoStop}%
\bibitem [{\citenamefont {Wang}\ \emph {et~al.}(2022)\citenamefont {Wang},
  \citenamefont {Rahman}, \citenamefont {Skaritka}, \citenamefont {Liu},
  \citenamefont {Biswas}, \citenamefont {Degen}, \citenamefont {Inacker},
  \citenamefont {Lambiase},\ and\ \citenamefont {Paniccia}}]{wang2022}%
  \BibitemOpen
  \bibfield  {author} {\bibinfo {author} {\bibfnamefont {E.}~\bibnamefont
  {Wang}}, \bibinfo {author} {\bibfnamefont {O.}~\bibnamefont {Rahman}},
  \bibinfo {author} {\bibfnamefont {J.}~\bibnamefont {Skaritka}}, \bibinfo
  {author} {\bibfnamefont {W.}~\bibnamefont {Liu}}, \bibinfo {author}
  {\bibfnamefont {J.}~\bibnamefont {Biswas}}, \bibinfo {author} {\bibfnamefont
  {C.}~\bibnamefont {Degen}}, \bibinfo {author} {\bibfnamefont
  {P.}~\bibnamefont {Inacker}}, \bibinfo {author} {\bibfnamefont
  {R.}~\bibnamefont {Lambiase}},\ and\ \bibinfo {author} {\bibfnamefont
  {M.}~\bibnamefont {Paniccia}},\ }\bibfield  {title} {\bibinfo {title} {High
  voltage dc gun for high intensity polarized electron source},\ }\href
  {https://doi.org/10.1103/PhysRevAccelBeams.25.033401} {\bibfield  {journal}
  {\bibinfo  {journal} {Phys. Rev. Accel. Beams}\ }\textbf {\bibinfo {volume}
  {25}},\ \bibinfo {pages} {033401} (\bibinfo {year} {2022})}\BibitemShut
  {NoStop}%
\bibitem [{\citenamefont {Cultrera}\ \emph {et~al.}(2020)\citenamefont
  {Cultrera}, \citenamefont {Galdi}, \citenamefont {Bae}, \citenamefont
  {Ikponmwen}, \citenamefont {Maxson},\ and\ \citenamefont
  {Bazarov}}]{cultrera2020}%
  \BibitemOpen
  \bibfield  {author} {\bibinfo {author} {\bibfnamefont {L.}~\bibnamefont
  {Cultrera}}, \bibinfo {author} {\bibfnamefont {A.}~\bibnamefont {Galdi}},
  \bibinfo {author} {\bibfnamefont {J.~K.}\ \bibnamefont {Bae}}, \bibinfo
  {author} {\bibfnamefont {F.}~\bibnamefont {Ikponmwen}}, \bibinfo {author}
  {\bibfnamefont {J.}~\bibnamefont {Maxson}},\ and\ \bibinfo {author}
  {\bibfnamefont {I.}~\bibnamefont {Bazarov}},\ }\bibfield  {title} {\bibinfo
  {title} {Long lifetime polarized electron beam production from negative
  electron affinity {GaAs} activated with {Sb}-{Cs}-{O}: Trade-offs between
  efficiency, spin polarization, and lifetime},\ }\href
  {https://doi.org/10.1103/PhysRevAccelBeams.23.023401} {\bibfield  {journal}
  {\bibinfo  {journal} {Phys. Rev. Accel. Beams}\ }\textbf {\bibinfo {volume}
  {23}},\ \bibinfo {pages} {023401} (\bibinfo {year} {2020})}\BibitemShut
  {NoStop}%
\bibitem [{\citenamefont {Bae}\ \emph {et~al.}(2018)\citenamefont {Bae},
  \citenamefont {Cultrera}, \citenamefont {DiGiacomo},\ and\ \citenamefont
  {Bazarov}}]{bae2018}%
  \BibitemOpen
  \bibfield  {author} {\bibinfo {author} {\bibfnamefont {J.~K.}\ \bibnamefont
  {Bae}}, \bibinfo {author} {\bibfnamefont {L.}~\bibnamefont {Cultrera}},
  \bibinfo {author} {\bibfnamefont {P.}~\bibnamefont {DiGiacomo}},\ and\
  \bibinfo {author} {\bibfnamefont {I.}~\bibnamefont {Bazarov}},\ }\bibfield
  {title} {\bibinfo {title} {Rugged spin-polarized electron sources based on
  negative electron affinity {GaAs} photocathode with robust
  $\mathrm{Cs}_2${Te} coating},\ }\href {https://doi.org/10.1063/1.5026701}
  {\bibfield  {journal} {\bibinfo  {journal} {Appl. Phys. Lett.}\ }\textbf
  {\bibinfo {volume} {112}},\ \bibinfo {pages} {154101} (\bibinfo {year}
  {2018})}\BibitemShut {NoStop}%
\bibitem [{\citenamefont {Mulhollan}\ and\ \citenamefont
  {Bierman}(2008)}]{mulhollan+bierman2008}%
  \BibitemOpen
  \bibfield  {author} {\bibinfo {author} {\bibfnamefont {G.~A.}\ \bibnamefont
  {Mulhollan}}\ and\ \bibinfo {author} {\bibfnamefont {J.~C.}\ \bibnamefont
  {Bierman}},\ }\bibfield  {title} {\bibinfo {title} {Enhanced chemical
  immunity for negative electron affinity {GaAs} photoemitters},\ }\href
  {https://doi.org/10.1116/1.2965816} {\bibfield  {journal} {\bibinfo
  {journal} {J. Vac. Sci. Technol. A}\ }\textbf {\bibinfo {volume} {26}},\
  \bibinfo {pages} {1195} (\bibinfo {year} {2008})}\BibitemShut {NoStop}%
\bibitem [{\citenamefont {Hagino}\ and\ \citenamefont
  {Nishida}(1969)}]{hagino+nishida1969}%
  \BibitemOpen
  \bibfield  {author} {\bibinfo {author} {\bibfnamefont {M.}~\bibnamefont
  {Hagino}}\ and\ \bibinfo {author} {\bibfnamefont {R.}~\bibnamefont
  {Nishida}},\ }\bibfield  {title} {\bibinfo {title} {Photoemission from
  {GaAs}-{Cs}-{Sb} ({Te})},\ }\href {https://doi.org/10.1143/jjap.8.123}
  {\bibfield  {journal} {\bibinfo  {journal} {Jpn. J. Appl. Phys.}\ }\textbf
  {\bibinfo {volume} {8}},\ \bibinfo {pages} {123} (\bibinfo {year}
  {1969})}\BibitemShut {NoStop}%
\bibitem [{\citenamefont {Sonnenberg}(1969{\natexlab{b}})}]{sonnenberg1969b}%
  \BibitemOpen
  \bibfield  {author} {\bibinfo {author} {\bibfnamefont {H.}~\bibnamefont
  {Sonnenberg}},\ }\bibfield  {title} {\bibinfo {title} {On photoemission from
  {GaAs}{\textendash}{Cs}{\textendash}{Sb}},\ }\href
  {https://doi.org/10.1143/jjap.8.806} {\bibfield  {journal} {\bibinfo
  {journal} {Jpn. J. Appl. Phys.}\ }\textbf {\bibinfo {volume} {8}},\ \bibinfo
  {pages} {806} (\bibinfo {year} {1969}{\natexlab{b}})}\BibitemShut {NoStop}%
\bibitem [{\citenamefont {Sugiyama}\ \emph {et~al.}(2011)\citenamefont
  {Sugiyama} \emph {et~al.}}]{sugiyama2011}%
  \BibitemOpen
  \bibfield  {author} {\bibinfo {author} {\bibfnamefont {H.}~\bibnamefont
  {Sugiyama}} \emph {et~al.},\ }\bibfield  {title} {\bibinfo {title} {A study
  of an electron affinity of cesium telluride thin film},\ }\href
  {https://doi.org/10.1088/1742-6596/298/1/012014} {\bibfield  {journal}
  {\bibinfo  {journal} {J. Phys. Conf. Ser.}\ }\textbf {\bibinfo {volume}
  {298}},\ \bibinfo {pages} {012014} (\bibinfo {year} {2011})}\BibitemShut
  {NoStop}%
\bibitem [{\citenamefont {Zhao}\ \emph {et~al.}(1993)\citenamefont {Zhao},
  \citenamefont {Zhou}, \citenamefont {Zhao},\ and\ \citenamefont
  {Xie}}]{zhao1993}%
  \BibitemOpen
  \bibfield  {author} {\bibinfo {author} {\bibfnamefont {L.}~\bibnamefont
  {Zhao}}, \bibinfo {author} {\bibfnamefont {Q.}~\bibnamefont {Zhou}}, \bibinfo
  {author} {\bibfnamefont {S.}~\bibnamefont {Zhao}},\ and\ \bibinfo {author}
  {\bibfnamefont {B.}~\bibnamefont {Xie}},\ }\bibfield  {title} {\bibinfo
  {title} {{Study of GaAs-Cs-Sb photoemitter}},\ }in\ \href
  {https://doi.org/10.1117/12.142006} {\emph {\bibinfo {booktitle}
  {Photoelectronic Detection and Imaging: Technology and Applications '93}}},\
  Vol.\ \bibinfo {volume} {1982},\ \bibinfo {editor} {edited by\ \bibinfo
  {editor} {\bibfnamefont {L.}~\bibnamefont {Zhou}}},\ \bibinfo {organization}
  {International Society for Optics and Photonics}\ (\bibinfo  {publisher}
  {SPIE},\ \bibinfo {year} {1993})\ pp.\ \bibinfo {pages} {140 --
  144}\BibitemShut {NoStop}%
\bibitem [{\citenamefont {Uchida}\ \emph {et~al.}(2014)\citenamefont {Uchida}
  \emph {et~al.}}]{uchida2014}%
  \BibitemOpen
  \bibfield  {author} {\bibinfo {author} {\bibfnamefont {K.}~\bibnamefont
  {Uchida}} \emph {et~al.},\ }\bibfield  {title} {{\selectlanguage
  {english}\bibinfo {title} {{A} {STUDY} {ON} {ROBUSTNESS} {OF} {NEA-GAAS}
  {PHOTOCATHODE}*}},\ }in\ \href
  {https://doi.org/https://doi.org/10.18429/JACoW-IPAC2014-MOPRI032}
  {{\selectlanguage {english}\emph {\bibinfo {booktitle} {Proc. 5th
  International Particle Accelerator Conference (IPAC'14)}}}},\ \bibinfo
  {series and number} {\bibinfo {series} {International Particle Accelerator
  Conference}\ No.~\bibinfo {number} {5}}\ (\bibinfo  {publisher} {JACoW},\
  \bibinfo {address} {Geneva, Switzerland},\ \bibinfo {year} {2014})\ pp.\
  \bibinfo {pages} {664--666}\BibitemShut {NoStop}%
\bibitem [{\citenamefont {Kuriki}\ \emph {et~al.}(2015)\citenamefont {Kuriki},
  \citenamefont {Kashiwagi}, \citenamefont {Seimiya},\ and\ \citenamefont
  {Uchida}}]{kuriki2015}%
  \BibitemOpen
  \bibfield  {author} {\bibinfo {author} {\bibfnamefont {M.}~\bibnamefont
  {Kuriki}}, \bibinfo {author} {\bibfnamefont {S.}~\bibnamefont {Kashiwagi}},
  \bibinfo {author} {\bibfnamefont {Y.}~\bibnamefont {Seimiya}},\ and\ \bibinfo
  {author} {\bibfnamefont {K.}~\bibnamefont {Uchida}},\ }\bibfield  {title}
  {{\selectlanguage {english}\bibinfo {title} {{G}a{A}s {P}hotocathode
  {A}ctivation with {C}s{T}e {T}hin {F}ilm}},\ }in\ \href
  {https://doi.org/https://doi.org/10.18429/JACoW-IPAC2015-TUPWA062}
  {{\selectlanguage {english}\emph {\bibinfo {booktitle} {Proc. 6th
  International Particle Accelerator Conference (IPAC'15)}}}},\ \bibinfo
  {series and number} {\bibinfo {series} {International Particle Accelerator
  Conference}\ No.~\bibinfo {number} {6}}\ (\bibinfo  {publisher} {JACoW},\
  \bibinfo {address} {Geneva, Switzerland},\ \bibinfo {year} {2015})\ pp.\
  \bibinfo {pages} {1567--1569}\BibitemShut {NoStop}%
\bibitem [{\citenamefont {Bae}\ \emph {et~al.}(2019)\citenamefont {Bae},
  \citenamefont {Bazarov}, \citenamefont {Cultrera}, \citenamefont {Galdi},
  \citenamefont {Ikponmwen},\ and\ \citenamefont {Maxson}}]{bae2019}%
  \BibitemOpen
  \bibfield  {author} {\bibinfo {author} {\bibfnamefont {J.~K.}\ \bibnamefont
  {Bae}}, \bibinfo {author} {\bibfnamefont {I.}~\bibnamefont {Bazarov}},
  \bibinfo {author} {\bibfnamefont {L.}~\bibnamefont {Cultrera}}, \bibinfo
  {author} {\bibfnamefont {A.}~\bibnamefont {Galdi}}, \bibinfo {author}
  {\bibfnamefont {F.}~\bibnamefont {Ikponmwen}},\ and\ \bibinfo {author}
  {\bibfnamefont {J.}~\bibnamefont {Maxson}},\ }\bibfield  {title}
  {{\selectlanguage {english}\bibinfo {title} {{Enhanced Robustness of
  {GaAs}-Based Photocathodes Activation by {Cs}, {Sb}, and $\mathrm{O}_2$}}},\
  }in\ \href {https://doi.org/10.18429/JACoW-NAPAC2019-MOPLH17}
  {{\selectlanguage {english}\emph {\bibinfo {booktitle} {Proc. 2019 North
  American Particle Accelerator Conference (NAPAC'19)}}}},\ \bibinfo {series
  and number} {\bibinfo {series} {North American Particle Accelerator
  Conference}\ No.~\bibinfo {number} {4}}\ (\bibinfo  {publisher} {JACoW
  Publishing, Geneva, Switzerland},\ \bibinfo {year} {2019})\ pp.\ \bibinfo
  {pages} {210--212}\BibitemShut {NoStop}%
\bibitem [{\citenamefont {Bae}\ \emph {et~al.}(2020)\citenamefont {Bae},
  \citenamefont {Galdi}, \citenamefont {Cultrera}, \citenamefont {Ikponmwen},
  \citenamefont {Maxson},\ and\ \citenamefont {Bazarov}}]{bae2020}%
  \BibitemOpen
  \bibfield  {author} {\bibinfo {author} {\bibfnamefont {J.~K.}\ \bibnamefont
  {Bae}}, \bibinfo {author} {\bibfnamefont {A.}~\bibnamefont {Galdi}}, \bibinfo
  {author} {\bibfnamefont {L.}~\bibnamefont {Cultrera}}, \bibinfo {author}
  {\bibfnamefont {F.}~\bibnamefont {Ikponmwen}}, \bibinfo {author}
  {\bibfnamefont {J.}~\bibnamefont {Maxson}},\ and\ \bibinfo {author}
  {\bibfnamefont {I.}~\bibnamefont {Bazarov}},\ }\bibfield  {title} {\bibinfo
  {title} {Improved lifetime of a high spin polarization superlattice
  photocathode},\ }\href {https://doi.org/10.1063/1.5139674} {\bibfield
  {journal} {\bibinfo  {journal} {J. Appl. Phys.}\ }\textbf {\bibinfo {volume}
  {127}},\ \bibinfo {pages} {124901} (\bibinfo {year} {2020})}\BibitemShut
  {NoStop}%
\bibitem [{\citenamefont {Rahman}\ \emph
  {et~al.}(2019{\natexlab{b}})\citenamefont {Rahman}, \citenamefont {Biswas},
  \citenamefont {Gaowei}, \citenamefont {Liu},\ and\ \citenamefont
  {Wang}}]{rahman2019b}%
  \BibitemOpen
  \bibfield  {author} {\bibinfo {author} {\bibfnamefont {O.}~\bibnamefont
  {Rahman}}, \bibinfo {author} {\bibfnamefont {J.}~\bibnamefont {Biswas}},
  \bibinfo {author} {\bibfnamefont {M.}~\bibnamefont {Gaowei}}, \bibinfo
  {author} {\bibfnamefont {W.}~\bibnamefont {Liu}},\ and\ \bibinfo {author}
  {\bibfnamefont {E.}~\bibnamefont {Wang}},\ }\bibfield  {title}
  {{\selectlanguage {english}\bibinfo {title} {{N}ew {A}ctivation {T}echniques
  for {H}igher {C}harge {L}ifetime from {G}a{A}s {P}hotocathodes}},\ }in\ \href
  {https://doi.org/doi:10.18429/JACoW-IPAC2019-TUPTS102} {{\selectlanguage
  {english}\emph {\bibinfo {booktitle} {Proc. 10th International Particle
  Accelerator Conference (IPAC'19)}}}},\ \bibinfo {series and number} {\bibinfo
  {series} {International Particle Accelerator Conference}\ No.~\bibinfo
  {number} {10}}\ (\bibinfo  {publisher} {JACoW Publishing},\ \bibinfo
  {address} {Geneva, Switzerland},\ \bibinfo {year} {2019})\ pp.\ \bibinfo
  {pages} {2157--2159},\ \bibinfo {note}
  {https://doi.org/10.18429/JACoW-IPAC2019-TUPTS102}\BibitemShut {NoStop}%
\bibitem [{\citenamefont {Biswas}\ \emph {et~al.}(2021)\citenamefont {Biswas},
  \citenamefont {Wang}, \citenamefont {Gaowei}, \citenamefont {Liu},
  \citenamefont {Rahman},\ and\ \citenamefont {Sadowski}}]{biswas2021}%
  \BibitemOpen
  \bibfield  {author} {\bibinfo {author} {\bibfnamefont {J.}~\bibnamefont
  {Biswas}}, \bibinfo {author} {\bibfnamefont {E.}~\bibnamefont {Wang}},
  \bibinfo {author} {\bibfnamefont {M.}~\bibnamefont {Gaowei}}, \bibinfo
  {author} {\bibfnamefont {W.}~\bibnamefont {Liu}}, \bibinfo {author}
  {\bibfnamefont {O.}~\bibnamefont {Rahman}},\ and\ \bibinfo {author}
  {\bibfnamefont {J.~T.}\ \bibnamefont {Sadowski}},\ }\bibfield  {title}
  {\bibinfo {title} {High quantum efficiency {GaAs} photocathodes activated
  with {Cs}, $\mathrm{O}_2$, and {Te}},\ }\href
  {https://doi.org/10.1063/5.0026839} {\bibfield  {journal} {\bibinfo
  {journal} {AIP Adv.}\ }\textbf {\bibinfo {volume} {11}},\ \bibinfo {pages}
  {025321} (\bibinfo {year} {2021})}\BibitemShut {NoStop}%
\bibitem [{\citenamefont {Mulhollan}(2010)}]{mulhollan2010}%
  \BibitemOpen
  \bibfield  {author} {\bibinfo {author} {\bibfnamefont {G.~A.}\ \bibnamefont
  {Mulhollan}},\ }\bibfield  {title} {\bibinfo {title} {Activation layer
  stabilization of high polarization photocathodes in sub-optimal rf gun
  environments},\ }\bibfield  {journal} {\bibinfo  {journal} {Scripta
  Metallurgica et Materialia}\ }\href {https://doi.org/10.2172/992578}
  {10.2172/992578} (\bibinfo {year} {2010})\BibitemShut {NoStop}%
\bibitem [{\citenamefont {Kurichiyanil}\ \emph {et~al.}(2019)\citenamefont
  {Kurichiyanil}, \citenamefont {Enders}, \citenamefont {Fritzsche},\ and\
  \citenamefont {Wagner}}]{kurichiyanil2019}%
  \BibitemOpen
  \bibfield  {author} {\bibinfo {author} {\bibfnamefont {N.}~\bibnamefont
  {Kurichiyanil}}, \bibinfo {author} {\bibfnamefont {J.}~\bibnamefont
  {Enders}}, \bibinfo {author} {\bibfnamefont {Y.}~\bibnamefont {Fritzsche}},\
  and\ \bibinfo {author} {\bibfnamefont {M.}~\bibnamefont {Wagner}},\
  }\bibfield  {title} {\bibinfo {title} {A test system for optimizing quantum
  efficiency and dark lifetime of {GaAs} photocathodes},\ }\href
  {https://doi.org/10.1088/1748-0221/14/08/p08025} {\bibfield  {journal}
  {\bibinfo  {journal} {J. Instrum.}\ }\textbf {\bibinfo {volume} {14}}\bibinfo
   {number} { (8)},\ \bibinfo {pages} {P08025}}\BibitemShut {NoStop}%
\bibitem [{\citenamefont {Herbert}\ \emph {et~al.}(2020)\citenamefont
  {Herbert}, \citenamefont {Enders}, \citenamefont {Poelker},\ and\
  \citenamefont {Hernandez-Garcia}}]{herbert2020}%
  \BibitemOpen
\bibfield  {number} {  }\bibfield  {author} {\bibinfo {author} {\bibfnamefont
  {M.}~\bibnamefont {Herbert}}, \bibinfo {author} {\bibfnamefont
  {J.}~\bibnamefont {Enders}}, \bibinfo {author} {\bibfnamefont
  {M.}~\bibnamefont {Poelker}},\ and\ \bibinfo {author} {\bibfnamefont
  {C.}~\bibnamefont {Hernandez-Garcia}},\ }\bibfield  {title} {\bibinfo {title}
  {{Lifetime Measurements of GaAs photocathodes at the Upgraded Injector Test
  Facility at Jefferson Lab}},\ }\href {https://doi.org/10.22323/1.379.0042}
  {\bibfield  {journal} {\bibinfo  {journal} {PoS}\ }\textbf {\bibinfo {volume}
  {PSTP2019}},\ \bibinfo {pages} {042} (\bibinfo {year} {2020})}\BibitemShut
  {NoStop}%
\bibitem [{\citenamefont {Kuriki}\ and\ \citenamefont
  {Masaki}(2019)}]{kuriki+masaki2019}%
  \BibitemOpen
  \bibfield  {author} {\bibinfo {author} {\bibfnamefont {M.}~\bibnamefont
  {Kuriki}}\ and\ \bibinfo {author} {\bibfnamefont {K.}~\bibnamefont
  {Masaki}},\ }\bibfield  {title} {\bibinfo {title} {Negative electron affinity
  {GaAs} cathode activation with {CsKTe} thin film},\ }\href
  {https://doi.org/10.1088/1742-6596/1350/1/012047} {\bibfield  {journal}
  {\bibinfo  {journal} {J. Phys. Conf. Ser.}\ }\textbf {\bibinfo {volume}
  {1350}},\ \bibinfo {pages} {012047} (\bibinfo {year} {2019})}\BibitemShut
  {NoStop}%
\bibitem [{\citenamefont {Bae}\ \emph {et~al.}(2022{\natexlab{a}})\citenamefont
  {Bae}, \citenamefont {Andorf}, \citenamefont {Bartnik}, \citenamefont
  {Galdi}, \citenamefont {Cultrera}, \citenamefont {Maxson},\ and\
  \citenamefont {Bazarov}}]{bae2022a}%
  \BibitemOpen
  \bibfield  {author} {\bibinfo {author} {\bibfnamefont {J.~K.}\ \bibnamefont
  {Bae}}, \bibinfo {author} {\bibfnamefont {M.}~\bibnamefont {Andorf}},
  \bibinfo {author} {\bibfnamefont {A.}~\bibnamefont {Bartnik}}, \bibinfo
  {author} {\bibfnamefont {A.}~\bibnamefont {Galdi}}, \bibinfo {author}
  {\bibfnamefont {L.}~\bibnamefont {Cultrera}}, \bibinfo {author}
  {\bibfnamefont {J.}~\bibnamefont {Maxson}},\ and\ \bibinfo {author}
  {\bibfnamefont {I.}~\bibnamefont {Bazarov}},\ }\bibfield  {title} {\bibinfo
  {title} {Operation of {Cs–Sb–O} activated {GaAs} in a high voltage {DC}
  electron gun at high average current},\ }\href
  {https://doi.org/10.1063/5.0100794} {\bibfield  {journal} {\bibinfo
  {journal} {AIP Adv.}\ }\textbf {\bibinfo {volume} {12}},\ \bibinfo {pages}
  {095017} (\bibinfo {year} {2022}{\natexlab{a}})}\BibitemShut {NoStop}%
\bibitem [{\citenamefont {Spicer}(1958)}]{spicer1958}%
  \BibitemOpen
  \bibfield  {author} {\bibinfo {author} {\bibfnamefont {W.~E.}\ \bibnamefont
  {Spicer}},\ }\bibfield  {title} {\bibinfo {title} {Photoemissive,
  photoconductive, and optical absorption studies of alkali-antimony
  compounds},\ }\href {https://doi.org/10.1103/PhysRev.112.114} {\bibfield
  {journal} {\bibinfo  {journal} {Phys. Rev.}\ }\textbf {\bibinfo {volume}
  {112}},\ \bibinfo {pages} {114} (\bibinfo {year} {1958})}\BibitemShut
  {NoStop}%
\bibitem [{\citenamefont {Spicer}\ and\ \citenamefont
  {Herrera-Gomez}(1993)}]{spicer+herrera-gomez1993}%
  \BibitemOpen
  \bibfield  {author} {\bibinfo {author} {\bibfnamefont {W.~E.}\ \bibnamefont
  {Spicer}}\ and\ \bibinfo {author} {\bibfnamefont {A.}~\bibnamefont
  {Herrera-Gomez}},\ }\bibfield  {title} {\bibinfo {title} {{Modern theory and
  applications of photocathodes}},\ }in\ \href
  {https://doi.org/10.1117/12.158575} {\emph {\bibinfo {booktitle}
  {Photodetectors and Power Meters}}},\ Vol.\ \bibinfo {volume} {2022},\
  \bibinfo {editor} {edited by\ \bibinfo {editor} {\bibfnamefont {K.~J.}\
  \bibnamefont {Kaufmann}}},\ \bibinfo {organization} {International Society
  for Optics and Photonics}\ (\bibinfo  {publisher} {SPIE},\ \bibinfo {year}
  {1993})\ pp.\ \bibinfo {pages} {18 -- 35}\BibitemShut {NoStop}%
\bibitem [{\citenamefont {Lampel}(1968)}]{lampel1968}%
  \BibitemOpen
  \bibfield  {author} {\bibinfo {author} {\bibfnamefont {G.}~\bibnamefont
  {Lampel}},\ }\bibfield  {title} {\bibinfo {title} {Nuclear dynamic
  polarization by optical electronic saturation and optical pumping in
  semiconductors},\ }\href {https://doi.org/10.1103/PhysRevLett.20.491}
  {\bibfield  {journal} {\bibinfo  {journal} {Phys. Rev. Lett.}\ }\textbf
  {\bibinfo {volume} {20}},\ \bibinfo {pages} {491} (\bibinfo {year}
  {1968})}\BibitemShut {NoStop}%
\bibitem [{\citenamefont {Parsons}(1969)}]{parsons1969}%
  \BibitemOpen
  \bibfield  {author} {\bibinfo {author} {\bibfnamefont {R.~R.}\ \bibnamefont
  {Parsons}},\ }\bibfield  {title} {\bibinfo {title} {Band-to-band optical
  pumping in solids and polarized photoluminescence},\ }\href
  {https://doi.org/10.1103/PhysRevLett.23.1152} {\bibfield  {journal} {\bibinfo
   {journal} {Phys. Rev. Lett.}\ }\textbf {\bibinfo {volume} {23}},\ \bibinfo
  {pages} {1152} (\bibinfo {year} {1969})}\BibitemShut {NoStop}%
\bibitem [{\citenamefont {Cardona}\ \emph {et~al.}(1988)\citenamefont
  {Cardona}, \citenamefont {Christensen},\ and\ \citenamefont
  {Fasol}}]{cardona1988}%
  \BibitemOpen
  \bibfield  {author} {\bibinfo {author} {\bibfnamefont {M.}~\bibnamefont
  {Cardona}}, \bibinfo {author} {\bibfnamefont {N.~E.}\ \bibnamefont
  {Christensen}},\ and\ \bibinfo {author} {\bibfnamefont {G.}~\bibnamefont
  {Fasol}},\ }\bibfield  {title} {\bibinfo {title} {Relativistic band structure
  and spin-orbit splitting of zinc-blende-type semiconductors},\ }\href
  {https://doi.org/10.1103/PhysRevB.38.1806} {\bibfield  {journal} {\bibinfo
  {journal} {Phys. Rev. B}\ }\textbf {\bibinfo {volume} {38}},\ \bibinfo
  {pages} {1806} (\bibinfo {year} {1988})}\BibitemShut {NoStop}%
\bibitem [{\citenamefont {Soboleva}(1974)}]{soboleva1974}%
  \BibitemOpen
  \bibfield  {author} {\bibinfo {author} {\bibfnamefont {N.~A.}\ \bibnamefont
  {Soboleva}},\ }\bibfield  {title} {\bibinfo {title} {A new class of electron
  emitters},\ }\href {https://doi.org/10.1070/pu1974v016n05abeh004152}
  {\bibfield  {journal} {\bibinfo  {journal} {Sov. Phys. Usp.}\ }\textbf
  {\bibinfo {volume} {16}},\ \bibinfo {pages} {726} (\bibinfo {year}
  {1974})}\BibitemShut {NoStop}%
\bibitem [{\citenamefont {Chubenko}\ \emph {et~al.}(2021)\citenamefont
  {Chubenko}, \citenamefont {Karkare}, \citenamefont {Dimitrov}, \citenamefont
  {Bae}, \citenamefont {Cultrera}, \citenamefont {Bazarov},\ and\ \citenamefont
  {Afanasev}}]{chubenko2021}%
  \BibitemOpen
  \bibfield  {author} {\bibinfo {author} {\bibfnamefont {O.}~\bibnamefont
  {Chubenko}}, \bibinfo {author} {\bibfnamefont {S.}~\bibnamefont {Karkare}},
  \bibinfo {author} {\bibfnamefont {D.~A.}\ \bibnamefont {Dimitrov}}, \bibinfo
  {author} {\bibfnamefont {J.~K.}\ \bibnamefont {Bae}}, \bibinfo {author}
  {\bibfnamefont {L.}~\bibnamefont {Cultrera}}, \bibinfo {author}
  {\bibfnamefont {I.}~\bibnamefont {Bazarov}},\ and\ \bibinfo {author}
  {\bibfnamefont {A.}~\bibnamefont {Afanasev}},\ }\bibfield  {title} {\bibinfo
  {title} {{Monte Carlo modeling of spin-polarized photoemission from p-doped
  bulk GaAs}},\ }\href {https://doi.org/10.1063/5.0060151} {\bibfield
  {journal} {\bibinfo  {journal} {J. Appl. Phys.}\ }\textbf {\bibinfo {volume}
  {130}},\ \bibinfo {pages} {063101} (\bibinfo {year} {2021})},\ \Eprint
  {https://arxiv.org/abs/https://pubs.aip.org/aip/jap/article-pdf/doi/10.1063/5.0060151/15264943/063101\_1\_online.pdf}
  {https://pubs.aip.org/aip/jap/article-pdf/doi/10.1063/5.0060151/15264943/063101\_1\_online.pdf}
  \BibitemShut {NoStop}%
\bibitem [{\citenamefont {Scheer}\ and\ \citenamefont {{Van
  Laar}}(1969)}]{scheer+laar1969}%
  \BibitemOpen
  \bibfield  {author} {\bibinfo {author} {\bibfnamefont {J.}~\bibnamefont
  {Scheer}}\ and\ \bibinfo {author} {\bibfnamefont {J.}~\bibnamefont {{Van
  Laar}}},\ }\bibfield  {title} {\bibinfo {title} {The influence of cesium
  adsorption on surface fermi level position in gallium arsenide},\ }\href
  {https://doi.org/https://doi.org/10.1016/0039-6028(69)90271-4} {\bibfield
  {journal} {\bibinfo  {journal} {Surf. Sci.}\ }\textbf {\bibinfo {volume}
  {18}},\ \bibinfo {pages} {130} (\bibinfo {year} {1969})}\BibitemShut
  {NoStop}%
\bibitem [{\citenamefont {Kampen}\ \emph {et~al.}(1998)\citenamefont {Kampen},
  \citenamefont {Eyckeler},\ and\ \citenamefont {Mönch}}]{kampen1998}%
  \BibitemOpen
  \bibfield  {author} {\bibinfo {author} {\bibfnamefont {T.~U.}\ \bibnamefont
  {Kampen}}, \bibinfo {author} {\bibfnamefont {M.}~\bibnamefont {Eyckeler}},\
  and\ \bibinfo {author} {\bibfnamefont {W.}~\bibnamefont {Mönch}},\
  }\bibfield  {title} {\bibinfo {title} {Electronic properties of
  cesium-covered {GaN}(0001) surfaces},\ }\href
  {https://doi.org/https://doi.org/10.1016/S0169-4332(97)00495-9} {\bibfield
  {journal} {\bibinfo  {journal} {Appl. Surf. Sci.}\ }\textbf {\bibinfo
  {volume} {123-124}},\ \bibinfo {pages} {28} (\bibinfo {year}
  {1998})}\BibitemShut {NoStop}%
\bibitem [{\citenamefont {Alperovich}\ \emph {et~al.}(1995)\citenamefont
  {Alperovich}, \citenamefont {Paulish},\ and\ \citenamefont
  {Terekhov}}]{alperovich1995}%
  \BibitemOpen
  \bibfield  {author} {\bibinfo {author} {\bibfnamefont {V.}~\bibnamefont
  {Alperovich}}, \bibinfo {author} {\bibfnamefont {A.}~\bibnamefont
  {Paulish}},\ and\ \bibinfo {author} {\bibfnamefont {A.}~\bibnamefont
  {Terekhov}},\ }\bibfield  {title} {\bibinfo {title} {Unpinned behavior of the
  electronic properties of a p-{GaAs(Cs,O)} surface at room temperature},\
  }\href {https://doi.org/https://doi.org/10.1016/0039-6028(95)00290-1}
  {\bibfield  {journal} {\bibinfo  {journal} {Surf. Sci.}\ }\textbf {\bibinfo
  {volume} {331-333}},\ \bibinfo {pages} {1250} (\bibinfo {year} {1995})},\
  \bibinfo {note} {proc. 14th European Conference on Surface
  Science}\BibitemShut {NoStop}%
\bibitem [{\citenamefont {Madey}\ and\ \citenamefont
  {Yates}(1971)}]{madey+yates1971}%
  \BibitemOpen
  \bibfield  {author} {\bibinfo {author} {\bibfnamefont {T.~E.}\ \bibnamefont
  {Madey}}\ and\ \bibinfo {author} {\bibfnamefont {J.}~\bibnamefont {Yates},
  \bibfnamefont {John~T.}},\ }\bibfield  {title} {\bibinfo {title}
  {Electron-stimulated desorption and work function studies of clean and
  cesiated (110) gaas},\ }\href {https://doi.org/10.1116/1.1316348} {\bibfield
  {journal} {\bibinfo  {journal} {J. Vac. Sci. Technol.}\ }\textbf {\bibinfo
  {volume} {8}},\ \bibinfo {pages} {39} (\bibinfo {year} {1971})}\BibitemShut
  {NoStop}%
\bibitem [{\citenamefont {Milton}\ and\ \citenamefont
  {Baer}(1971)}]{milton+baer1971}%
  \BibitemOpen
  \bibfield  {author} {\bibinfo {author} {\bibfnamefont {A.~F.}\ \bibnamefont
  {Milton}}\ and\ \bibinfo {author} {\bibfnamefont {A.~D.}\ \bibnamefont
  {Baer}},\ }\bibfield  {title} {\bibinfo {title} {Interfacial barrier of
  heterojunction photocathodes},\ }\href {https://doi.org/10.1063/1.1659897}
  {\bibfield  {journal} {\bibinfo  {journal} {J. Appl. Phys.}\ }\textbf
  {\bibinfo {volume} {42}},\ \bibinfo {pages} {5095} (\bibinfo {year}
  {1971})}\BibitemShut {NoStop}%
\bibitem [{\citenamefont {Burt}\ and\ \citenamefont
  {Heine}(1978)}]{burt+heine1978}%
  \BibitemOpen
  \bibfield  {author} {\bibinfo {author} {\bibfnamefont {M.~G.}\ \bibnamefont
  {Burt}}\ and\ \bibinfo {author} {\bibfnamefont {V.}~\bibnamefont {Heine}},\
  }\bibfield  {title} {\bibinfo {title} {The theory of the workfunction of
  caesium suboxides and caesium films},\ }\href
  {https://doi.org/10.1088/0022-3719/11/5/016} {\bibfield  {journal} {\bibinfo
  {journal} {J. Phys. C}\ }\textbf {\bibinfo {volume} {11}},\ \bibinfo {pages}
  {961} (\bibinfo {year} {1978})}\BibitemShut {NoStop}%
\bibitem [{\citenamefont {Biswas}\ \emph {et~al.}(2020)\citenamefont {Biswas},
  \citenamefont {Cen}, \citenamefont {Gaowei}, \citenamefont {Rahman},
  \citenamefont {Liu}, \citenamefont {Tong},\ and\ \citenamefont
  {Wang}}]{biswas2020}%
  \BibitemOpen
  \bibfield  {author} {\bibinfo {author} {\bibfnamefont {J.}~\bibnamefont
  {Biswas}}, \bibinfo {author} {\bibfnamefont {J.}~\bibnamefont {Cen}},
  \bibinfo {author} {\bibfnamefont {M.}~\bibnamefont {Gaowei}}, \bibinfo
  {author} {\bibfnamefont {O.}~\bibnamefont {Rahman}}, \bibinfo {author}
  {\bibfnamefont {W.}~\bibnamefont {Liu}}, \bibinfo {author} {\bibfnamefont
  {X.}~\bibnamefont {Tong}},\ and\ \bibinfo {author} {\bibfnamefont
  {E.}~\bibnamefont {Wang}},\ }\bibfield  {title} {\bibinfo {title} {Revisiting
  heat treatment and surface activation of gaas photocathodes: In situ studies
  using scanning tunneling microscopy and photoelectron spectroscopy},\
  }\bibfield  {journal} {\bibinfo  {journal} {J. Appl. Phys.}\ }\textbf
  {\bibinfo {volume} {128}},\ \href {https://doi.org/10.1063/5.0008969}
  {10.1063/5.0008969} (\bibinfo {year} {2020}),\ \bibinfo {note}
  {045308}\BibitemShut {NoStop}%
\bibitem [{\citenamefont {Sonnenberg}(1970)}]{sonnenberg1970}%
  \BibitemOpen
  \bibfield  {author} {\bibinfo {author} {\bibfnamefont {H.}~\bibnamefont
  {Sonnenberg}},\ }\bibfield  {title} {\bibinfo {title}
  {Negative-electron-affinity photoemitters},\ }\href
  {https://doi.org/10.1109/JSSC.1970.1050126} {\bibfield  {journal} {\bibinfo
  {journal} {IEEE J. Solid-State Circuits}\ }\textbf {\bibinfo {volume} {5}},\
  \bibinfo {pages} {272} (\bibinfo {year} {1970})}\BibitemShut {NoStop}%
\bibitem [{\citenamefont {Topping}\ and\ \citenamefont
  {Chapman}(1927)}]{topping+chapman1927}%
  \BibitemOpen
  \bibfield  {author} {\bibinfo {author} {\bibfnamefont {J.}~\bibnamefont
  {Topping}}\ and\ \bibinfo {author} {\bibfnamefont {S.}~\bibnamefont
  {Chapman}},\ }\bibfield  {title} {\bibinfo {title} {On the mutual potential
  energy of a plane network of doublets},\ }\href
  {https://doi.org/10.1098/rspa.1927.0025} {\bibfield  {journal} {\bibinfo
  {journal} {Proc. R. Soc. London A}\ }\textbf {\bibinfo {volume} {114}},\
  \bibinfo {pages} {67} (\bibinfo {year} {1927})},\ \Eprint
  {https://arxiv.org/abs/https://royalsocietypublishing.org/doi/pdf/10.1098/rspa.1927.0025}
  {https://royalsocietypublishing.org/doi/pdf/10.1098/rspa.1927.0025}
  \BibitemShut {NoStop}%
\bibitem [{\citenamefont {Clemens}\ \emph {et~al.}(1978)\citenamefont
  {Clemens}, \citenamefont {{Von Wienskowski}},\ and\ \citenamefont
  {Mönch}}]{clemens1978}%
  \BibitemOpen
  \bibfield  {author} {\bibinfo {author} {\bibfnamefont {H.}~\bibnamefont
  {Clemens}}, \bibinfo {author} {\bibfnamefont {J.}~\bibnamefont {{Von
  Wienskowski}}},\ and\ \bibinfo {author} {\bibfnamefont {W.}~\bibnamefont
  {Mönch}},\ }\bibfield  {title} {\bibinfo {title} {On the interaction of
  cesium with cleaved gaas(110) and ge(111) surfaces: Work function
  measurements and adsorption site model},\ }\href
  {https://doi.org/https://doi.org/10.1016/0039-6028(78)90238-8} {\bibfield
  {journal} {\bibinfo  {journal} {Surf. Sci.}\ }\textbf {\bibinfo {volume}
  {78}},\ \bibinfo {pages} {648} (\bibinfo {year} {1978})}\BibitemShut
  {NoStop}%
\bibitem [{\citenamefont {Kampen}\ and\ \citenamefont
  {M\"onch}(1992)}]{kampen+moench1992}%
  \BibitemOpen
  \bibfield  {author} {\bibinfo {author} {\bibfnamefont {T.~U.}\ \bibnamefont
  {Kampen}}\ and\ \bibinfo {author} {\bibfnamefont {W.}~\bibnamefont
  {M\"onch}},\ }\bibfield  {title} {\bibinfo {title} {Hydrogen-induced
  variations of the ionization energy on gaas(110) surfaces},\ }\href
  {https://doi.org/10.1103/PhysRevB.46.13309} {\bibfield  {journal} {\bibinfo
  {journal} {Phys. Rev. B}\ }\textbf {\bibinfo {volume} {46}},\ \bibinfo
  {pages} {13309} (\bibinfo {year} {1992})}\BibitemShut {NoStop}%
\bibitem [{\citenamefont {Hannay}\ and\ \citenamefont
  {Smyth}(1946)}]{hannay1946}%
  \BibitemOpen
  \bibfield  {author} {\bibinfo {author} {\bibfnamefont {N.~B.}\ \bibnamefont
  {Hannay}}\ and\ \bibinfo {author} {\bibfnamefont {C.~P.}\ \bibnamefont
  {Smyth}},\ }\bibfield  {title} {\bibinfo {title} {The dipole moment of
  hydrogen fluoride and the ionic character of bonds},\ }\href
  {https://doi.org/10.1021/ja01206a003} {\bibfield  {journal} {\bibinfo
  {journal} {J. Am. Chem. Soc.}\ }\textbf {\bibinfo {volume} {68}},\ \bibinfo
  {pages} {171} (\bibinfo {year} {1946})}\BibitemShut {NoStop}%
\bibitem [{\citenamefont {Dowell}\ \emph {et~al.}(2006)\citenamefont {Dowell},
  \citenamefont {King}, \citenamefont {Kirby}, \citenamefont {Schmerge},\ and\
  \citenamefont {Smedley}}]{dowell2006}%
  \BibitemOpen
  \bibfield  {author} {\bibinfo {author} {\bibfnamefont {D.~H.}\ \bibnamefont
  {Dowell}}, \bibinfo {author} {\bibfnamefont {F.~K.}\ \bibnamefont {King}},
  \bibinfo {author} {\bibfnamefont {R.~E.}\ \bibnamefont {Kirby}}, \bibinfo
  {author} {\bibfnamefont {J.~F.}\ \bibnamefont {Schmerge}},\ and\ \bibinfo
  {author} {\bibfnamefont {J.~M.}\ \bibnamefont {Smedley}},\ }\bibfield
  {title} {\bibinfo {title} {In situ cleaning of metal cathodes using a
  hydrogen ion beam},\ }\href {https://doi.org/10.1103/PhysRevSTAB.9.063502}
  {\bibfield  {journal} {\bibinfo  {journal} {Phys. Rev. Accel. Beams}\
  }\textbf {\bibinfo {volume} {9}},\ \bibinfo {pages} {063502} (\bibinfo {year}
  {2006})}\BibitemShut {NoStop}%
\bibitem [{\citenamefont {Akinlami}\ and\ \citenamefont
  {Ashamu}(2013)}]{akinlami2013}%
  \BibitemOpen
  \bibfield  {author} {\bibinfo {author} {\bibfnamefont {J.~O.}\ \bibnamefont
  {Akinlami}}\ and\ \bibinfo {author} {\bibfnamefont {A.}~\bibnamefont
  {Ashamu}},\ }\bibfield  {title} {\bibinfo {title} {Optical properties of
  gaas},\ }\href {https://doi.org/10.1088/1674-4926/34/3/032002} {\bibfield
  {journal} {\bibinfo  {journal} {J. Semicond.}\ }\textbf {\bibinfo {volume}
  {34}},\ \bibinfo {pages} {032002} (\bibinfo {year} {2013})}\BibitemShut
  {NoStop}%
\bibitem [{\citenamefont {Yang}\ \emph {et~al.}(1992)\citenamefont {Yang},
  \citenamefont {Ciullo}, \citenamefont {Guidi},\ and\ \citenamefont
  {Tecchio}}]{yang1992}%
  \BibitemOpen
  \bibfield  {author} {\bibinfo {author} {\bibfnamefont {B.}~\bibnamefont
  {Yang}}, \bibinfo {author} {\bibfnamefont {G.}~\bibnamefont {Ciullo}},
  \bibinfo {author} {\bibfnamefont {V.}~\bibnamefont {Guidi}},\ and\ \bibinfo
  {author} {\bibfnamefont {L.}~\bibnamefont {Tecchio}},\ }\bibfield  {title}
  {\bibinfo {title} {Monte carlo simulation of a gaas electron source},\ }\href
  {https://doi.org/10.1088/0022-3727/25/12/022} {\bibfield  {journal} {\bibinfo
   {journal} {J. Phys. D}\ }\textbf {\bibinfo {volume} {25}},\ \bibinfo {pages}
  {1834} (\bibinfo {year} {1992})}\BibitemShut {NoStop}%
\bibitem [{\citenamefont {Uebbing}\ and\ \citenamefont
  {James}(1970)}]{uebbing+james1970}%
  \BibitemOpen
  \bibfield  {author} {\bibinfo {author} {\bibfnamefont {J.~J.}\ \bibnamefont
  {Uebbing}}\ and\ \bibinfo {author} {\bibfnamefont {L.~W.}\ \bibnamefont
  {James}},\ }\bibfield  {title} {\bibinfo {title} {Behavior of cesium oxide as
  a low work‐function coating},\ }\href {https://doi.org/10.1063/1.1658489}
  {\bibfield  {journal} {\bibinfo  {journal} {J. Appl. Phys.}\ }\textbf
  {\bibinfo {volume} {41}},\ \bibinfo {pages} {4505} (\bibinfo {year}
  {1970})}\BibitemShut {NoStop}%
\bibitem [{\citenamefont {Su}\ \emph {et~al.}(1983)\citenamefont {Su},
  \citenamefont {Spicer},\ and\ \citenamefont {Lindau}}]{su1983}%
  \BibitemOpen
  \bibfield  {author} {\bibinfo {author} {\bibfnamefont {C.~Y.}\ \bibnamefont
  {Su}}, \bibinfo {author} {\bibfnamefont {W.~E.}\ \bibnamefont {Spicer}},\
  and\ \bibinfo {author} {\bibfnamefont {I.}~\bibnamefont {Lindau}},\
  }\bibfield  {title} {\bibinfo {title} {Photoelectron spectroscopic
  determination of the structure of (cs,o) activated gaas (110) surfaces},\
  }\href {https://doi.org/10.1063/1.332166} {\bibfield  {journal} {\bibinfo
  {journal} {J. Appl. Phys.}\ }\textbf {\bibinfo {volume} {54}},\ \bibinfo
  {pages} {1413} (\bibinfo {year} {1983})}\BibitemShut {NoStop}%
\bibitem [{\citenamefont {Gregory}\ \emph {et~al.}(1974)\citenamefont
  {Gregory}, \citenamefont {Spicer}, \citenamefont {Ciraci},\ and\
  \citenamefont {Harrison}}]{gregory1974}%
  \BibitemOpen
  \bibfield  {author} {\bibinfo {author} {\bibfnamefont {P.~E.}\ \bibnamefont
  {Gregory}}, \bibinfo {author} {\bibfnamefont {W.~E.}\ \bibnamefont {Spicer}},
  \bibinfo {author} {\bibfnamefont {S.}~\bibnamefont {Ciraci}},\ and\ \bibinfo
  {author} {\bibfnamefont {W.~A.}\ \bibnamefont {Harrison}},\ }\bibfield
  {title} {\bibinfo {title} {{Surface state band on GaAs (110) face}},\ }\href
  {https://doi.org/10.1063/1.1655570} {\bibfield  {journal} {\bibinfo
  {journal} {Appl. Phys. Lett.}\ }\textbf {\bibinfo {volume} {25}},\ \bibinfo
  {pages} {511} (\bibinfo {year} {1974})}\BibitemShut {NoStop}%
\bibitem [{\citenamefont {Kahn}\ \emph {et~al.}(1978)\citenamefont {Kahn},
  \citenamefont {So}, \citenamefont {Mark}, \citenamefont {Duke},\ and\
  \citenamefont {Meyer}}]{kahn1978}%
  \BibitemOpen
  \bibfield  {author} {\bibinfo {author} {\bibfnamefont {A.}~\bibnamefont
  {Kahn}}, \bibinfo {author} {\bibfnamefont {E.}~\bibnamefont {So}}, \bibinfo
  {author} {\bibfnamefont {P.}~\bibnamefont {Mark}}, \bibinfo {author}
  {\bibfnamefont {C.~B.}\ \bibnamefont {Duke}},\ and\ \bibinfo {author}
  {\bibfnamefont {R.~J.}\ \bibnamefont {Meyer}},\ }\bibfield  {title} {\bibinfo
  {title} {{Surface and near‐surface atomic structure of GaAs (110)}},\
  }\href {https://doi.org/10.1116/1.569697} {\bibfield  {journal} {\bibinfo
  {journal} {J. Vac. Sci. Technol.}\ }\textbf {\bibinfo {volume} {15}},\
  \bibinfo {pages} {1223} (\bibinfo {year} {1978})}\BibitemShut {NoStop}%
\bibitem [{\citenamefont {Manghi}\ \emph {et~al.}(1984)\citenamefont {Manghi},
  \citenamefont {Calandra}, \citenamefont {Bertoni},\ and\ \citenamefont
  {Molinary}}]{manghi1984}%
  \BibitemOpen
  \bibfield  {author} {\bibinfo {author} {\bibfnamefont {F.}~\bibnamefont
  {Manghi}}, \bibinfo {author} {\bibfnamefont {C.}~\bibnamefont {Calandra}},
  \bibinfo {author} {\bibfnamefont {C.}~\bibnamefont {Bertoni}},\ and\ \bibinfo
  {author} {\bibfnamefont {E.}~\bibnamefont {Molinary}},\ }\bibfield  {title}
  {\bibinfo {title} {Electronic properties of the cs-gaas(110) interface at
  monolayer coverage},\ }\href {https://doi.org/10.1016/0039-6028(84)90635-6}
  {\bibfield  {journal} {\bibinfo  {journal} {Surf. Sci.}\ }\textbf {\bibinfo
  {volume} {136}},\ \bibinfo {pages} {629} (\bibinfo {year}
  {1984})}\BibitemShut {NoStop}%
\bibitem [{\citenamefont {Schailey}\ and\ \citenamefont
  {Ray}(1999)}]{schailey+ray1999}%
  \BibitemOpen
  \bibfield  {author} {\bibinfo {author} {\bibfnamefont {R.}~\bibnamefont
  {Schailey}}\ and\ \bibinfo {author} {\bibfnamefont {A.~K.}\ \bibnamefont
  {Ray}},\ }\bibfield  {title} {\bibinfo {title} {{An ab initio cluster study
  of chemisorption of atomic Cs on Ga-rich GaAs (100) (2×1), (2×2), and
  $\beta$(4×2) surfaces}},\ }\href {https://doi.org/10.1063/1.480203}
  {\bibfield  {journal} {\bibinfo  {journal} {J. Chem. Phys.}\ }\textbf
  {\bibinfo {volume} {111}},\ \bibinfo {pages} {8628} (\bibinfo {year}
  {1999})}\BibitemShut {NoStop}%
\bibitem [{\citenamefont {Feenstra}\ and\ \citenamefont
  {Stroscio}(1993)}]{feenstra+stroscio1993}%
  \BibitemOpen
  \bibfield  {author} {\bibinfo {author} {\bibfnamefont {R.}~\bibnamefont
  {Feenstra}}\ and\ \bibinfo {author} {\bibfnamefont {J.~A.}\ \bibnamefont
  {Stroscio}},\ }\bibfield  {title} {\bibinfo {title} {5.3. gallium arsenide},\
  }in\ \href {https://doi.org/10.1016/S0076-695X(08)60012-5} {\emph {\bibinfo
  {booktitle} {Scanning Tunneling Microscopy}}},\ \bibinfo {series} {Methods in
  Experimental Physics}, Vol.~\bibinfo {volume} {27},\ \bibinfo {editor}
  {edited by\ \bibinfo {editor} {\bibfnamefont {J.~A.}\ \bibnamefont
  {Stroscio}}\ and\ \bibinfo {editor} {\bibfnamefont {W.~J.}\ \bibnamefont
  {Kaiser}}}\ (\bibinfo  {publisher} {Academic Press},\ \bibinfo {year}
  {1993})\ pp.\ \bibinfo {pages} {251--276}\BibitemShut {NoStop}%
\bibitem [{\citenamefont {Fisher}\ \emph {et~al.}(1972)\citenamefont {Fisher},
  \citenamefont {Enstrom}, \citenamefont {Escher},\ and\ \citenamefont
  {Williams}}]{fisher1972}%
  \BibitemOpen
  \bibfield  {author} {\bibinfo {author} {\bibfnamefont {D.~G.}\ \bibnamefont
  {Fisher}}, \bibinfo {author} {\bibfnamefont {R.~E.}\ \bibnamefont {Enstrom}},
  \bibinfo {author} {\bibfnamefont {J.~S.}\ \bibnamefont {Escher}},\ and\
  \bibinfo {author} {\bibfnamefont {B.~F.}\ \bibnamefont {Williams}},\
  }\bibfield  {title} {\bibinfo {title} {Photoelectron surface escape
  probability of (ga,in)as : Cs–o in the 0.9 to ~ 1.6 $\mu$m range},\ }\href
  {https://doi.org/10.1063/1.1661817} {\bibfield  {journal} {\bibinfo
  {journal} {J. Appl. Phys.}\ }\textbf {\bibinfo {volume} {43}},\ \bibinfo
  {pages} {3815} (\bibinfo {year} {1972})}\BibitemShut {NoStop}%
\bibitem [{\citenamefont {Bakin}\ \emph {et~al.}(2015)\citenamefont {Bakin},
  \citenamefont {Toropetsky}, \citenamefont {Scheibler}, \citenamefont
  {Terekhov}, \citenamefont {Jones}, \citenamefont {Militsyn},\ and\
  \citenamefont {Noakes}}]{bakin2015}%
  \BibitemOpen
  \bibfield  {author} {\bibinfo {author} {\bibfnamefont {V.~V.}\ \bibnamefont
  {Bakin}}, \bibinfo {author} {\bibfnamefont {K.~V.}\ \bibnamefont
  {Toropetsky}}, \bibinfo {author} {\bibfnamefont {H.~E.}\ \bibnamefont
  {Scheibler}}, \bibinfo {author} {\bibfnamefont {A.~S.}\ \bibnamefont
  {Terekhov}}, \bibinfo {author} {\bibfnamefont {L.~B.}\ \bibnamefont {Jones}},
  \bibinfo {author} {\bibfnamefont {B.~L.}\ \bibnamefont {Militsyn}},\ and\
  \bibinfo {author} {\bibfnamefont {T.~C.~Q.}\ \bibnamefont {Noakes}},\
  }\bibfield  {title} {\bibinfo {title} {p-gaas(cs,o)-photocathodes:
  Demarcation of domains of validity for practical models of the activation
  layer},\ }\href {https://doi.org/10.1063/1.4919447} {\bibfield  {journal}
  {\bibinfo  {journal} {Appl. Phys. Lett.}\ }\textbf {\bibinfo {volume}
  {106}},\ \bibinfo {pages} {183501} (\bibinfo {year} {2015})}\BibitemShut
  {NoStop}%
\bibitem [{\citenamefont {Sun}\ \emph {et~al.}(2009)\citenamefont {Sun},
  \citenamefont {Kirby}, \citenamefont {Maruyama}, \citenamefont {Mulhollan},
  \citenamefont {Bierman},\ and\ \citenamefont {Pianetta}}]{sun2009}%
  \BibitemOpen
  \bibfield  {author} {\bibinfo {author} {\bibfnamefont {Y.}~\bibnamefont
  {Sun}}, \bibinfo {author} {\bibfnamefont {R.~E.}\ \bibnamefont {Kirby}},
  \bibinfo {author} {\bibfnamefont {T.}~\bibnamefont {Maruyama}}, \bibinfo
  {author} {\bibfnamefont {G.~A.}\ \bibnamefont {Mulhollan}}, \bibinfo {author}
  {\bibfnamefont {J.~C.}\ \bibnamefont {Bierman}},\ and\ \bibinfo {author}
  {\bibfnamefont {P.}~\bibnamefont {Pianetta}},\ }\bibfield  {title} {\bibinfo
  {title} {The surface activation layer of gaas negative electron affinity
  photocathode activated by cs, li, and nf3},\ }\href
  {https://doi.org/10.1063/1.3257730} {\bibfield  {journal} {\bibinfo
  {journal} {Appl. Phys. Lett.}\ }\textbf {\bibinfo {volume} {95}},\ \bibinfo
  {pages} {174109} (\bibinfo {year} {2009})}\BibitemShut {NoStop}%
\bibitem [{\citenamefont {Herbert}\ \emph {et~al.}(2018)\citenamefont {Herbert}
  \emph {et~al.}}]{herbert2018}%
  \BibitemOpen
  \bibfield  {author} {\bibinfo {author} {\bibfnamefont {M.}~\bibnamefont
  {Herbert}} \emph {et~al.},\ }\bibfield  {title} {\bibinfo {title} {{I}nverted
  geometry photo-electron gun research and development at {TU} {D}armstadt},\
  }in\ \href {https://doi.org/10.18429/JACoW-IPAC2018-THPMK101} {\emph
  {\bibinfo {booktitle} {Proc. 9th International Particle Accelerator
  Conference (IPAC'18)}}},\ \bibinfo {series and number} {\bibinfo {series}
  {International Particle Accelerator Conference}\ No.~\bibinfo {number} {9}}\
  (\bibinfo  {publisher} {JACoW Publishing},\ \bibinfo {year} {2018})\ pp.\
  \bibinfo {pages} {4545--4547}\BibitemShut {NoStop}%
\bibitem [{\citenamefont {Herbert}(2022)}]{herbert2022}%
  \BibitemOpen
  \bibfield  {author} {\bibinfo {author} {\bibfnamefont {M.}~\bibnamefont
  {Herbert}},\ }{\selectlanguage {english}\emph {\bibinfo {title} {Electron
  emission from {GaAs} photocathodes using conventional and {Li}-enhanced
  activation procedures}}},\ \href
  {https://doi.org/https://doi.org/10.26083/tuprints-00020707} {\bibinfo {type}
  {Doctoral dissertation}},\ \bibinfo  {school} {Technische Universit{\"a}t
  Darmstadt}, \bibinfo {address} {Darmstadt} (\bibinfo {year}
  {2022})\BibitemShut {NoStop}%
\bibitem [{\citenamefont {Herbert}\ \emph {et~al.}(2023)\citenamefont
  {Herbert}, \citenamefont {Eggert}, \citenamefont {Enders}, \citenamefont
  {Engart}, \citenamefont {Fritzsche},\ and\ \citenamefont
  {Wende}}]{herbert2023a}%
  \BibitemOpen
  \bibfield  {author} {\bibinfo {author} {\bibfnamefont {M.}~\bibnamefont
  {Herbert}}, \bibinfo {author} {\bibfnamefont {T.}~\bibnamefont {Eggert}},
  \bibinfo {author} {\bibfnamefont {J.}~\bibnamefont {Enders}}, \bibinfo
  {author} {\bibfnamefont {M.}~\bibnamefont {Engart}}, \bibinfo {author}
  {\bibfnamefont {Y.}~\bibnamefont {Fritzsche}},\ and\ \bibinfo {author}
  {\bibfnamefont {V.}~\bibnamefont {Wende}},\ }\bibfield  {title} {\bibinfo
  {title} {{Automated Activation Procedure for {GaAs} Photocathodes at
  {Photo-CATCH}}},\ }in\ \href {https://doi.org/10.22323/1.433.0003} {\emph
  {\bibinfo {booktitle} {Proc. 19th Workshop on Polarized Sources, Targets and
  Polarimetry {\textemdash} PoS(PSTP2022)}}},\ Vol.\ \bibinfo {volume} {433}\
  (\bibinfo {year} {2023})\ p.\ \bibinfo {pages} {003}\BibitemShut {NoStop}%
\bibitem [{\citenamefont {Herbert}(2023)}]{herbert2023b}%
  \BibitemOpen
  \bibfield  {author} {\bibinfo {author} {\bibfnamefont {M.}~\bibnamefont
  {Herbert}},\ }\bibfield  {title} {\bibinfo {title} {Negative
  electron-affinity activation procedures for {GaAs} photocathodes at
  {Photo-CATCH}},\ }in\ \href
  {https://doi.org/doi:10.18429/jacow-ipac2023-tupa035} {\emph {\bibinfo
  {booktitle} {Proc. 14th International Particle Accelerator Conference
  (IPAC'23)}}},\ \bibinfo {series and number} {\bibinfo {series} {IPAC'23 -
  14th International Particle Accelerator Conference}\ No.~\bibinfo {number}
  {14}}\ (\bibinfo  {publisher} {JACoW Publishing, Geneva, Switzerland},\
  \bibinfo {year} {2023})\ pp.\ \bibinfo {pages} {1397--1399}\BibitemShut
  {NoStop}%
\bibitem [{\citenamefont {Richter}(1996)}]{richter1996}%
  \BibitemOpen
  \bibfield  {author} {\bibinfo {author} {\bibfnamefont {A.}~\bibnamefont
  {Richter}},\ }\bibfield  {title} {\bibinfo {title} {Operational experience at
  the s-dalinac},\ }in\ \href {http://tubiblio.ulb.tu-darmstadt.de/2551/}
  {\emph {\bibinfo {booktitle} {Proc. 5th European Particle Accelerator
  Conference (EPAC'96)}}},\ \bibinfo {series and number} {\bibinfo {series}
  {European Particle Accelerator Conference}\ No.~\bibinfo {number} {5}}\
  (\bibinfo  {publisher} {{IOP} Publishing},\ \bibinfo {year} {1996})\ pp.\
  \bibinfo {pages} {110--114}\BibitemShut {NoStop}%
\bibitem [{\citenamefont {Pietralla}(2018)}]{pietralla2018}%
  \BibitemOpen
  \bibfield  {author} {\bibinfo {author} {\bibfnamefont {N.}~\bibnamefont
  {Pietralla}},\ }\bibfield  {title} {\bibinfo {title} {The institute of
  nuclear physics at the tu darmstadt},\ }\href
  {https://doi.org/10.1080/10619127.2018.1463013} {\bibfield  {journal}
  {\bibinfo  {journal} {Nuclear Physics News}\ }\textbf {\bibinfo {volume}
  {28}},\ \bibinfo {pages} {4} (\bibinfo {year} {2018})}\BibitemShut {NoStop}%
\bibitem [{\citenamefont {Poltoratska}\ \emph {et~al.}(2011)\citenamefont
  {Poltoratska} \emph {et~al.}}]{poltoratska2011}%
  \BibitemOpen
  \bibfield  {author} {\bibinfo {author} {\bibfnamefont {Y.}~\bibnamefont
  {Poltoratska}} \emph {et~al.},\ }\bibfield  {title} {\bibinfo {title} {Status
  and recent developments at the polarized-electron injector of the
  superconducting darmstadt electron linear accelerator s-{DALINAC}},\ }\href
  {https://doi.org/10.1088/1742-6596/298/1/012002} {\bibfield  {journal}
  {\bibinfo  {journal} {J. Phys. Conf. Ser.}\ }\textbf {\bibinfo {volume}
  {298}},\ \bibinfo {pages} {012002} (\bibinfo {year} {2011})}\BibitemShut
  {NoStop}%
\bibitem [{\citenamefont {Arnold}\ \emph {et~al.}(2020)\citenamefont {Arnold},
  \citenamefont {Birkhan}, \citenamefont {Pforr}, \citenamefont {Pietralla},
  \citenamefont {Schlie\ss{}mann}, \citenamefont {Steinhorst},\ and\
  \citenamefont {Hug}}]{arnold2020}%
  \BibitemOpen
  \bibfield  {author} {\bibinfo {author} {\bibfnamefont {M.}~\bibnamefont
  {Arnold}}, \bibinfo {author} {\bibfnamefont {J.}~\bibnamefont {Birkhan}},
  \bibinfo {author} {\bibfnamefont {J.}~\bibnamefont {Pforr}}, \bibinfo
  {author} {\bibfnamefont {N.}~\bibnamefont {Pietralla}}, \bibinfo {author}
  {\bibfnamefont {F.}~\bibnamefont {Schlie\ss{}mann}}, \bibinfo {author}
  {\bibfnamefont {M.}~\bibnamefont {Steinhorst}},\ and\ \bibinfo {author}
  {\bibfnamefont {F.}~\bibnamefont {Hug}},\ }\bibfield  {title} {\bibinfo
  {title} {First operation of the superconducting darmstadt linear electron
  accelerator as an energy recovery linac},\ }\href
  {https://doi.org/10.1103/PhysRevAccelBeams.23.020101} {\bibfield  {journal}
  {\bibinfo  {journal} {Phys. Rev. Accel. Beams}\ }\textbf {\bibinfo {volume}
  {23}},\ \bibinfo {pages} {020101} (\bibinfo {year} {2020})}\BibitemShut
  {NoStop}%
\bibitem [{\citenamefont {Schliessmann}\ \emph {et~al.}(2023)\citenamefont
  {Schliessmann}, \citenamefont {Arnold}, \citenamefont {Juergensen},
  \citenamefont {Pietralla}, \citenamefont {Dutine}, \citenamefont {Fischer},
  \citenamefont {Grewe}, \citenamefont {Steinhorst}, \citenamefont {Stobbe},\
  and\ \citenamefont {Weih}}]{schliessmann2023}%
  \BibitemOpen
  \bibfield  {author} {\bibinfo {author} {\bibfnamefont {F.}~\bibnamefont
  {Schliessmann}}, \bibinfo {author} {\bibfnamefont {M.}~\bibnamefont
  {Arnold}}, \bibinfo {author} {\bibfnamefont {L.}~\bibnamefont {Juergensen}},
  \bibinfo {author} {\bibfnamefont {N.}~\bibnamefont {Pietralla}}, \bibinfo
  {author} {\bibfnamefont {M.}~\bibnamefont {Dutine}}, \bibinfo {author}
  {\bibfnamefont {M.}~\bibnamefont {Fischer}}, \bibinfo {author} {\bibfnamefont
  {R.}~\bibnamefont {Grewe}}, \bibinfo {author} {\bibfnamefont
  {M.}~\bibnamefont {Steinhorst}}, \bibinfo {author} {\bibfnamefont
  {L.}~\bibnamefont {Stobbe}},\ and\ \bibinfo {author} {\bibfnamefont
  {S.}~\bibnamefont {Weih}},\ }\bibfield  {title} {\bibinfo {title}
  {Realization of a multi-turn energy recovery accelerator},\ }\bibfield
  {journal} {\bibinfo  {journal} {Nature Physics}\ }\href
  {https://doi.org/10.1038/s41567-022-01856-w} {10.1038/s41567-022-01856-w}
  (\bibinfo {year} {2023})\BibitemShut {NoStop}%
\bibitem [{\citenamefont {Meier}\ \emph {et~al.}(2023)\citenamefont {Meier},
  \citenamefont {Arnold}, \citenamefont {Enders},\ and\ \citenamefont
  {Pietralla}}]{meier2023}%
  \BibitemOpen
  \bibfield  {author} {\bibinfo {author} {\bibfnamefont {M.}~\bibnamefont
  {Meier}}, \bibinfo {author} {\bibfnamefont {M.}~\bibnamefont {Arnold}},
  \bibinfo {author} {\bibfnamefont {J.}~\bibnamefont {Enders}},\ and\ \bibinfo
  {author} {\bibfnamefont {N.}~\bibnamefont {Pietralla}},\ }\bibfield  {title}
  {\bibinfo {title} {Development of a setup for laser-compton backscattering at
  the s-dalinac},\ }in\ \href {https://doi.org/10.18429/JACoW-IPAC2023-TUPL168}
  {\emph {\bibinfo {booktitle} {Proc. 14th International Particle Accelerator
  Conference (IPAC'23)}}},\ \bibinfo {series and number} {\bibinfo {series}
  {International Particle Accelerator Conference}\ No.~\bibinfo {number} {14}}\
  (\bibinfo  {publisher} {JACoW Publishing, Geneva, Switzerland},\ \bibinfo
  {year} {2023})\ pp.\ \bibinfo {pages} {2139--2142}\BibitemShut {NoStop}%
\bibitem [{epi()}]{epics2023}%
  \BibitemOpen
  \href@noop {} {\bibinfo {title} {{EPICS} - {Experimental Physics and
  Industrial Control System}}},\ \bibinfo {note}
  {\url{https://epics-controls.org/}, last accessed: 11.08.2023}\BibitemShut
  {NoStop}%
\bibitem [{css()}]{css2023}%
  \BibitemOpen
  \href@noop {} {\bibinfo {title} {{CSS} - control system studio}},\ \bibinfo
  {note} {\url{https://controlsystemstudio.org/}, last accessed:
  11.08.2023}\BibitemShut {NoStop}%
\bibitem [{\citenamefont {Kurichiyanil}(2017)}]{kurichiyanil2017}%
  \BibitemOpen
  \bibfield  {author} {\bibinfo {author} {\bibfnamefont {N.}~\bibnamefont
  {Kurichiyanil}},\ }\emph {\bibinfo {title} {Design and construction of a test
  stand for photocathode research and experiments}},\ \href
  {http://tuprints.ulb.tu-darmstadt.de/5903/} {\bibinfo {type} {Doctoral
  dissertation}},\ \bibinfo  {school} {Technische Universit{\"a}t Darmstadt},
  \bibinfo {address} {Darmstadt} (\bibinfo {year} {2017})\BibitemShut {NoStop}%
\bibitem [{\citenamefont {Aulenbacher}\ \emph {et~al.}(2005)\citenamefont
  {Aulenbacher}, \citenamefont {Arz}, \citenamefont {Barday},\ and\
  \citenamefont {Tioukine}}]{aulenbacher2005}%
  \BibitemOpen
  \bibfield  {author} {\bibinfo {author} {\bibfnamefont {K.}~\bibnamefont
  {Aulenbacher}}, \bibinfo {author} {\bibfnamefont {G.}~\bibnamefont {Arz}},
  \bibinfo {author} {\bibfnamefont {R.}~\bibnamefont {Barday}},\ and\ \bibinfo
  {author} {\bibfnamefont {V.}~\bibnamefont {Tioukine}},\ }\bibinfo {title}
  {Photocathode life time research at {MAMI}},\ in\ \href
  {https://doi.org/10.1142/9789812701909_0196} {\emph {\bibinfo {booktitle}
  {Proc. 16th International Spin Physics Symposium and Workshop on Polarized
  Electron Sources and Polarimeters (Spin 2004)}}}\ (\bibinfo {year} {2005})\
  pp.\ \bibinfo {pages} {975--979}\BibitemShut {NoStop}%
\bibitem [{\citenamefont {Bae}\ \emph {et~al.}(2022{\natexlab{b}})\citenamefont
  {Bae}, \citenamefont {Andorf}, \citenamefont {Bazarov}, \citenamefont
  {Cultrera}, \citenamefont {Galdi},\ and\ \citenamefont {Maxson}}]{bae2022b}%
  \BibitemOpen
  \bibfield  {author} {\bibinfo {author} {\bibfnamefont {J.}~\bibnamefont
  {Bae}}, \bibinfo {author} {\bibfnamefont {M.}~\bibnamefont {Andorf}},
  \bibinfo {author} {\bibfnamefont {I.}~\bibnamefont {Bazarov}}, \bibinfo
  {author} {\bibfnamefont {L.}~\bibnamefont {Cultrera}}, \bibinfo {author}
  {\bibfnamefont {A.}~\bibnamefont {Galdi}},\ and\ \bibinfo {author}
  {\bibfnamefont {J.}~\bibnamefont {Maxson}},\ }\bibfield  {title}
  {{\selectlanguage {english}\bibinfo {title} {{Optimizing Activation Recipe
  with Cs, Te, O for GaAs-Based Photocathodes}}},\ }in\ \href
  {https://doi.org/10.18429/JACoW-IPAC2022-MOPOMS004} {{\selectlanguage
  {english}\emph {\bibinfo {booktitle} {Proc. 13th International Particle
  Accelerator Conference (IPAC'22)}}}},\ \bibinfo {series and number} {\bibinfo
  {series} {International Particle Accelerator Conference}\ No.~\bibinfo
  {number} {13}}\ (\bibinfo  {publisher} {JACoW Publishing, Geneva,
  Switzerland},\ \bibinfo {year} {2022})\ pp.\ \bibinfo {pages}
  {628--630}\BibitemShut {NoStop}%
\bibitem [{\citenamefont {Rabinzohn}\ \emph {et~al.}(1984)\citenamefont
  {Rabinzohn}, \citenamefont {Gautherin}, \citenamefont {Agius},\ and\
  \citenamefont {Cohen}}]{rabinzohn1984}%
  \BibitemOpen
  \bibfield  {author} {\bibinfo {author} {\bibfnamefont {P.}~\bibnamefont
  {Rabinzohn}}, \bibinfo {author} {\bibfnamefont {G.}~\bibnamefont
  {Gautherin}}, \bibinfo {author} {\bibfnamefont {B.}~\bibnamefont {Agius}},\
  and\ \bibinfo {author} {\bibfnamefont {C.}~\bibnamefont {Cohen}},\ }\bibfield
   {title} {\bibinfo {title} {Cleaning of si and {GaAs} crystal surfaces by ion
  bombardment in the 50{\textendash}1500 {eV} range: Influence of bombarding
  energy and sample temperature on damage and incorporation},\ }\href
  {https://doi.org/10.1149/1.2115726} {\bibfield  {journal} {\bibinfo
  {journal} {J. Electrochem. Soc.}\ }\textbf {\bibinfo {volume} {131}},\
  \bibinfo {pages} {905} (\bibinfo {year} {1984})}\BibitemShut {NoStop}%
\bibitem [{\citenamefont {Jalin}\ \emph {et~al.}(1973)\citenamefont {Jalin},
  \citenamefont {Hagemann},\ and\ \citenamefont {Botter}}]{jalin1973}%
  \BibitemOpen
  \bibfield  {author} {\bibinfo {author} {\bibfnamefont {R.}~\bibnamefont
  {Jalin}}, \bibinfo {author} {\bibfnamefont {R.}~\bibnamefont {Hagemann}},\
  and\ \bibinfo {author} {\bibfnamefont {R.}~\bibnamefont {Botter}},\
  }\bibfield  {title} {\bibinfo {title} {{Absolute electron impact ionization
  cross sections of Li in the energy range from 100 to 2000 eV}},\ }\href
  {https://doi.org/10.1063/1.1680119} {\bibfield  {journal} {\bibinfo
  {journal} {J. Chem. Phys.}\ }\textbf {\bibinfo {volume} {59}},\ \bibinfo
  {pages} {952} (\bibinfo {year} {1973})}\BibitemShut {NoStop}%
\bibitem [{\citenamefont {McFarland}\ and\ \citenamefont
  {Kinney}(1965)}]{mcfarland1965}%
  \BibitemOpen
  \bibfield  {author} {\bibinfo {author} {\bibfnamefont {R.~H.}\ \bibnamefont
  {McFarland}}\ and\ \bibinfo {author} {\bibfnamefont {J.~D.}\ \bibnamefont
  {Kinney}},\ }\bibfield  {title} {\bibinfo {title} {Absolute cross sections of
  lithium and other alkali metal atoms for ionization by electrons},\ }\href
  {https://doi.org/10.1103/PhysRev.137.A1058} {\bibfield  {journal} {\bibinfo
  {journal} {Phys. Rev.}\ }\textbf {\bibinfo {volume} {137}},\ \bibinfo {pages}
  {A1058} (\bibinfo {year} {1965})}\BibitemShut {NoStop}%
\bibitem [{\citenamefont {Kim}\ and\ \citenamefont
  {Rudd}(1994)}]{kim+rudd1994}%
  \BibitemOpen
  \bibfield  {author} {\bibinfo {author} {\bibfnamefont {Y.-K.}\ \bibnamefont
  {Kim}}\ and\ \bibinfo {author} {\bibfnamefont {M.~E.}\ \bibnamefont {Rudd}},\
  }\bibfield  {title} {\bibinfo {title} {Binary-encounter-dipole model for
  electron-impact ionization},\ }\href
  {https://doi.org/10.1103/PhysRevA.50.3954} {\bibfield  {journal} {\bibinfo
  {journal} {Phys. Rev. A}\ }\textbf {\bibinfo {volume} {50}},\ \bibinfo
  {pages} {3954} (\bibinfo {year} {1994})}\BibitemShut {NoStop}%
\bibitem [{\citenamefont {Märk}(1975)}]{maerk1975}%
  \BibitemOpen
  \bibfield  {author} {\bibinfo {author} {\bibfnamefont {T.~D.}\ \bibnamefont
  {Märk}},\ }\bibfield  {title} {\bibinfo {title} {{Cross section for single
  and double ionization of N2 and O2 molecules by electron impact from
  threshold up to 170 eV}},\ }\href {https://doi.org/10.1063/1.431864}
  {\bibfield  {journal} {\bibinfo  {journal} {J. Chem. Phys.}\ }\textbf
  {\bibinfo {volume} {63}},\ \bibinfo {pages} {3731} (\bibinfo {year}
  {1975})}\BibitemShut {NoStop}%
\bibitem [{\citenamefont {Rodway}\ and\ \citenamefont
  {Allenson}(1986)}]{rodway+allenson1986}%
  \BibitemOpen
  \bibfield  {author} {\bibinfo {author} {\bibfnamefont {D.~C.}\ \bibnamefont
  {Rodway}}\ and\ \bibinfo {author} {\bibfnamefont {M.~B.}\ \bibnamefont
  {Allenson}},\ }\bibfield  {title} {\bibinfo {title} {In situ surface study of
  the activating layer on {GaAs} (cs, o) photocathodes},\ }\href
  {https://doi.org/10.1088/0022-3727/19/7/024} {\bibfield  {journal} {\bibinfo
  {journal} {J. Phys. D}\ }\textbf {\bibinfo {volume} {19}},\ \bibinfo {pages}
  {1353} (\bibinfo {year} {1986})}\BibitemShut {NoStop}%
\bibitem [{\citenamefont {Stocker}(1975)}]{stocker1975}%
  \BibitemOpen
  \bibfield  {author} {\bibinfo {author} {\bibfnamefont {B.~J.}\ \bibnamefont
  {Stocker}},\ }\bibfield  {title} {\bibinfo {title} {Aes and leed study of the
  activation of gaas-cs-o negative electron affinity surfaces},\ }\href
  {https://doi.org/10.1016/0039-6028(75)90197-1} {\bibfield  {journal}
  {\bibinfo  {journal} {Surf. Sci.}\ }\textbf {\bibinfo {volume} {47}},\
  \bibinfo {pages} {501} (\bibinfo {year} {1975})}\BibitemShut {NoStop}%
\bibitem [{\citenamefont {Togawa}\ \emph {et~al.}(1998)\citenamefont {Togawa},
  \citenamefont {Nakanishi}, \citenamefont {Baba}, \citenamefont {Furuta},
  \citenamefont {Horinaka}, \citenamefont {Ida}, \citenamefont {Kurihara},
  \citenamefont {Matsumoto}, \citenamefont {Matsuyama}, \citenamefont {Mizuta},
  \citenamefont {Okumi}, \citenamefont {Omori}, \citenamefont {Suzuki},
  \citenamefont {Takeuchi}, \citenamefont {Wada}, \citenamefont {Wada},\ and\
  \citenamefont {Yoshioka}}]{togawa1998}%
  \BibitemOpen
  \bibfield  {author} {\bibinfo {author} {\bibfnamefont {K.}~\bibnamefont
  {Togawa}}, \bibinfo {author} {\bibfnamefont {T.}~\bibnamefont {Nakanishi}},
  \bibinfo {author} {\bibfnamefont {T.}~\bibnamefont {Baba}}, \bibinfo {author}
  {\bibfnamefont {F.}~\bibnamefont {Furuta}}, \bibinfo {author} {\bibfnamefont
  {H.}~\bibnamefont {Horinaka}}, \bibinfo {author} {\bibfnamefont
  {T.}~\bibnamefont {Ida}}, \bibinfo {author} {\bibfnamefont {Y.}~\bibnamefont
  {Kurihara}}, \bibinfo {author} {\bibfnamefont {H.}~\bibnamefont {Matsumoto}},
  \bibinfo {author} {\bibfnamefont {T.}~\bibnamefont {Matsuyama}}, \bibinfo
  {author} {\bibfnamefont {M.}~\bibnamefont {Mizuta}}, \bibinfo {author}
  {\bibfnamefont {S.}~\bibnamefont {Okumi}}, \bibinfo {author} {\bibfnamefont
  {T.}~\bibnamefont {Omori}}, \bibinfo {author} {\bibfnamefont
  {C.}~\bibnamefont {Suzuki}}, \bibinfo {author} {\bibfnamefont
  {Y.}~\bibnamefont {Takeuchi}}, \bibinfo {author} {\bibfnamefont
  {K.}~\bibnamefont {Wada}}, \bibinfo {author} {\bibfnamefont {K.}~\bibnamefont
  {Wada}},\ and\ \bibinfo {author} {\bibfnamefont {M.}~\bibnamefont
  {Yoshioka}},\ }\bibfield  {title} {\bibinfo {title} {Surface charge limit in
  nea superlattice photocathodes of polarized electron source},\ }\href
  {https://doi.org/10.1016/S0168-9002(98)00552-X} {\bibfield  {journal}
  {\bibinfo  {journal} {Nucl. Instrum. Methods, Sect. A}\ }\textbf {\bibinfo
  {volume} {414}},\ \bibinfo {pages} {431} (\bibinfo {year}
  {1998})}\BibitemShut {NoStop}%
\bibitem [{\citenamefont {Agostini}\ \emph {et~al.}(2021)\citenamefont
  {Agostini} \emph {et~al.}}]{agostini2021}%
  \BibitemOpen
  \bibfield  {author} {\bibinfo {author} {\bibfnamefont {P.}~\bibnamefont
  {Agostini}} \emph {et~al.},\ }\bibfield  {title} {\bibinfo {title} {The large
  hadron–electron collider at the hl-lhc},\ }\href
  {https://doi.org/10.1088/1361-6471/abf3ba} {\bibfield  {journal} {\bibinfo
  {journal} {J. Phys. G}\ }\textbf {\bibinfo {volume} {48}},\ \bibinfo {pages}
  {110501} (\bibinfo {year} {2021})}\BibitemShut {NoStop}%
\bibitem [{\citenamefont {Dunham}\ \emph {et~al.}(2007)\citenamefont {Dunham},
  \citenamefont {Sinclair}, \citenamefont {Bazarov}, \citenamefont {Li},
  \citenamefont {Liu},\ and\ \citenamefont {Smolenski}}]{dunham2007}%
  \BibitemOpen
  \bibfield  {author} {\bibinfo {author} {\bibfnamefont {B.~M.}\ \bibnamefont
  {Dunham}}, \bibinfo {author} {\bibfnamefont {C.~K.}\ \bibnamefont
  {Sinclair}}, \bibinfo {author} {\bibfnamefont {I.~V.}\ \bibnamefont
  {Bazarov}}, \bibinfo {author} {\bibfnamefont {Y.}~\bibnamefont {Li}},
  \bibinfo {author} {\bibfnamefont {X.}~\bibnamefont {Liu}},\ and\ \bibinfo
  {author} {\bibfnamefont {K.~W.}\ \bibnamefont {Smolenski}},\ }\bibfield
  {title} {\bibinfo {title} {Performance of a very high voltage photoemission
  electron gun for a high brightness, high average current erl injector},\ }in\
  \href {https://doi.org/10.1109/PAC.2007.4441037} {\emph {\bibinfo {booktitle}
  {2007 Particle Accelerator Conference (PAC'07)}}}\ (\bibinfo {year} {2007})\
  pp.\ \bibinfo {pages} {1224--1226}\BibitemShut {NoStop}%
\bibitem [{\citenamefont {Schade}\ \emph {et~al.}(1972)\citenamefont {Schade},
  \citenamefont {Nelson},\ and\ \citenamefont {Kressel}}]{schade1972}%
  \BibitemOpen
  \bibfield  {author} {\bibinfo {author} {\bibfnamefont {H.}~\bibnamefont
  {Schade}}, \bibinfo {author} {\bibfnamefont {H.}~\bibnamefont {Nelson}},\
  and\ \bibinfo {author} {\bibfnamefont {H.}~\bibnamefont {Kressel}},\
  }\bibfield  {title} {\bibinfo {title} {Novel {GaAs–(AlGa)As} cold‐cathode
  structure and factors affecting extended operation},\ }\href
  {https://doi.org/10.1063/1.1653986} {\bibfield  {journal} {\bibinfo
  {journal} {Appl. Phys. Lett.}\ }\textbf {\bibinfo {volume} {20}},\ \bibinfo
  {pages} {385} (\bibinfo {year} {1972})}\BibitemShut {NoStop}%
\bibitem [{\citenamefont {Iijima}\ \emph {et~al.}(2010)\citenamefont {Iijima},
  \citenamefont {Shonaka}, \citenamefont {Kuriki}, \citenamefont {Kubo},\ and\
  \citenamefont {Masumoto}}]{iijima2010}%
  \BibitemOpen
  \bibfield  {author} {\bibinfo {author} {\bibfnamefont {H.}~\bibnamefont
  {Iijima}}, \bibinfo {author} {\bibfnamefont {C.}~\bibnamefont {Shonaka}},
  \bibinfo {author} {\bibfnamefont {M.}~\bibnamefont {Kuriki}}, \bibinfo
  {author} {\bibfnamefont {D.}~\bibnamefont {Kubo}},\ and\ \bibinfo {author}
  {\bibfnamefont {Y.}~\bibnamefont {Masumoto}},\ }\bibfield  {title} {\bibinfo
  {title} {A study of lifetime of {NEA}-{GaAs} photocathode at various
  temperatures},\ }in\ \href {https://epaper.kek.jp/IPAC10/papers/tupe086.pdf}
  {\emph {\bibinfo {booktitle} {Proc. 1th International Particle Accelerator
  Conference (IPAC'10)}}},\ \bibinfo {series and number} {\bibinfo {series}
  {International Particle Accelerator Conference}\ No.~\bibinfo {number} {1}}\
  (\bibinfo  {publisher} {JACoW Publishing},\ \bibinfo {year} {2010})\ pp.\
  \bibinfo {pages} {2323--2325}\BibitemShut {NoStop}%
\end{thebibliography}%

\end{document}